\documentclass{aa}  

\pdfoutput=1
\usepackage{longtable}
\usepackage{natbib,afterpage}
\usepackage{color}
\usepackage{amssymb,times}
\usepackage[english]{babel}
\usepackage{subfigure}
\usepackage{ifpdf}
\usepackage{morefloats}
\usepackage[rightcaption]{sidecap}

\newif\ifpdf \ifx\pdfoutput\undefined\pdffalse\else\pdftrue\fi
\ifpdf\pdfoutput=1
	\usepackage{graphicx}
        \usepackage{epstopdf}
	\DeclareGraphicsExtensions{.pdf,.png,.gif,.jpg}
	\else \usepackage[dvips]{color,graphicx} \fi

\def\Msun{\hbox{M$_{\odot}$}}               
\def\Lsun{\hbox{L$_{\odot}$}}               
\def\Rstar{\hbox{R$_{\star}$}}              
\def\Teff{\hbox{$\rm{T}_{\rm eff}$}}            
\def\arcsec{\hbox{$^{\prime\prime}$}}

\def\deg{\hbox{$^\circ$}}       
\def\degree{\hbox{$^\circ$}}

\begin{document} 
\selectlanguage{english}
\newcommand{\red}{\textcolor[rgb]{1,0,0}}
\newcommand{\blue}{\textcolor[rgb]{0,0,1}}

   \title{ALMA data suggest the presence of a spiral structure in the inner wind of CW~Leo}


   \author{
   L. Decin\inst{1}
          \and
          A. M. S. Richards\inst{2}
	  \and
	  D. Neufeld\inst{3}
	  \and
	  W. Steffen\inst{4}
	  \and
	  G. Melnick\inst{5}
	  \and
	  R. Lombaert\inst{1}
          }

  \offprints{Leen.Decin@ster.kuleuven.be}

  \institute{
  Instituut voor Sterrenkunde, Katholieke Universiteit Leuven, Celestijnenlaan 200D, 3001 Leuven, Belgium \\ 
  \email{Leen.Decin@ster.kuleuven.be}
  \and
 JBCA, Department Physics and Astronomy, University of Manchester, Manchester M13 9PL, UK\\
 \and
 The Johns Hopkins University, Baltimore, MD, 21218, USA\\
 \and
 Instituto de Astronom\'{\i}a, Universidad Nacional Aut\'onoma de M\'exico, Apdo.\ Postal 877, 22800, Ensenada, BC, Mexico\\
 \and
 Harvard-Smithsonian Center for Astrophysics, 60 Garden Street, MS 66, Cambridge, MA 02138, USA}

   \date{Received date; accepted date}

 
  \abstract
   {Evolved low-mass stars lose a significant fraction of their mass through a stellar wind. While the overall morphology of the stellar wind structure during the Asymptotic Giant Branch (AGB) phase is thought to be roughly spherically symmetric, the morphology changes dramatically during the post-AGB and planetary nebula phase during which often bipolar and multi-polar structures are observed. }
   {We aim to study the inner wind structure of the closest well-known AGB star CW~Leo.  Different diagnostics probing different geometrical scales have pointed toward a non-homogeneous mass-loss process for this star: dust clumps are observed at milli-arcsec scale, a bipolar structure is seen at arcsecond-scale and multi-concentric shells are detected beyond 1\arcsec.}
   {We present the first ALMA Cycle~0 band~9 data around 650\,GHz (450\,$\mu$m) tracing the inner wind of CW~Leo. The full-resolution data have a spatial resolution of 0\farcs42$\times$0\farcs24, allowing us to study the morpho-kinematical structure of CW~Leo within $\sim$6\arcsec.}
   {We have detected 25 molecular emission lines in four spectral windows. The emission of all but one line is spatially resolved. The dust and molecular lines are centered around the continuum peak position, assumed to be dominated by stellar emission. The dust emission has an asymmetric distribution with a central peak flux density of $\sim$2\,Jy. The molecular emission lines trace different regions in the wind acceleration region and suggest that the wind velocity increases rapidly from about 5\,\Rstar\ almost reaching  the terminal velocity at $\sim$11\,\Rstar. The images prove that vibrational lines are excited close to the stellar surface and that SiO is a parent molecule. The channel maps for the brighter lines show a complex structure; specifically for the $^{13}$CO J=6-5 line different arcs are detected within 
the first few arcseconds. The curved structure present in the position-velocity (PV) map of the $^{13}$CO J=6-5 line 
can be explained by a spiral structure in the inner wind of CW~Leo, probably induced by a binary companion. From modeling the ALMA data, we deduce that the potential orbital axis for the binary system lies at a position angle of $\sim$10--20\deg\ to the North-East and that the spiral structure is seen almost edge-on. We infer an orbital period of 55\,yr and a binary separation of 25\,au (or $\sim$8.2\,\Rstar). We tentatively estimate that the companion is an unevolved low-mass main-sequence star.}
   {A scenario of a binary-induced spiral shell can explain the correlated structure seen in the ALMA PV images of CW~Leo. Moreover, this scenario can also explain many other observational signatures seen at different spatial scales and in different wavelength regions, such as the bipolar structure and the almost-concentric shells. The ALMA data hence provide us for the first time with the crucial kinematical link between the dust clumps seen at milli-arcsecond scale and the almost concentric arcs seen at arcsecond scale.}

   
\titlerunning{Edge-on spiral detected with ALMA in the inner wind of CW~Leo}
  \maketitle

\section{Introduction}\label{Sec:introduction}

CW~Leo  (a.k.a. IRC\,+10216) is the nearest carbon-rich Asymptotic Giant Branch (AGB) star at a distance around 120 to 150\,pc \citep[][and reference therein]{Menshchikov2001A&A...368..497M, Groenewegen2012A&A...543L...8G}. During the AGB-phase, the evolution is determined by mass loss, with the mass-loss rate being significantly larger than the nuclear burning rate \citep{Lagadec2008MNRAS.383..399L}. The general accepted idea is that mass-loss proceeds via a 2-steps process: (1)\, pulsations lift photospheric layers to temperatures where it is cool and dense enough for molecules to nucleate and form dust species, (2)\, radiation pressure on the dust grains pushes the material outwards and a wind is generated. 
 However, even for the nearest carbon-rich AGB star it is still not well established if the dominant behaviour of the mass-loss is via homogeneous, isotropic processes or if small-scale irregularities/instabilities dictate the overall appearance. 
 
 Specifically for CW Leo, it has been shown that at \textit{milli-arcsecond scale} dust clumps formed in the dust formation region are moving due to the expanding wind of the star, with clump velocity motions varying between 7.9 and 17.5\,km/s \citep{Tuthill2000ApJ...543..284T, Weigelt2002A&A...392..131W, Menut2007MNRAS.376L...6M}. Formation of new dust clumps at irregular time intervals might explain the fading of the illumination of some dust features during some periods. At \textit{arcsecond-scale}, a bipolar morphology is seen in optical data with a bipolar axis at a position angle estimated around 8\deg\ \citep{Mauron2000A&A...359..707M} to 22\deg\ \citep{Leao2006A&A...455..187L} (with respect to the North). This feature probably indicates that scattering from stellar photons is more efficient in the polar direction. Evidence was provided from optical and near-infrared data for an equatorial density enhancement in the form of a geometrically thick disk \citep{Dyck1987PASP...99...99D} or dusty 
torus \citep{Skinner1998MNRAS.300L..29S, Murakawa2005A&A...436..601M, Jeffers2014} seen almost edge-on. 
 \textit{Beyond 1\arcsec}, multiple, almost concentric, shells (or arcs) are detected on top of the smooth extended envelope \citep{Mauron1999A&A...349..203M, Mauron2000A&A...359..707M, Leao2006A&A...455..187L, Dinh2008ApJ...678..303D, Fong2003ApJ...582L..39F, Decin2011A&A...534A...1D}. The nested shells are composed of thinner, elongated arcs, with a typical width of $\sim$1.6\arcsec\ and separated by $\sim$3\arcsec-20\arcsec, corresponding to $\sim$150--1000\,yr. These partial shells have a typical angular extent of $\sim$30\deg\ to 90\deg\  and are detected up to a distance of 320\arcsec\ \citep{Decin2011A&A...534A...1D}. Around 450\arcsec, Herschel data have revealed the presence of a bow shock \citep{Ladjal2010A&A...518L.141L}.
 
  In this paper, we present the first ALMA band 9 observations of the inner envelope of CW~Leo at a spatial resolution of $\sim$0.2\arcsec\ \citep[$\sim$10\,\Rstar, with 1\Rstar$\sim 5 \times 10^{13}$\,cm;][]{DeBeck2012A&A...539A.108D}. 
 In Sec.~\ref{Sec:ALMA_data} we describe the ALMA observations and data reduction including imaging and image fidelity. Sec.~\ref{Sec:astronometry} reports on the astrometry and continuum properties. The ALMA spectra are described in Sec.~\ref{Sec:spectral_results} and the channel maps and PV-diagrams are shown in Sec.~\ref{Sec:data_results}. A qualitative interpretation of the ALMA $^{13}$CO J=6-5 channel maps is given in Sec.~\ref{Sec:qualitative}. Preliminary morpho-kinematical modeling of the ALMA data is presented in Sec.~\ref{Sec:Shape}.  The results are discussed in Sec.~\ref{Sec:Discussion} and the conclusions are given in Sec.~\ref{Sec:conclusions}.


\section{ALMA observations and data reduction} \label{Sec:ALMA_data}

  \subsection{ALMA observations}
  IRC\,+10216 was observed by ALMA on 1 December 2012 for proposal code
2011.0.00277.S (Cycle~0).  The total time on IRC\,+10216 was 17.5\,min spread over 35\,min. The
precipitable water vapour was 0.55\,mm.  The 18 antennas provided good
data, with minimum and maximum baselines of 25 -- 340\,m.

Four 1.875-GHz frequency bands (spectral windows, spw)
were used for science, centred at 643.6645, 646.4827, 658.0587 and
661.1205\,GHz (ALMA Band 9). Each spectral window was divided into 3840 channels of 488\,kHz. After Hanning smoothing, 
the effective velocity resolution is 0.455\,km s$^{-1}$.  The usual ALMA instrumental corrections were applied including
those derived from water vapour radiometry and system temperature
measurements.  The Cycle 0 flux scale was derived with respect to Titan. The
flux density scale is formally accurate to 20\% in Band 9; the accuracy of our data might even be 
better, as the flux densities derived for other calibration sources were consistent with values measured during
other observations.  The compact, bright QSO 3C279 and J0854+201 were
used as bandpass calibration and phase reference sources,
respectively; both were also used to calibrate the pointing. 

  \subsection{Data reduction} \label{Sec:data_reduction}
  
  Normal data reduction procedures were performed in CASA \citep{McMullin2007ASPC..376..127M, 2011ascl.soft07013I} \footnote{Credit: International consortium of scientists based at the National
Radio Astronomical Observatory (NRAO), the European Southern
Observatory (ESO), the National Astronomical Observatory of Japan
(NAOJ), the CSIRO Australia Telescope National Facility (CSIRO/ATNF),
and the Netherlands Institute for Radio Astronomy (ASTRON) under the
guidance of NRAO.}.  Inspection of the phase reference source data showed that the phase of
the first few scans could be connected smoothly, without wrapping, but
there were apparent discontinuities in the second half of the data.
We applied the phase-reference solutions to the target data and
inspected the first half of the corrected data to identify the
line-free channels.  We used these to make a continuum image, which provided a model for
self-calibration of all the target continuum data. The residual
scatter in the final self-calibration solutions was about 5\deg\ in
phase and 10\% in amplitude.  Using the expressions from \citet{Perley1999ASPC..180..275P},
this corresponds to a dynamic range limitation of $\sim$150--300, depending
on whether the errors are correlated between scans. We applied the solutions to all the channels of the target data. Finally, we made a preliminary cube to refine the selection of
continuum channels, subtracted these and made the final line and
continuum images. 

We used CASA to Fourier transform
   and clean each channel to produce image cubes for all spectral
   windows. During this process, we corrected for the motion of the Earth relative to the Local Standard of Rest.
    For the identified
   lines, we extracted the channels covering that line and set the
   velocity to zero at its rest frequency. The noise in the image cubes is 0.2 -- 0.3\,Jy beam$^{-1}$ per 488\,kHz
channel\footnote{The requested sensitivity of our Cycle~0 proposal was 0.04\,Jy beam$^{-1}$ per km, which would have required six times as
 long on-target.}. The noise is worst near the 658\,GHz atmospheric water line
and is also higher in bright channels due to dynamic range
limitations, as well as being slightly increased by continuum
subtraction and by residual calibration errors.

The short duration of the observations provided sparse visibility
plane coverage, most noticeable at the highest spectral resolution.
The full-resolution images were made with partial uniform weighting to
give a synthesized beam 0\farcs42$\times$0\farcs24 for a position angle of 117\deg\ (measured
anticlockwise from north). This is similar to the natural beam
position angle, but improves the resolution without increasing the
noise appreciably\footnote{Natural weighting simply takes each data
point at the sampled position in the visibility plane; uniform
weighting interpolates into the missing spacings and the intermediate
scheme used here gives more weight depending on the local sample
density.}. A beam of
0\farcs4$\times$0\farcs4 was used for the 3-channel averaged images which have a resolution of $\sim$0.7\,km/s and a noise value of 0.13--0.18\,Jy beam$^{-1}$ (see Sect.~\ref{Sec:channel_maps}). 

The shortest baselines in our observations impose a limit of
$\sim3$\arcsec\ on the size of any structure which can be imaged
accurately. Emission on scales 3\arcsec--6\arcsec\ is imperfectly sampled and
emission that is smoothly extended over $>6$\arcsec\ is undetectable, even
if it is much brighter than our limit ($\sim1$ Jy beam$^{-1}$ at full
resolution) for compact emission. 

These limitations (dynamic range and sparse visibility plane coverage)
produce two kinds of artifacts. The dynamic range limitations are
apparent as residual sidelobes of the synthesized beam: the intensity is
proportional to the channel peak and it is placed symmetrically w.r.t.\ the channel peak at a constant (scaled
with frequency) spacing.  We took care to check that any apparently symmetric
   structures were not in fact co-incident with sidelobes. The effects of poorly sampled emission are
more complicated, being the convolution of the actual distribution
with the sidelobes. If the true distribution would for example be a disc $>$6\arcsec\
across in a given channel, the inner $\sim$3\arcsec\ would be faithfully
imaged but at larger separations negative and positive ring-like
artifacts (known as the `cereal bowl' effect) might be visible.
Nevertheless,  smaller and larger rings (or patches of emission) seen in the channel maps  with a size in the  3\arcsec--6\arcsec\ range
result from actual structures with a smaller or larger size, respectively, which allows one to at least partly reconstruct the actual morphological structures.

\subsection{Position measurements} \label{Sec:pos_measurement}

We measured the positions and extents of emission by fitting 2D
Gaussian components to the continuum peak and to the zeroth moment of
each line (see Sect.~\ref{Sec:astronometry}).  The noise-based uncertainty (for a
sparsely-filled snapshot) is given by the beam size divided by the
signal-to-noise ratio (before self-calibration), and gives the uncertainty
in comparing positions within the same data set.  However, if
the actual flux distribution is non-Gaussian, the uncertainties are larger.

\section{Imaging results} \label{Sec:astronometry}

In this section, we discuss the astrometry and stellar and dust properties of CW~Leo as deduced from the ALMA images in band 9.

\subsection{Astrometry}

The position of the 650\,GHz continuum peak is 09:47:57.4553
+13:16:43.749 (J2000), epoch 2012.92.  Four factors affect the
astrometry. Uncertainties in the antenna positions and the phase
reference source coordinates are less than a few mas.  The noise-based
uncertainty (for a sparsely-filled snapshot) is given by the beam size
divided by the signal-to-noise ratio (before self-calibration),
$\sim$$0\farcs02$ for the continuum peak. The uncertainty in measuring the extent of
emission is $\sim$2 times the position uncertainty. The phase reference source is
14$^{\circ}$ from IRC\,+10216 and the timescales of phase changes due to
the atmosphere leads to the dominant uncertainty of $0\farcs25$.
 \citet{Menten2012A&A...543A..73M} used the VLA in 2006.16 to measure the position and
proper motion of IRC\,+10216 which would give 09:47:57.4417
+13:16:43.896 in 2012.92 (uncertainty $0\farcs012$).  The discrepancy
from the position measured by ALMA is ($0\farcs20$, $-0\farcs15$),
i.e.\ within the uncertainties.

The position fitted to the zeroth moment of each line apart from
$^{13}$CO is within 30 mas (the combined position uncertainties) of
the continuum peak; the mean position of all these lines is within 6
mas.  The $^{13}$CO peak is offset by ($-$67, $-$37) mas, but this is
probably due to the relatively poor fit of a Gaussian component to the
complex and extended CO emission (see Sec.~\ref{Sec:channel_maps}).

\subsection{Stellar and dust properties} \label{Sec:dust}

\begin{figure}[htp]
 \includegraphics[width=0.48\textwidth]{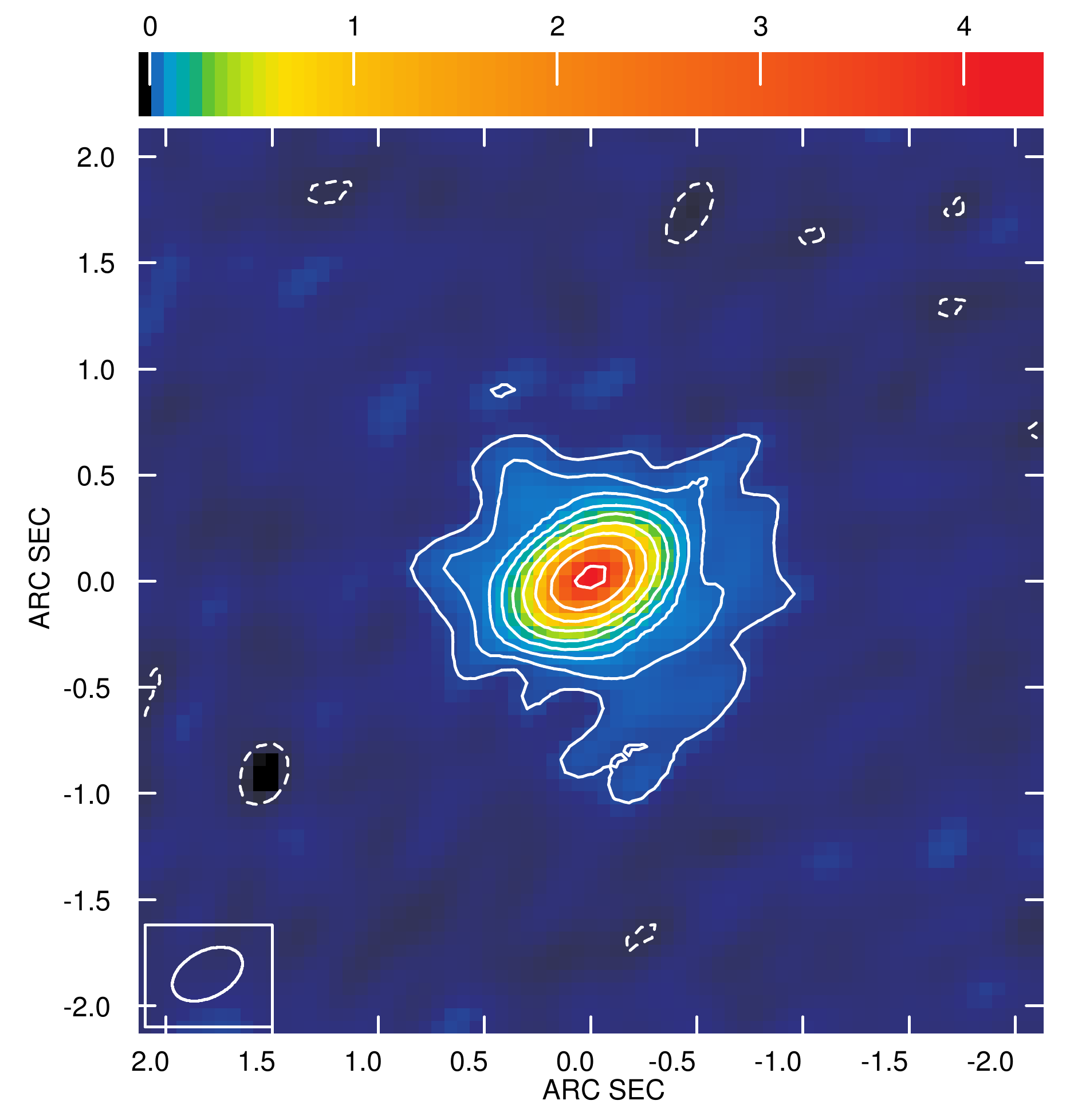}
  \caption{Dust continuum image of CW~Leo for a beam of $0\farcs42\times0\farcs24$ (shown as an ellipse in the bottom left corner). The contour levels are at [$-$1, 1, 2, 4, 8, 16, 32, 64, 128, 256] $\times$ 15\,mJy beam$^{-1}$ (3\,${\sigma_{\mathrm{rms}}}$).}
 \label{Fig:dust_continuum}
\end{figure}

The total flux density within the $3\sigma$ contour is 5.66\,Jy, see Fig.~\ref{Fig:dust_continuum}.
The continuum peak is 4.362\,Jy beam$^{-1}$ with a noise uncertainty of
0.023\,Jy beam$^{-1}$ and a calibration uncertainty $<$0.87\,Jy\,beam$^{-1}$. The size (deconvolved from the beam) is $120\pm4$\,mas by
$108\pm5$\,mas at a position angle (PA) of $128\pm20^{\circ}$.  The ALMA total flux density at 650\,GHz is slightly higher, although within the uncertainty limits, than the continuum emission flux density derived by
\citet{Young2004ApJ...616L..51Y} using the SMA at 680\,GHz, who found a compact unresolved component in their $\sim$2\arcsec\ full width at half maximum (FWHM) beam with a flux density of $3.9\pm1.2$\,Jy. 

The stellar photospheric emission contributes part of the total flux density at 650\,GHz. Following \citet{Jenness2002MNRAS.336...14J}, the ALMA data were taken near maximum light (at phase $\varphi$\,=\,0.98), which would imply a stellar luminosity of $\sim$15\,600\,\Lsun\ according to \citet{DeBeck2012A&A...539A.108D}. For an effective temperature of 2330\,K \citep{DeBeck2012A&A...539A.108D}, this yields a stellar diameter of 48\,mas. 
\citet{Groenewegen1997A&A...322L..21G} gives a stellar diameter of 70.2\,mas at 243\,GHz for an effective temperature of
$T_{\mathrm{eff}}$ of 2000\,K. \citet{Menten2012A&A...543A..73M}  found a diameter of 83\,mas at 43\,GHz. Thus, the star is likely to be unresolved at our spatial resolution.

We used two approaches to estimate the stellar flux contribution at 650\,GHz. First, using the IRAM Plateau de Bure Interferometer (PdBI) \citet{Lucas1999IAUS..191..305L} found a `point source' at 89\,GHz (3.5\,mm) and 242\,GHz (1.3\,mm) with flux densities of $65\pm7$\,mJy and $487\pm70$\,mJy, respectively. They identified the point source as being the photospheric emission. Using a spectral index of $1.96$ ($\pm0.04$)  as derived by \citet{Menten2006}, this yields a stellar contribution of 3.3\,Jy at 650\,GHz. The uncertainty on this value is estimated around 20\% and arises from the variability of the pulsating star and the uncertainties on the spectral index.
Second, the stellar contribution at 650\,GHz can also be calculated from the effective temperature and stellar diameter, under the assumption of a blackbody spectrum. Using the values obtained by \citet{DeBeck2012A&A...539A.108D} at maximum luminosity (see previous paragraph), the stellar flux at 650\,GHz is 1.3\,Jy, while the values derived by \citet{Groenewegen1997A&A...322L..21G} yield a stellar flux of 2.3\,Jy. 
The difference in stellar flux estimates mainly arises from the uncertainties in the angular diameter; the uncertainty on the effective temperature only gives on uncertainty in the flux of 0.2\,Jy.
These values for the stellar contribution are a factor $\sim$1.4--2.5 lower than the value obtained by extrapolating the `point-source' values of \citet{Lucas1999IAUS..191..305L}. This might mean that part of the 'point-source' flux densities at 89\,GHz (3.5\,mm) and 243\,GHz (1.3\,mm) might not come from the central star, resulting in too high an estimate for the stellar contribution at 650\,GHz when using the data of \citet{Lucas1999IAUS..191..305L}. Following \citet{Reid1997ApJ...476..327R, Groenewegen1997A&A...317..503G, Groenewegen1997A&A...322L..21G},  we suggest that free-free emission at wavelengths beyond 800\,$\mu$m might (at least partly) contribute to the high flux density values of \citet{Lucas1999IAUS..191..305L}. 

The estimated stellar flux density at 650\,GHz of $\sim$2.3\,Jy \citep[adopting the parameters of][]{Groenewegen1997A&A...322L..21G} is significantly lower than the flux density of 5.66\,Jy obtained from the ALMA data. Since free-free emission is negligible at 650\,GHz \citep{Groenewegen1997A&A...317..503G}, we conclude that the extended emission seen in the ALMA data is  due to dust emission alone. 

\begin{figure}[htp]
 \includegraphics[width=.48\textwidth]{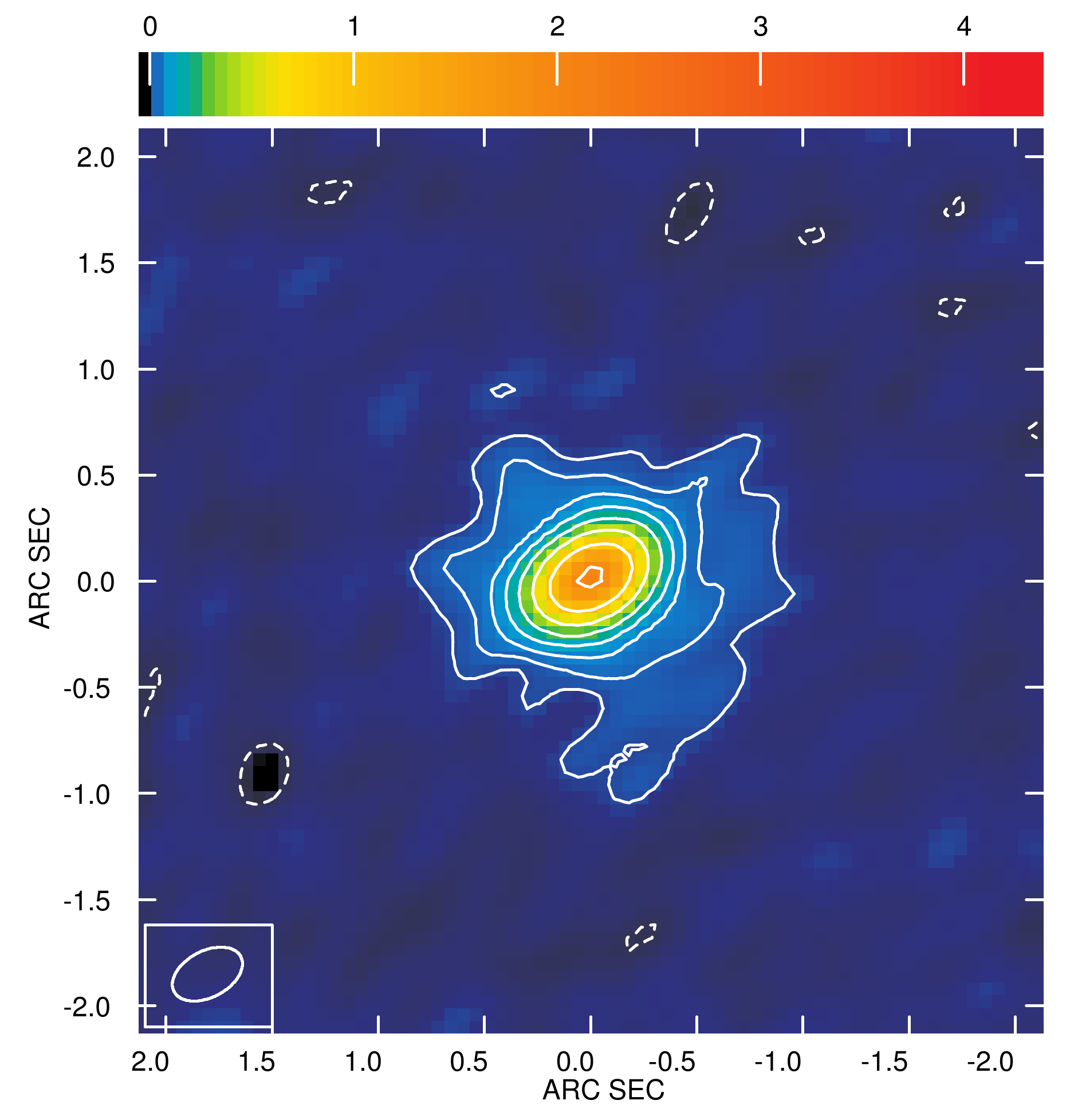}
  \caption{Flux density after subtracting a 2.3\,Jy star. The contour levels are at [$-$1, 1, 2, 4, 8, 16, 32, 64, 128] $\times$ 15\,mJy beam$^{-1}$ (3\,${\sigma_{\mathrm{rms}}}$). }
 \label{Fig:Flux-4}
\end{figure}

We estimated the 650\,GHz emission from dust by subtracting a 2.3\,Jy point source, convolved with the natural synthesised beam, from the continuum image, leaving a residual 3.36\,Jy (see Fig.~\ref{Fig:Flux-4}). The main sources of uncertainty are the 20\% uncertainty on the ALMA data and the estimated stellar flux density. The residual emission is extended  and clearly shows an asymmetric distribution, with a central peak flux density of $\sim$2\,Jy. The lowest 3-sigma
contour in Fig.~\ref{Fig:Flux-4} corresponds to a position uncertainty $\sim$1/3 of the beam size,
so although the apparent Southern bifurcation may be unreliable the
overall extension is real. The inner emission appears to be elongated in a direction similar to the $\sim$128\deg\ position angle of the equatorial
density enhancement detected in scattered light and polarimetry data \citep{Skinner1998MNRAS.300L..29S, Murakawa2005A&A...436..601M, Jeffers2014}. However, this is close to the direction of
elongation of the natural synthesised beam, and may be coincidence.
The faintest emission is, nonetheless, elongated in an almost
orthogonal direction, at a position angle of $\sim$20\deg\ and 200\deg, corresponding to
the biconical cone detected by \citet{Skinner1998MNRAS.300L..29S, Mauron2000A&A...359..707M} (see Sect.~\ref{Sec:qualitative}).

The dust nucleation zone starts around 5\,\Rstar\ \citep[or 100-150\,mas,][]{Decin2010A&A...518L.143D}. This is comparable to the resolution of these images, but we can not distinguish faint details so close to the strong stellar peak. The
outer radius of the dust detectable by ALMA is set by the radius at
which it becomes too diffuse or too cool to be detectable. 
Subtracting any realistic estimate for the
stellar contribution (1.3--3.3\,Jy) leaves an approximately flat or centrally-peaked flux
distribution within the inner $\sim$160\,mas radius, suggesting that the dust could be close to optically thick. In the case that the stellar contribution would be $\ga$4\,Jy, the dust emission is characterized by a central hollow.

Using the continuum image after subtracting a 2.3\,Jy star, assuming that the dust is optically thin and for an opacity $\kappa_\lambda$ at 650\,GHz of 90\,cm$^2$/g \citep{Mennella1998ApJ...496.1058M} and a dust temperature distributed as derived by \citet{Decin2010Natur.467...64D}, we obtain a dust mass within 0.8\arcsec\ of 3--6$\times$10$^{-7}$\,\Msun, corresponding to a dust mass loss rate of 1.1--2.2$\times$10$^{-8}$\,\Msun/yr.  This estimate for the dust mass-loss rate is a factor $\sim$4 lower than derived by \citet{Decin2010Natur.467...64D} from modeling the spectral energy distribution. The large error margin is due to the flux scale uncertainty, the unknown stellar contribution, and the unknown composition of the dust and geometry in the inner regions. Assuming a constant dust velocity equal to the terminal velocity of 14.5\,km/s, this yields a dust optical depth of $\sim$0.1. Since the dust velocity might be overestimated in the wind acceleration region, the derived value for the 
dust optical depth might be an underestimate by a factor of a few. Future multi-wavelength high-resolution ALMA observations will provide a spectral index map  and thus a better measurement of the optical depth.

\section{Spectral results} \label{Sec:spectral_results}

In this section, we present the ALMA spectra of CW Leo in band~9 around 650\,GHz (Sect.~\ref{Sec:ALMA_spectra}). For each detected emission line, we determine the line strength, line width and spatial extension. The line strengths are compared to the line strengths as deduced from Herschel/HIFI observations in the same frequency window (Sect.~\ref{Sec:Herschel_ALMA}). The line widths and spatial extension are used to determine the velocity structure of CW~Leo in the inner wind region, where the wind is accelerated from the sound velocity to the terminal velocity (Sect.~\ref{Sec:vgas}).

  \subsection{ALMA spectra of CW~Leo in band 9} \label{Sec:ALMA_spectra}
  
   \begin{figure*}[htp]
     \begin{minipage}[t]{.48\textwidth}
        \centerline{\resizebox{1.2\textwidth}{!}{\includegraphics[angle=180]{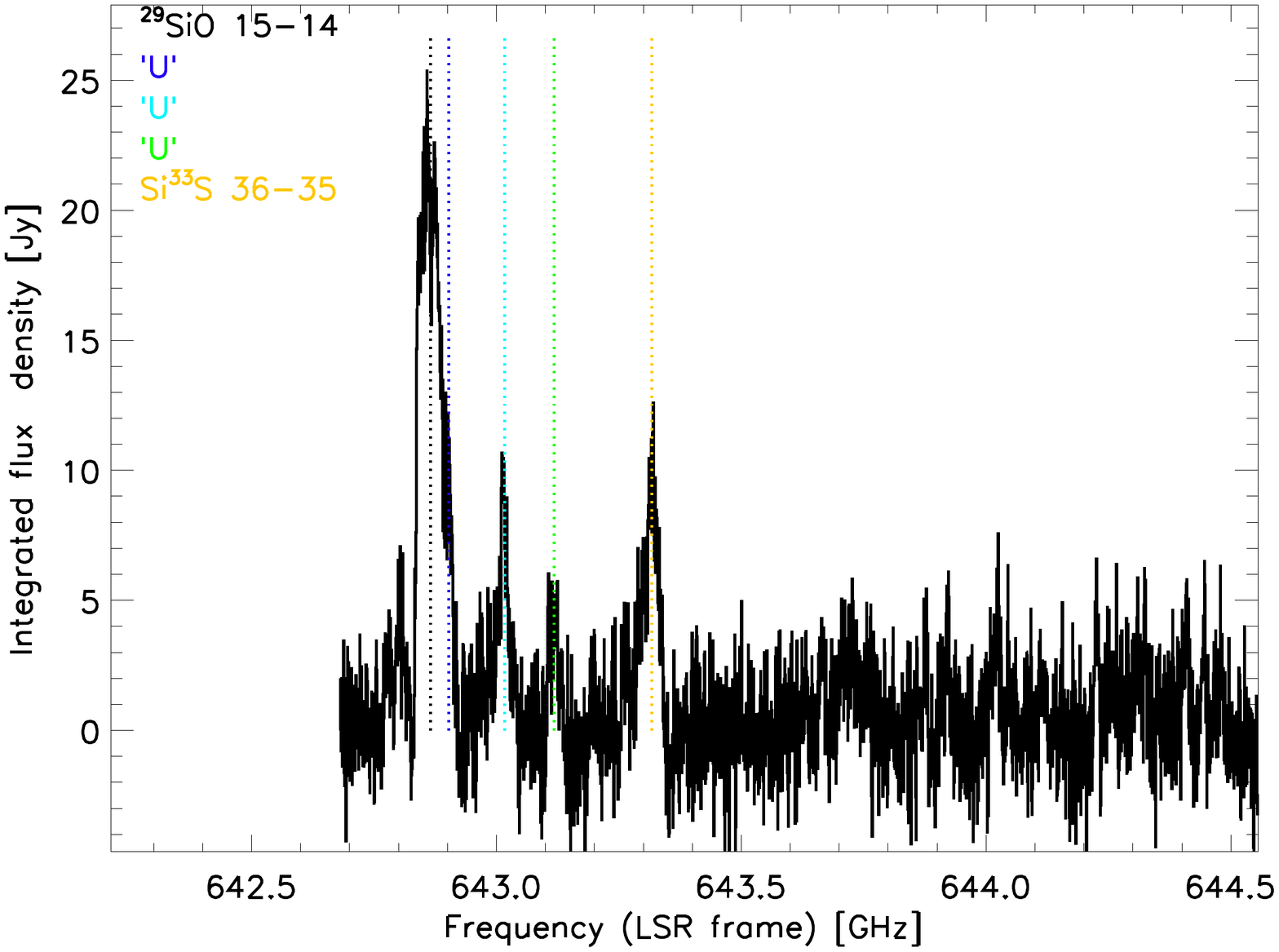}}}
    \end{minipage}
    \hfill
    \begin{minipage}[t]{.48\textwidth}
\centerline{\resizebox{1.2\textwidth}{!}{\includegraphics[angle=180]{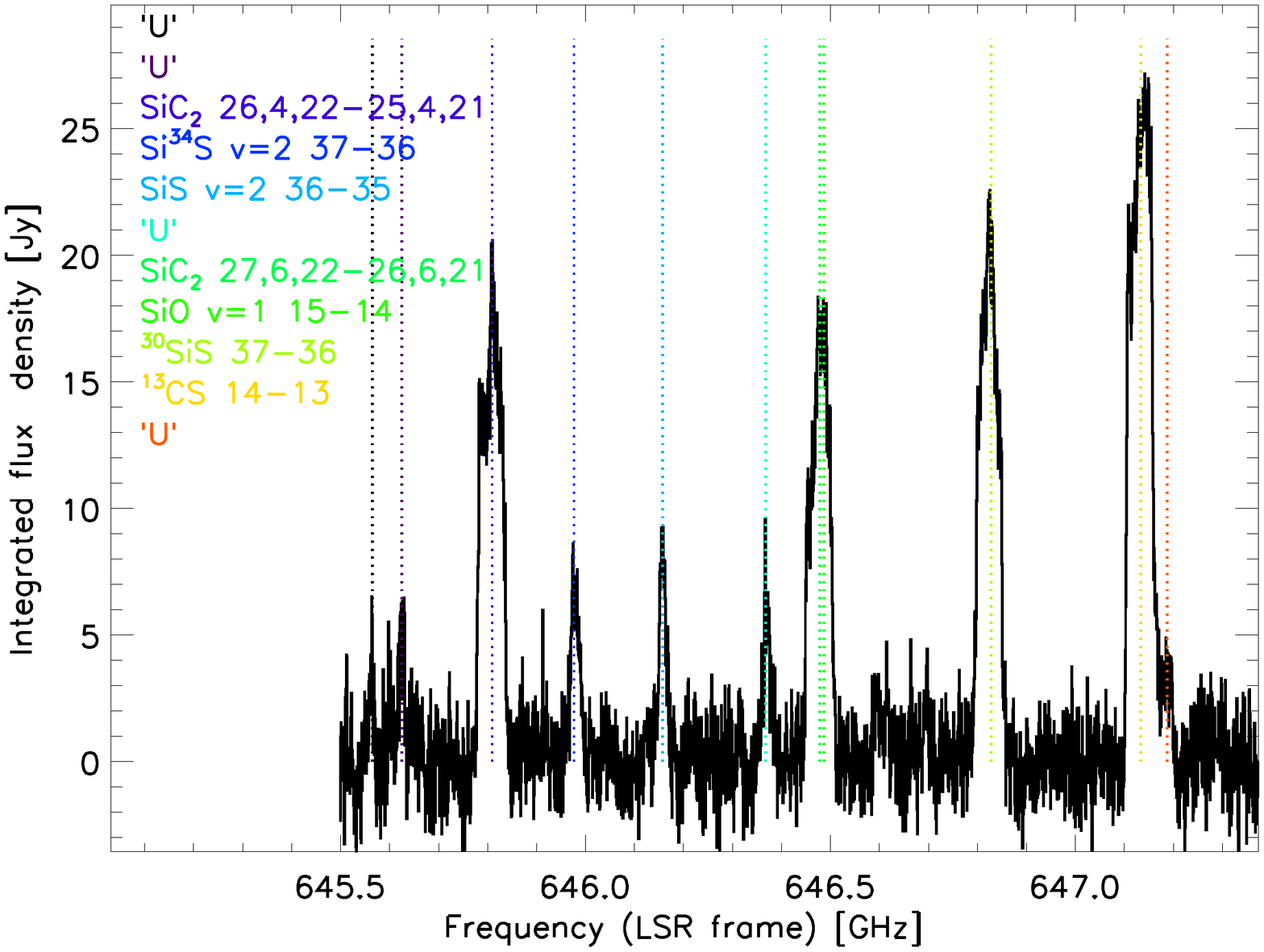}}}
     \end{minipage}
     \begin{minipage}[t]{.48\textwidth}
    \vspace*{-1cm}
       \centerline{\resizebox{1.2\textwidth}{!}{\includegraphics[angle=180]{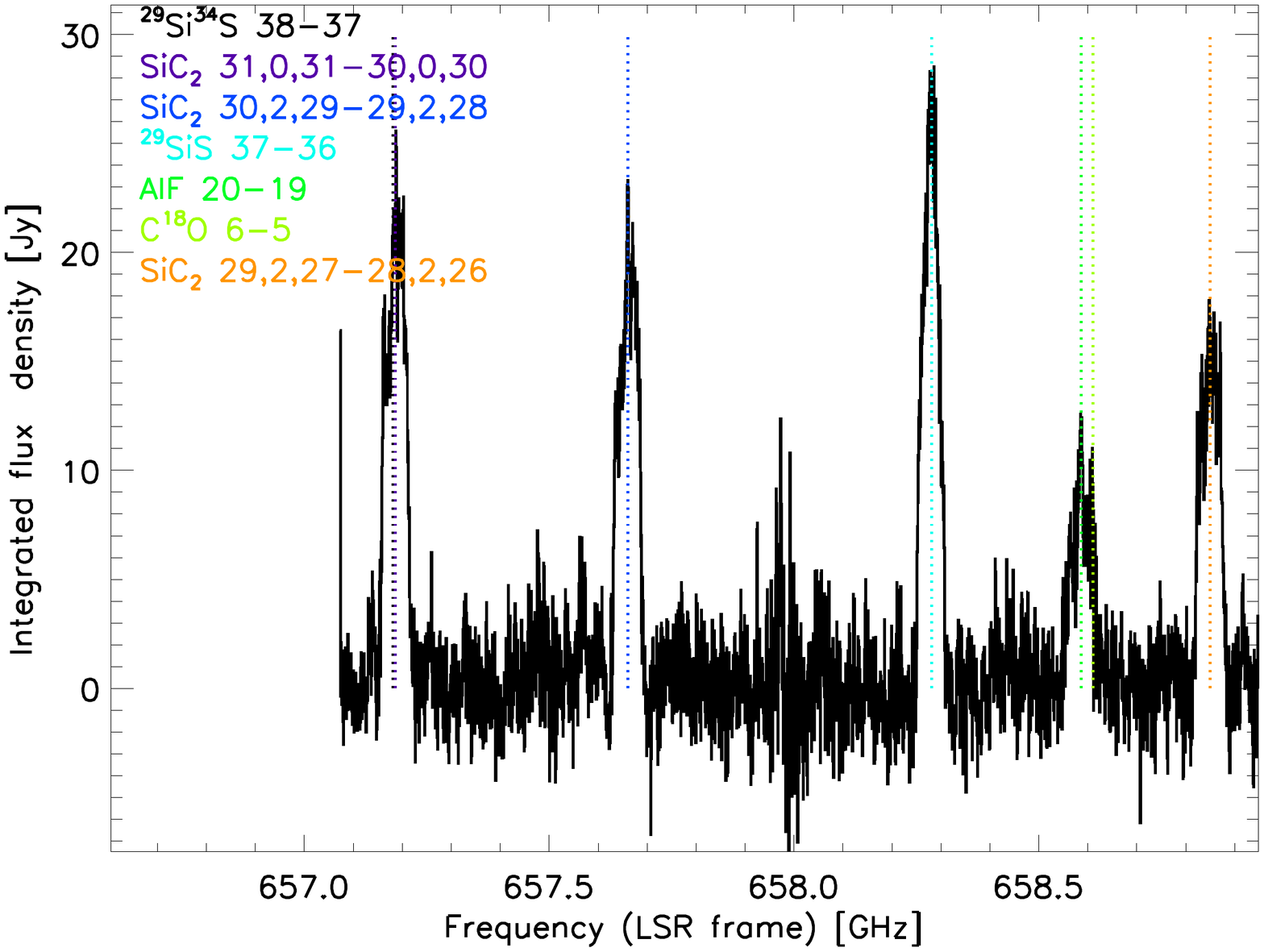}}}
    \end{minipage}
    \hfill
    \begin{minipage}[t]{.48\textwidth}
   \vspace*{-1cm}
 \centerline{\resizebox{1.2\textwidth}{!}{\includegraphics[angle=180]{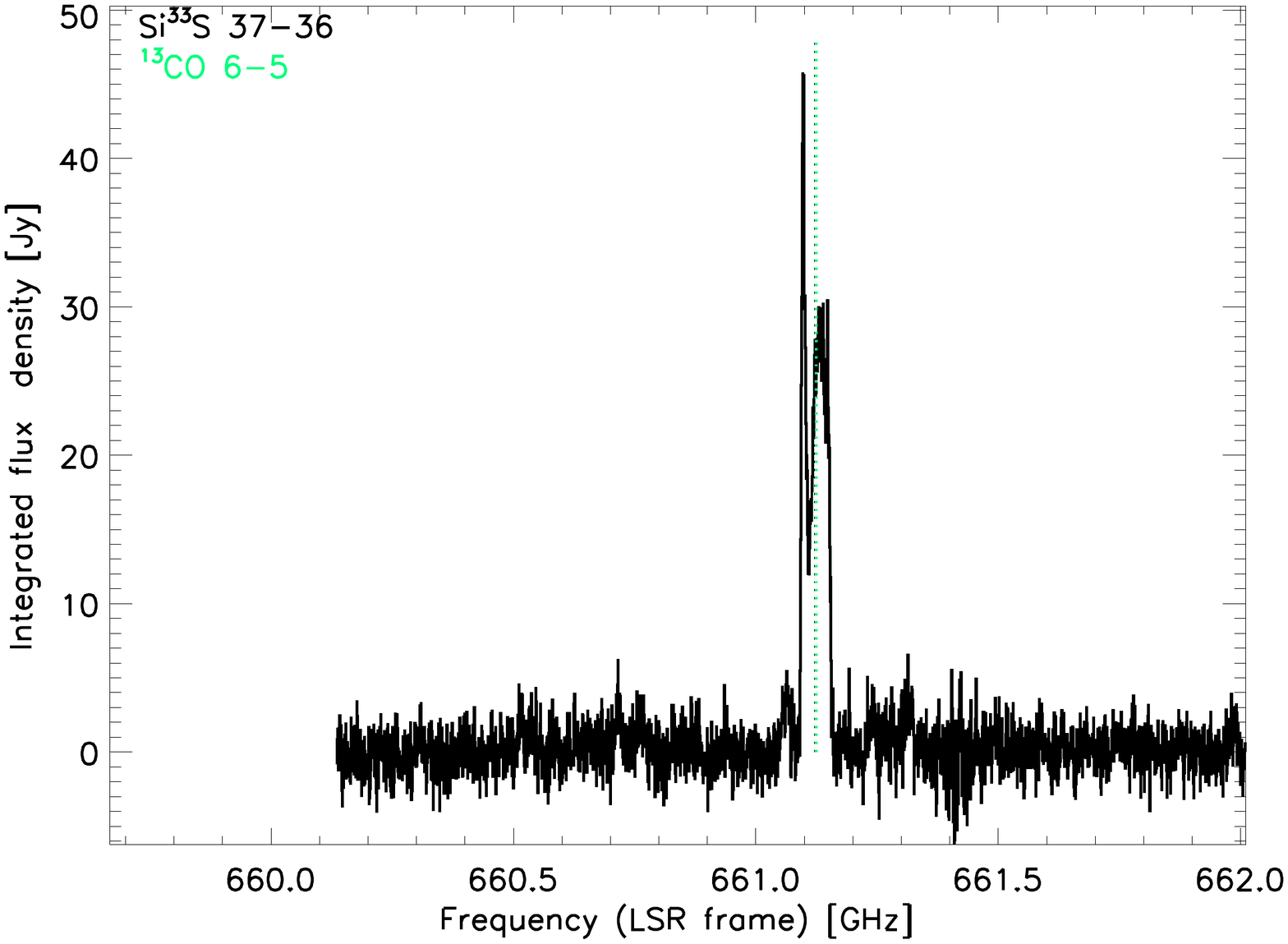}}}
    \end{minipage}
    \vspace*{-2ex}
    \caption{Integrated flux density of the 4 spectral setups extracted in a circle of 1.2\arcsec\ in radius centred on the continuum peak. The molecular lines are indicated in color; unidentified lines are marked as 'U'. Note that the noise around 658\,GHz is due to the strong atmospheric opacities at this frequency. The frequency values are corrected for the Earth's motion.}
    \label{Fig:spectra_ALMA}
\end{figure*}

Fig.~\ref{Fig:spectra_ALMA} shows the spectra of the four spectral bands extracted using a circular aperture of 1.2\arcsec\ centered on the dust continuum peak. In the ALMA data, 25 molecular line transitions are detected, of which 7 are unidentified (see Table~\ref{Table:FWHM}). The identified emission lines arise from $^{13}$CO, C$^{18}$O, SiO, $^{29}$SiO, SiS, $^{29}$SiS, $^{30}$SiS, Si$^{33}$S, Si$^{34}$S, $^{29}$Si$^{34}$S, SiC$_2$, $^{13}$CS, and AlF.

  To determine the  velocity structure in the wind (from the half width of the emission lines at continuum level) and the central line frequency, each line has been fitted using the soft-parabola line profile \citep[Eq.~1 in][]{Olofsson1993ApJS...87..267O} (see Table~\ref{Table:FWHM}). The soft-parabola fit to the ALMA data yields a mean difference between the rest frequencies and the ALMA frequencies of $-$57.4\,MHz or $-26.41$\,km/s, in accordance with the local standard of rest velocity derived by \citet{DeBeck2012A&A...539A.108D} from the high-spectral resolution Herschel/HIFI data. 
  The  half width of the emission lines, $v_{\rm{line}}$, is determined by the expansion velocity, $v_{\rm{exp}}(r)$, and turbulent velocity, $v_{\rm{turb}}(r)$, with $r$ being the radial distance. The turbulent velocity in stellar winds is not well established.  \citet{Keady1988ApJ...326..832K} propose a turbulent velocity in the inner wind region of CW~Leo of $\sim$5\,km/s and a terminal turbulent velocity of $\sim$1.0\,km/s; \citet{DeBeck2012A&A...539A.108D} derived a value of 1.5\,km/s from a fit to high-resolution Herschel/HIFI CO lines.  
  The derived half 
line widths, $v_{\rm{line}}$, range between $\sim$3.5\,km/s and $\sim$15\,km/s. The velocity structure as derived from the ALMA data is discussed in detail in Sect.~\ref{Sec:vgas}.

\begin{figure}[htp]
 \includegraphics[width=.48\textwidth]{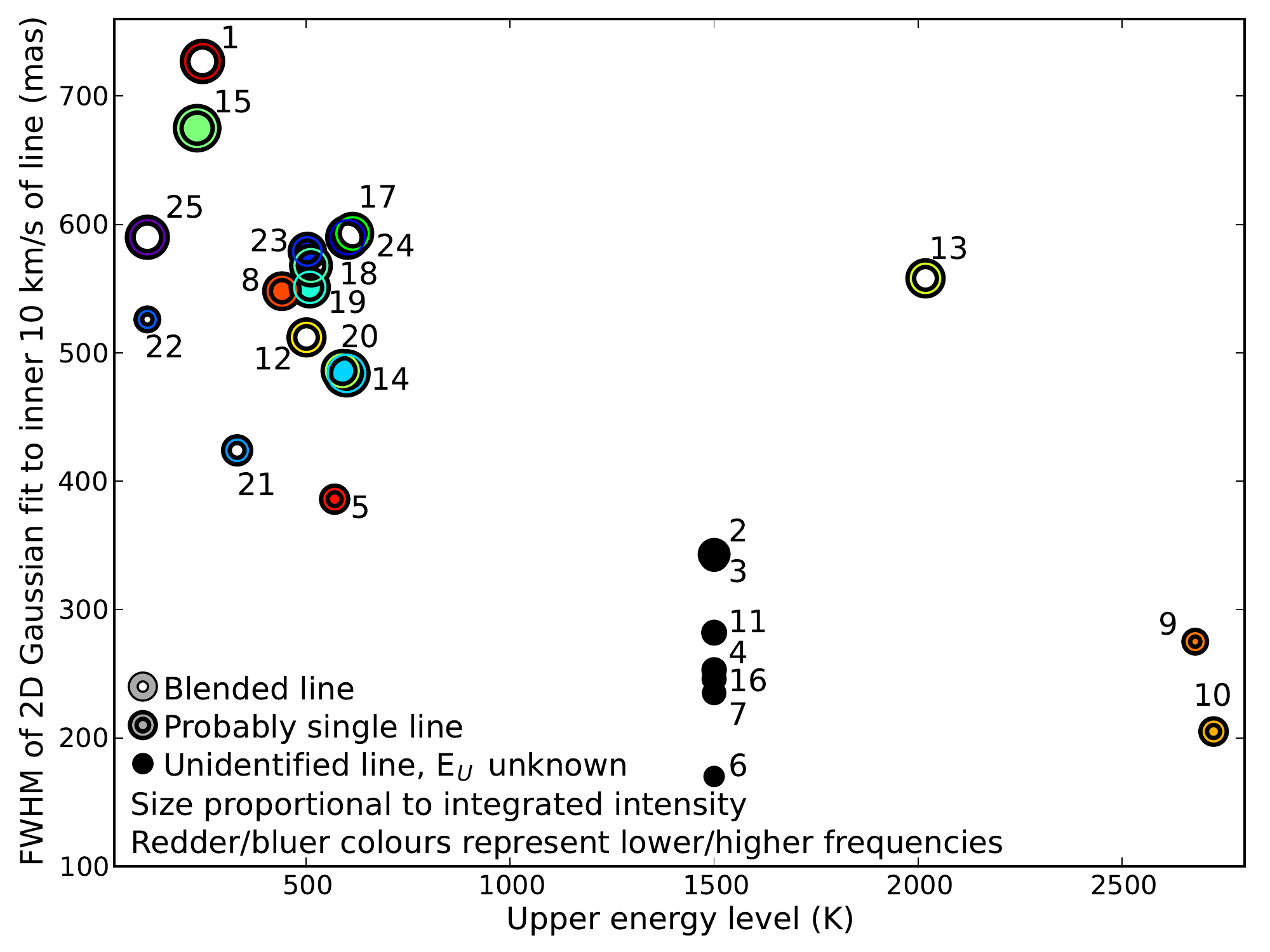}
 \caption{Spatial FWHM in function of the upper state energy level. Lines are identified by their number given in the first column of Table~\ref{Table:FWHM}. The size of the symbols is proportional to the integrated intensity; unidentified lines are put at an artifical upper state energy of 1500\,K. }
 \label{Fig:Eupper_FWHM}
 \end{figure}

 To determine the spatial extension, we have computed the first moment for the inner 10\,km\,s$^{-1}$ of each line and have measured the spatial FWHM by fitting a 2-D Gaussian component and deconvolving the restoring beam (see Table~\ref{Table:FWHM}). We only perform the spatial fitting to the inner 10\,km/s to avoid line blending and since for an expanding shell, the emission at line center would be most spatially extended, whilst the extreme velocity emission comes from closer to the line of sight to the star. All lines have a high
 enough signal-to-noise ratio to be spatially resolved apart from the
 faintest unidentified line ('U' line) at 645.507\,GHz. They are all compact enough for a good
 fit except for the $^{13}$CO, for which its apparent spatial FWHM, $s_{\mathrm{FWHM}}$,
 represents only the inner peak.  There is
 likely also to be extended C$^{18}$O emission below our detection
 threshold. Table~\ref{Table:FWHM} also lists the (not deconvolved) largest angular size (LAS) of the central emission brighter than 2\,Jy km/s per beam (approximately 3$\sigma_{\mathrm{rms}}$). In the case of $^{13}$CO J=6-5 there is also non-contiguous emission extended throughout the field of view (see Sect.~\ref{Sec:channel_maps}). As illustrated in Fig.~\ref{Fig:Eupper_FWHM} the extension scales inversely with the upper state energy, $E_{\mathrm{upper}}$. The anomalies in this plot are due to blending and  the fact that the ALMA data resolve out a lot of the $^{13}$CO flux (see Sect.~\ref{Sec:Herschel_ALMA}).

\begin{table*}[htp]
    \caption{Spectral and spatial extent of the detected lines. First column designates a number to each transition, second and third columns list the molecular line, fourth column the line rest frequency $\nu_0$, fifth column the upper energy level $E_{\rm{upper}}$, sixth column the integrated flux density of the fitted Gaussian component $S_{\rm{int}}$ and its uncertainty between parentheses, seventh column the spatial FWHM $s_{\rm{FWHM}}$ and its uncertainty, eighth column the LAS  above 2\,Jy km/s per beam, ninth column the spectral FWHM, and the last column the half line width at continuum level $v_{\rm{line}}$ (indicative of the expansion and turbulent velocity) and its uncertainty.}
    \label{Table:FWHM}
\begin{tabular}{llllcrccll}
\hline 
 \rule[0mm]{0mm}{5mm} & {molecule} & \multicolumn{1}{c}{transition} & \multicolumn{1}{c}{$\nu_0$} & $E_{\rm{upper}}$ & \multicolumn{1}{c}{$S_{\mathrm{int}}$}  & $s_{\mathrm{FWHM}}$  & LAS & \multicolumn{1}{c}{FWHM}  & \multicolumn{1}{c}{$v_{\rm{line}}$} \\
 \rule[-3mm]{0mm}{3mm} & & & \multicolumn{1}{c}{[GHz]} & \multicolumn{1}{c}{[K]} & [Jy\,MHz]  & [mas]  & [arcsec] & \multicolumn{1}{c}{[km/s]} & \multicolumn{1}{c}{[km/s]} \\
 \hline
\rule[0mm]{0mm}{5mm}1 & $^{29}$SiO &15-14         & 642.8080$^*$     	 & 246 & 413 (9) & 727 (15) & 2.0	& 25.78$^*$ 	& 14.36 (0.26)$^*$\\
2 &'U'	&   		 & 642.845$^*$		 & --  &  167 (4) & 343 (14) & 1.2	& 25.78$^*$ 	& 14.36 (0.26)$^*$\\
3 & 'U'	&		 & 642.96		 & --  &  124 (2) & 341 (18) & 1.3	& 6.52		& 5.13 (0.70)\\
4 & 'U'	& 		 & 643.06		 & --  &  75 (2) & 253 (26) & 1.2	& 10.00		& 7.92 (0.17)\\
5 & Si$^{33}$S & 36-35         & 643.2608 	         & 570 &  143 (2) & 386 (14) & 1.3	& 17.81 	& 15.19 (0.80)\\
6 & 'U'	&    		 & 645.507		 & --  &  39 (2) & $<$216 & 0.7	& \ \ 5.33	& \ \ 3.57 (0.92)\\
7 & 'U'	& 		 & 645.567		 & --  &  64 (2) & 235 (36) & 0.9	& \ \ 7.60	& \ \ 4.95 (0.80)\\
8 & SiC$_2$ & 26,4,22-25,4,21  & 645.7526   	 & 441 & 287 (9) & 548 (18) & 2.0	& 20.66 	& 14.63 (0.03)\\
9 & Si$^{34}$S & v=2 37-36     & 645.9198   	 & 2679 &  101 (2) & 275 (23) & 1.3	& 10.20		& \ \ 5.10 (0.18)\\
10 & SiS & v=2 36-35            & 646.1000    	 & 2724 &  133 (2) & 205 (19) & 1.1	& \ \ 7.89	& \ \ 3.94 (0.12)\\
11 & 'U'& 			 & 646.31		 & --  &  79 (2) & 282 (29) & 1.5	& \ \ 5.49	& \ \ 3.61 (0.98)\\
12 & SiC$_2$ & 27,6,22-26,6,21  & 646.4211$^*$	 & 501 & 295 (6) & 512 (16) & 1.7	& 19.15$^*$	& 15.37 (0.28)$^*$\\
13 & SiO & v=1 15-14            & 646.4296$^*$	 & 2018 & 300 (9) & 558 (18) & 2.0	& 19.15$^*$	& 15.37 (0.28)$^*$\\
14 & $^{30}$SiS & 37-36         & 646.7703    	 & 589 & 375 (6) & 486 (13) & 1.8	& 19.54		& 14.09 (0.17)\\
15 & $^{13}$CS & 14-13          & 647.0762    	& 233 & 492 (13) & 675 (16) & 1.9	& 22.10		& 15.29 (0.08)\\
16 & 'U'& 			 & 647.13		 & --  &  71 (2) & 246 (39) & 1.1	& 10.80		& \ \ 7.46 (0.95)\\
17 & $^{29}$Si$^{34}$S & 38-37  & 657.1243$^*$	& 614 & 364 (13) & 593 (23) & 2.1	& 20.56$^*$	& 15.41 (0.18)$^*$\\
18 & SiC$_2$ & 31,0,31-30,0,30  & 657.1290$^*$	 & 512 & 377 (13) & 568 (23) & 2.2	& 20.56$^*$	& 15.41 (0.18)$^*$\\
19 & SiC$_2$ & 30,2,29-29,2,28  & 657.6034     	 & 509 & 340 (13) & 551 (24) & 2.0	& 19.70		& 14.40 (0.09)\\
20 & $^{29}$SiS & 37-36         & 658.2241  		 & 599 & 486 (11) & 484 (13) & 2.1	& 16.73		& 14.74 (0.36)\\
21 & AlF & 20-19                & 658.5295$^*$ 	 & 331 &  158 (6) & 424 (28) & 1.8	& 22.06$^*$	& 17.08 (0.46)$^*$\\
22 & C$^{18}$O & 6-5            & 658.5533$^*$	 & 111 &  100 (9) & 526 (54) & 1.5	& 22.06$^*$	& 17.08 (0.46)$^*$\\
23 & SiC$_2$ & 29,2,27-28,2,26  & 658.7924    	 & 503 & 280 (13) & 579 (27) & 2.0	& 19.77		& 13.57 (0.04)\\
24 & Si$^{33}$S & 37-36         & 661.0665$^{\dagger}$ & 602 & 407 (13) & 590 (21) & 2.8	& 24.69$^{\dagger}$& 13.56 (0.01)$^{\dagger}$\\
25 & $^{13}$CO & 6-5            & 661.0673$^{\dagger}$ & 111 & 407 (13) & 590 (21) & 2.8	& 24.69	$^{\dagger}$& 13.56 (0.01)$^{\dagger}$\\
\hline
\end{tabular}
\tablefoot{
$^*$ blended within 2--20 km s$^{-1}$;
$^{\dagger}$ blended within 0.4 km s$^{-1}$
}
\end{table*}

  \subsection{Comparison between ALMA and Herschel} \label{Sec:Herschel_ALMA}
  
  The frequency range observed with ALMA was also observed by the Herschel/HIFI instrument \citep{deGraauw2010A&A...518L...6D}. A survey in all HIFI bands was carried out on May 11--15 2010. The data reduction and first results are described in \citet{Cernicharo2010A&A...521L...8C}. The HIFI beam size around 640\,GHz is 33.2\arcsec\ and the main-beam antenna efficiency is 0.75 \citep{Roelfsema2012A&A...537A..17R}. The data were taken in double beam-switching mode at a frequency resolution
of 1.1\,MHz. To increase the SNR of the weaker lines, some spectral regions were rebinned.  The rms noise after averaging all scans is given in Table~\ref{Table_ALMA_Herschel}. To convert from antenna temperatures (in K) to fluxes in Jy, we have used the point-source sensitivity of 466\,Jy/K \citep{Roelfsema2012A&A...537A..17R}. The source size of all transitions listed in Table~\ref{Table_ALMA_Herschel} can be approximated with a point-source within the Herschel beam, with the exception of the $^{13}$CO J=6-5 line. Based on observations with the JCMT and CSO \citep{Crosas1997ApJ...483..913C} and our modeling of the $^{13}$CO emission presented in \citet{Decin2010A&A...518L.143D} and \cite{DeBeck2012A&A...539A.108D}, we derive a source size of $\sim$10\arcsec\ for the $^{13}$CO J=6-5 line, yielding a 3\% correction w.r.t.\ point-source sensitivity.

\begin{table*}[htp]
\caption{Comparison between the Herschel and ALMA integrated line fluxes. The first and second columns list the molecule and its transition, the third , fourth and fifth column give the Herschel antenna peak temperature, rms and SNR; the Herschel and ALMA integrated line fluxes are tabulated in the sixth column; comments on blends as seen in the Herschel and ALMA data sets are given in the last two columns. The integrated line fluxes plagued by weak line blends are given within parentheses. A `$-$' indicates the cases of strong line blending where only detailed modeling can reveal the contribution of the individual components. The Herschel point-source sensitivity of 466 Jy/K \citep{Roelfsema2012A&A...537A..17R} has been used for all lines. For $^{13}$CO J=6-5, the estimated source size of 10\arcsec\ \citep{Crosas1997ApJ...483..913C} was taken into account.}
\label{Table_ALMA_Herschel}
\setlength{\tabcolsep}{2mm}
{\small{
\begin{tabular}{lllllllll}
\hline \hline
\rule[0mm]{0mm}{5mm}Molecule &Transition&Herschel&Herschel&Herschel&Herschel&ALMA&comments&comments \\
&&$T_{\rm{a}}^{\rm{peak}}$&rms&SNR&$F_\nu\,d\nu$&$F_\nu\,d\nu$&Herschel &ALMA \\
\rule[-3mm]{0mm}{3mm}&&(K)&(mK)&&(Jy MHz)&(Jy MHz)&data&data \\
\hline
\rule[0mm]{0mm}{5mm}$^{29}$SiO&           15-14&0.065&9&7&(1408)&(834)&& blend `U'   \\
Si$^{33}$S &           36-35&0.047&8.9&5&210&271&low SNR&\\
SiC$_{2}$& 26,4,22-25,4,21&0.054&5.5&10&1168&816&&    \\
SiS v=2&           36-35&0.018&3.3&5&191&159&low SNR&\\
SiC$_{2}$& 27,6,22-26,6,21&$-$&4.5&$-$&$-$&$-$&blend SiO v=1& blend SiO v=1   \\
SiO v=1&           15-14&$-$&2.9&$-$&$-$&$-$&blend SiC$_2$& blend SiC$_2$   \\
$^{30}$SiS&37-36&0.055&4.9&11&1035&950&&\\
$^{13}$CS&           14-13&0.093&4.7&20&2012&1397&& weak blend `U'\\
$^{29}$Si$^{34}S$&           38-37&$-$&7&$-$&$-$&$-$&blend SiC$_2$&  blend SiC$_{2}$  \\
SiC$_{2}$& 31,0,31-30,0,30&$-$&5.1&$-$&$-$&$-$&blend $^{29}$Si$^{34}$S& blend $^{29}$Si$^{34}$S   \\
SiC$_{2}$&30,2,29-29,2,28&0.061&8.4&7&1297&883&&   \\
$^{29}$SiS&           37-36&0.075&6.2&12&1452&1059&&\\
AlF&           20-19&$-$&6.4&$-$&$-$&$-$&blend C$^{18}$O  &  blend C$^{18}$O   \\
C$^{18}$O&             6-5&$-$&9&$-$&$-$&$-$&blend AlF&  blend AlF   \\
SiC$_{2}$& 29,2,27-28,2,26&0.055&5.6&10&1281&735&&\\
$^{13}$CO&             6-5&1.235&7&176&32525&1366&weak blend Si$^{33}$S J=37-36&  blend Si$^{33}$S J=37-36 \\
\hline
\end{tabular}
}}
\end{table*}

Table~\ref{Table_ALMA_Herschel} lists the integrated line intensities measured with Herschel and ALMA. Note that the ALMA integrated line intensities are higher compared to the values listed in Table~\ref{Table:FWHM} due to the fact that the spatial fitting used for the values in Table~\ref{Table:FWHM} was restricted to the inner 10\,km/s (see previous section).
As can be seen from Table~\ref{Table_ALMA_Herschel}, only a few lines are not plagued by blends with other molecular line transitions. The half line width at continuum level, $v_{\rm{line}}$, of all common unblended lines is in agreement for both instrument. The integrated flux density of the non-blended lines measured from
the Herschel data is typically 1.2 -- 1.7 times larger than that
measured using ALMA (note that the overall ALMA flux
scale is uncertain by up to 20\%).  The ratio is lower for the Si$^{33}$S J=36-35
transition but this line has a low signal-to-noise ratio in the Herschel data, and
the ratio is much higher for $^{13}$CO J=6-5 line.
The 20--50\% additional flux detected by Herschel compared with ALMA can
be explained by two factors.

Firstly, there are observational differences. Herschel is
fundamentally more sensitive to large-scale emission.  Its 5$\sigma$
sensitivity for these observations, per 0.7\,km s$^{-1}$, was $\sim$20\,Jy beam$^{-1}$, compared with $<$0.2\.Jy beam$^{-1}$ for ALMA. However, the Herschel beam is $\sim$$10^4$ times the area of the ALMA
synthesized beam.  A 20\,Jy point source would be very bright in the
ALMA data but, spread out over an area more than 33\farcs2 in diameter,
the surface brightness per ALMA beam would be below the detection
limit.  In addition, even brighter emission which is smooth on scales
$>3$'' would be poorly imaged by ALMA, and not detected at all,
however bright, if it was $>6$'' as explained in Sect.~\ref{Sec:data_reduction}.  Our
results are consistent with this, as the fraction of the Herschel flux
detected by ALMA is higher for higher excitation
temperature lines and lower for more spatially-extended species and we
detect a similar fraction of the flux with ALMA for similar
transitions. Specifically,
for the transitions listed in Table~\ref{Table_ALMA_Herschel}, the estimated emission
region for the SiS isotopologues (J=37-36 or J=36-35)
in the ground-vibrational state is $<6$\farcs5 \citep[based on modeling by][]{Decin2010A&A...518L.143D}. For the $^{13}$CS J=14-13 line the main
emission region is within $\sim7$\arcsec\ \citep[based on the $^{12}$CS J=7-6 and $^{13}$CS J=7-6 line observed by][]{Williams1992A&A...266..365W, Patel2011ApJS..193...17P} and the $^{12}$CS J=14-13 SMA data of \citet{Young2004ApJ...616L..51Y} even indicate an emission region within $\sim$4\arcsec. As mentioned
above, the main emission region for the $^{13}$CO J=6-5 transition
is $\sim10$\arcsec\ and the ALMA observations have resolved-out at least
95\% of the flux. No flux will be resolved out for the high excitation
SiS v=2 J=36-35 line with an upper state energy
of 2724\,K; the difference between ALMA and Herschel integrated line flux values is within the calibration uncertainties of both instruments.

Secondly, the variability of thermal line emission due to the variability of the central star. Following \citet{Jenness2002MNRAS.336...14J}, the Herschel and ALMA data were taken at phase 0.46 and  0.98, respectively. The changes in the radiation field during the pulsation period can influence the molecular excitation. The infrared radiation field is expected to mainly
affect molecules with strong vibrational transitions, with stronger modulations for lines excited close to the star. Using Herschel data, \citet{Teyssier2013} have recently shown that the $^{12}$CS J=14-13 has a maximum amplification factor in the line intensities of $\sim$1.3, while for SiS J=37-36 and J=36-35 this is $\sim$1.2. The low excitation CO lines seem quite insensitive to a change in the infrared radiation field. 

Predicting if a line will have a higher or lower amplification factor depending on the variability phase at the time of the observations is, however, complex. The ALMA data are taken near 
 light maximum and the Herschel data near light minimum, which indicates that higher excitation levels might be more populated at the time of the ALMA observations. However, the exact effect is dependent on the transition rate probabilities with the connecting levels.

 \subsection{Gas velocity structure in the inner wind region} \label{Sec:vgas}

\begin{figure*}[htp]
   \begin{minipage}[t]{.48\textwidth}
        \centerline{\resizebox{\textwidth}{!}{\includegraphics[angle=90]{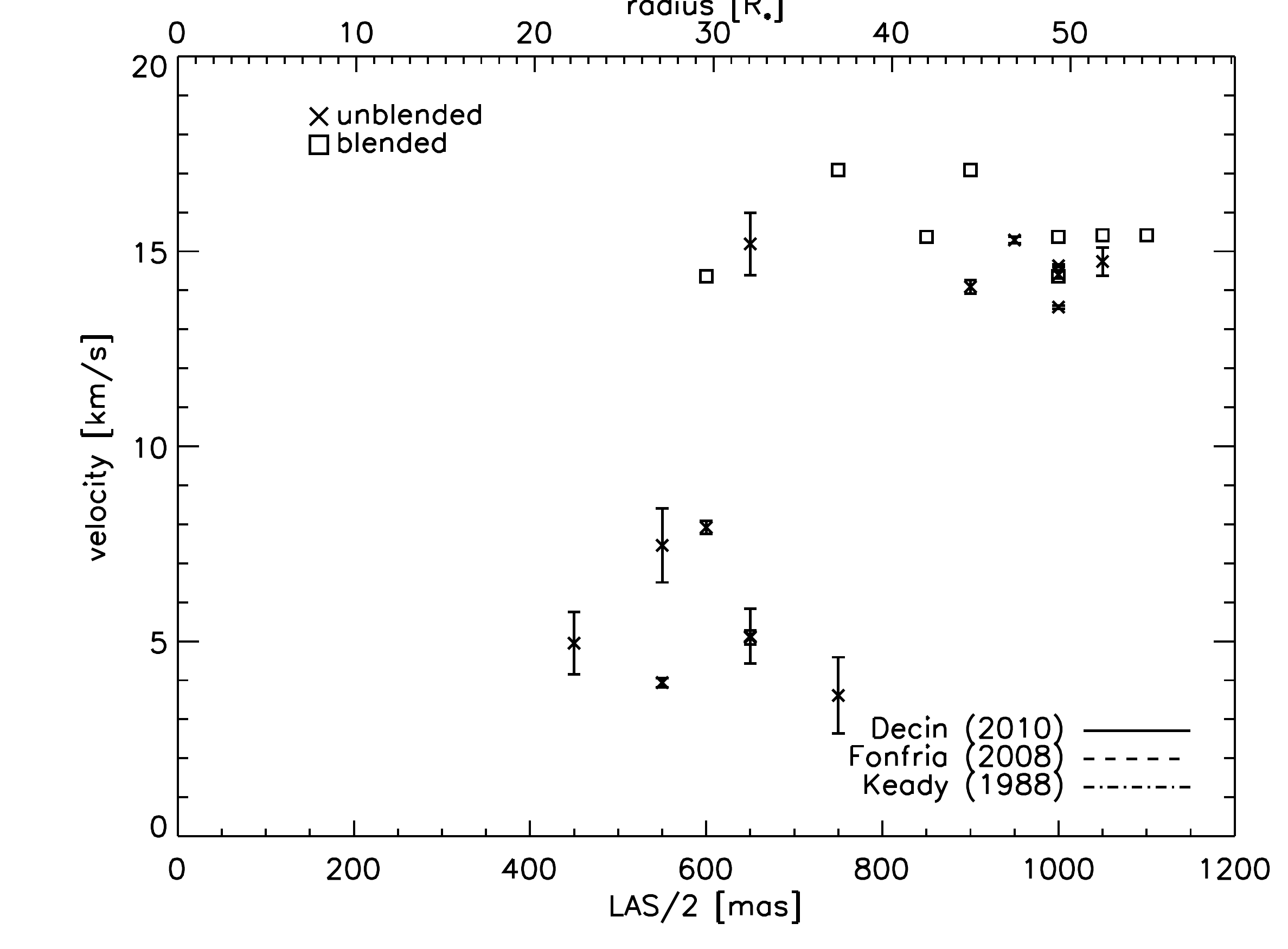}}}
    \end{minipage}
    \hfill
    \begin{minipage}[t]{.48\textwidth}
        \centerline{\resizebox{\textwidth}{!}{\includegraphics[angle=90]{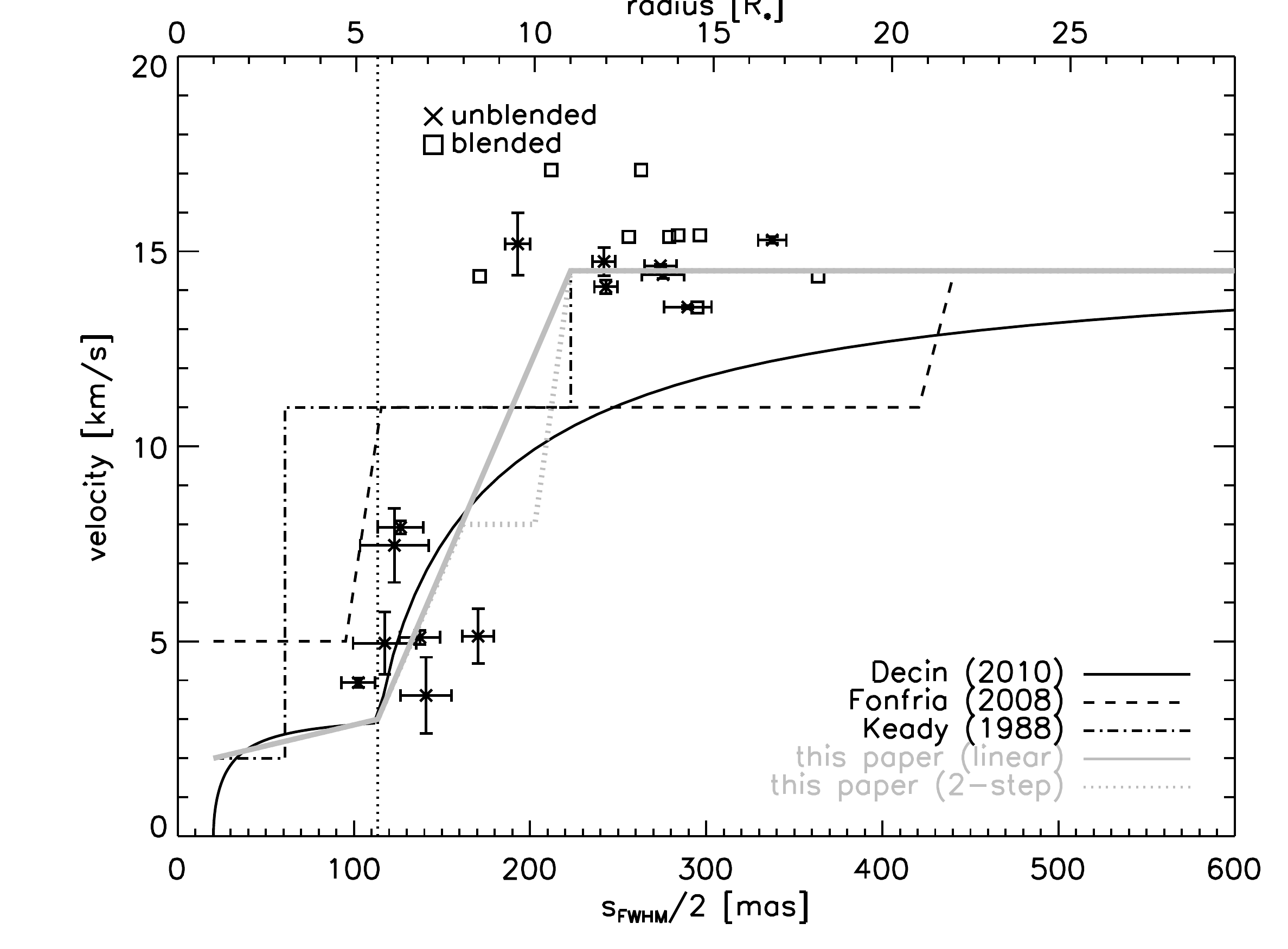}}}
    \end{minipage}
 \caption{Measured velocities [$v_{\rm{line}}$ in km/s] versus half of the LAS (representing the not-deconvolved maximum detectable scale of each transition) in the left panel and half of the spatial FWHM $s_{\rm{FWHM}}$ (representing the dominant line formation region) in the right panel. Note that the value for the LAS is restricted by the sensitivity limit of the observations and the fact that large-scale flux might be resolved out (see Sect.~\ref{Sec:data_reduction}). The blended lines are indicated with a squared box, the unblended lines with a cross (and include the error bars). In the right panel, the full black line shows the velocity structure as derived by \citet{Decin2010Natur.467...64D} from solving the momentum equation; the dashed line indicates the velocity structure as derived by \citet{Fonfria2008ApJ...673..445F} from modelling  different $J$ ro-vibrational transitions of C$_2$H$_2$, the dash-dotted line the velocity structure derived by \citet{Keady1988ApJ...326..832K} from the analysis of 
the near-infrared ro-vibrational CO spectrum, and the thick grey lines the acceleration derived from the ALMA data. The vertical dotted line indicates the dust nucleation region around 5.6\,\Rstar, where the sound velocity is $\sim$3\,km/s. }
\label{Fig:vexp_LAS_sFWHM}
\end{figure*}    

Table~\ref{Table:FWHM} and Fig.~\ref{Fig:vexp_LAS_sFWHM} show that the expansion velocity of the lines falls into
two groups. All but two of the identified lines trace a velocity larger than 13.5\,km/s at a LAS radius (representing the not-deconvolved maximum detectable scale) beyond 600\,mas radius (or 30\,\Rstar) and a spatial FWHM (illustrating the main line formation region) beyond 200\,mas (10\,\Rstar), indicating that the wind has reached its terminal velocity  within $\sim$10\,\Rstar\ from the central star, in agreement with the results of \citet{Keady1988ApJ...326..832K}. The other group has line width velocities $<$8\,km/s within 800\,mas LAS radius (40\,\Rstar).  This contains
the two identified lines in v=2 excited vibrational states and all unblended 'U' lines, suggesting that these also come from
high-excitation states. The narrowest line is the ‘U’ line at 646.31\,GHz with a velocity of only 3.6\,km/s, slightly higher than the sound velocity of $\sim$3\,km/s at the base of the dust nucleation region.  

Plotting the derived velocities in function of the spatial FWHM (see Fig.~\ref{Fig:vexp_LAS_sFWHM}) shows that the measured line width velocities rapidly increase from the sound velocity to $\sim$8\,km/s between $\sim$100 to 200\,mas (5--10\,\Rstar).  Assuming that the turbulent velocity does not vary substantially in this small region, this indicates that carbonaceous material condenses around 5\,\Rstar, resulting in a rapid acceleration of the gas until a velocity of $\sim$8\,km/s.
In the four ALMA spectral windows, no lines are detected with a velocity between 8 and 13.5\,km/s, but the data suggest a rapid increase of the gas velocity from $\sim$8\,km/s to $\sim$14.5\,km/s around $\sim$10--12\,\Rstar.
This acceleration profile is not captured by the solution of the wind momentum equation derived by \citet{Decin2010Natur.467...64D}, who assumed instantaneous condensation of  amorphous carbon, iron, silicon carbide and magnesium sulfide  dust species at a radius of 5.6\,\Rstar. \citet{Keady1988ApJ...326..832K} suggested a 2-steps acceleration process from an analysis of high-resolution near-infrared ro-vibrational CO spectra. They suggested an increase around 3 and 11\,\Rstar, with the first increase up to a velocity of 11\,km/s being due to the condensation of carbon and a second increase that might be linked to dust condensation involving Mg and/or S. This idea was used and adapted by \citet{Fonfria2008ApJ...673..445F} who modeled a large 
set of C$_2$H$_2$ lines. They suggested a strong velocity gradient from 5 to 11\,km/s around 4.7-5.7\,\Rstar\ and a second velocity increase to 14.5\,km/s around 20.75-21.75\,\Rstar\ (see Fig.~\ref{Fig:vexp_LAS_sFWHM}). 
Based on the ALMA band 9 data, the wind acceleration profile can be further refined. The data can be fit  assuming a linear velocity increase from around 2--3\,km/s at $\sim$5\,\Rstar\ reaching terminal velocity around 11\,\Rstar (see full grey line in the right panel of Fig.~\ref{Fig:vexp_LAS_sFWHM}). Assuming  a 2-steps acceleration process, we obtain
\begin{equation}
 v(r) = \left\{
\begin{array}{rl}
2-3{\rm{\,km/s}}, & \textnormal{$1\le r/\textnormal{R$_*$} <5.6$}\\
3-8{\rm{\,km/s }}, & \textnormal{$5.6\le r/\textnormal{R$_*$} <8$}\\
8{\rm{\,km/s }}, & \textnormal{$8\le r/\textnormal{R$_*$} <10$}\\
8-14.5{\rm{\,km/s}}, & \textnormal{$10\le r/\textnormal{R$_*$} <11$}\\
14.5 {\rm{\,km/s}}, & \textnormal{$11\le r/\textnormal{R$_*$}$}\,,
\end{array}
\right.
\label{Eq:velocity}
\end{equation}
i.e.\ a first jump around 5.6\,\Rstar\ and a second one around 10\,\Rstar\ (see dotted grey line in the right panel of Fig.~\ref{Fig:vexp_LAS_sFWHM}). To improve upon the solution of the momentum equation as derived by \citet{Decin2010Natur.467...64D}, different possibilities exist. (1)\,Either the dust absorption efficiencies need to increase in the near-infrared, where the bulk of the stellar photons are emitted. (2)\, Or larger grains ($\ga$0.2\,$\mu$m) should be formed more efficiently close to the star, thus enhancing the wind acceleration via scattering processes \citep{Hofner2008A&A...491L...1H}. (3)\,A third possibility might exist in adapting the velocity at the start of the dust condensation region, which is now assumed to be the sound velocity \citep{Decin2006A&A...456..549D}. However, in that case, the dust-to-gas ratio would need to be decreased in order not to overshoot the observed terminal velocity.

 \section{Channel maps and PV diagrams} \label{Sec:data_results}
 
 In Sect.~\ref{Sec:channel_maps} and Sect.~\ref{Sec:PV}, we show and discuss the channel maps and position-velocity (PV) diagrams of 1 representative high-excitation line with an upper state energy (E$_{\rm upper}$) around 2700\,K, of 1 medium excitation line (E$_{\rm upper}\sim$450\,K) and of 1 lower excitation line (E$_{\rm upper}\sim$100\,K). As will be seen, the vibrationally-excited line traces the wind acceleration region, molecules present in the inner envelope with excitation energies around a few hundred K are slightly resolved with the ALMA beam of $\sim$0.2\arcsec, while the $^{13}$CO J=6-5 line shows a complex emission pattern.

\subsection{Channel maps and morphology} \label{Sec:channel_maps}

\begin{figure*}[htp]
\sidecaption
 \includegraphics[width=12cm]{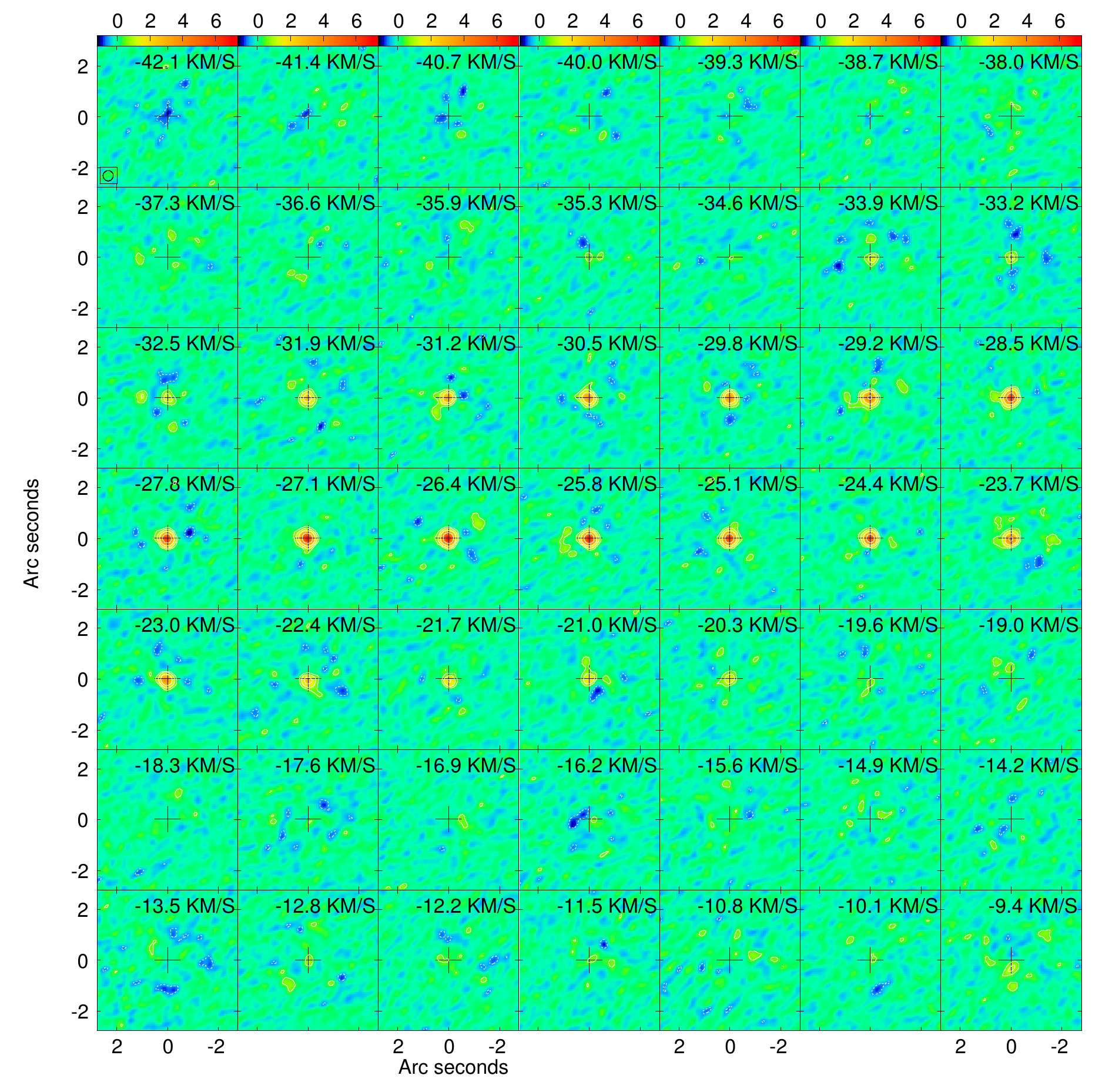}
 \caption{SiS v=2 J=36-35 channel map, averaged over 3 channels with a natural weighting and for a circular beam of 0.4\arcsec. North is up, East is left; the black cross indicates the center of the continuum map. The flux density units are Jy/beam. The contours are at ($-$1,1,2,4,8\ldots)$\times$0.5\,Jy/beam ($\sim$3\,$\sigma_{\rm{rms}}$).
 This high-excitation line has a width of $\sim$9\,km/s and the line formation region is not resolved with the current ALMA beam. The contrast in the figure is best visible on screen.}
 \label{Fig:SiSv2_J36_channel}
\end{figure*}

As expected, the channel map of a very high-excitation line of a  molecule present in the inner wind region just above the photosphere, e.g.\ SiS v=2 J=36-35 (Fig.~\ref{Fig:SiSv2_J36_channel}), shows that the emission of this line comes from the very inner regions of the envelope, close to the stellar surface, where the wind material has a velocity around 4\,km/s and the gas kinetic temperature is typically around 900-1300\,K. The morphology in the region $<$1\arcsec (or diameter of 50\,\Rstar) is still more or less circular (for a circular beam of 0\farcs4).

\begin{figure*}[htp]
\sidecaption
 \includegraphics[width=12cm]{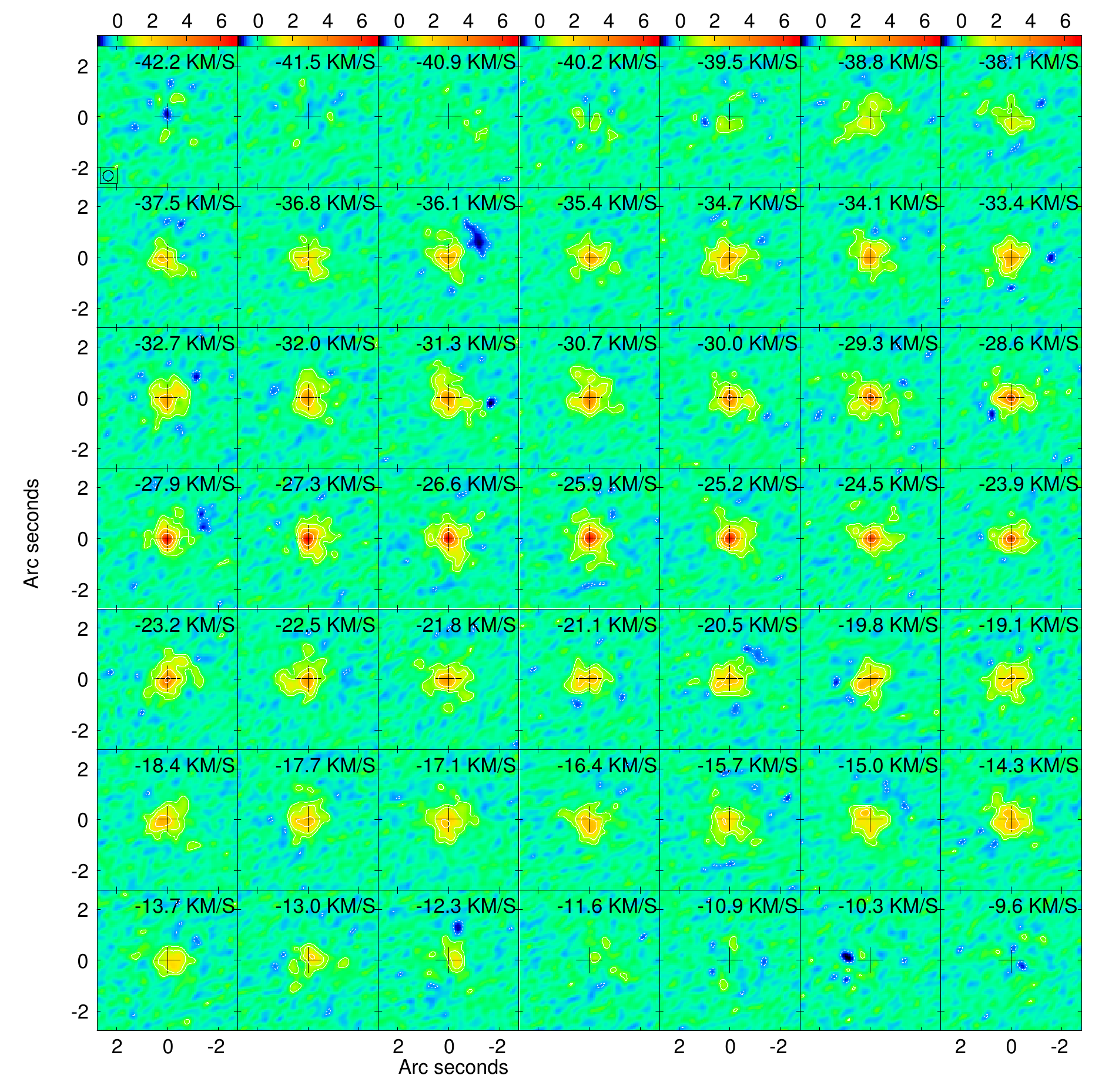}
 \caption{SiC$_2$ 26(4,22)-25(4,21) channel map, averaged over 3 channels with a natural weighting and for a circular beam of 0.4\arcsec. North is up, East is left; the black cross indicates the center of the continuum map. The flux density units are Jy/beam. The contours are at ($-$1,1,2,4,8\ldots)$\times$0.5\,Jy/beam ($\sim$3\,$\sigma_{\rm{rms}}$). The ALMA instrument resolves the line formation region. At different velocities the flux densities display a non-spherical structure. The contrast in the figure is best visible on screen.}
 \label{Fig:SiC2_J26_channel}
\end{figure*}

The channel maps of molecular transitions formed in the inner wind region and with upper state energies around a few hundred K, such as the SiC$_2$ 26(4,22)-25(4,21) line displayed in Fig.~\ref{Fig:SiC2_J26_channel}, show that the emission of this line is extended and displays a non-homogeneous  distribution of the molecular density. The typical spatial FWHM of these lines is around a few hundred milli-arcsec (see Table~\ref{Table:FWHM}) and these lines are formed in a region of the wind where the velocity has already reached the terminal velocity.
 
\begin{figure*}[htp]
\sidecaption
 \includegraphics[width=12cm]{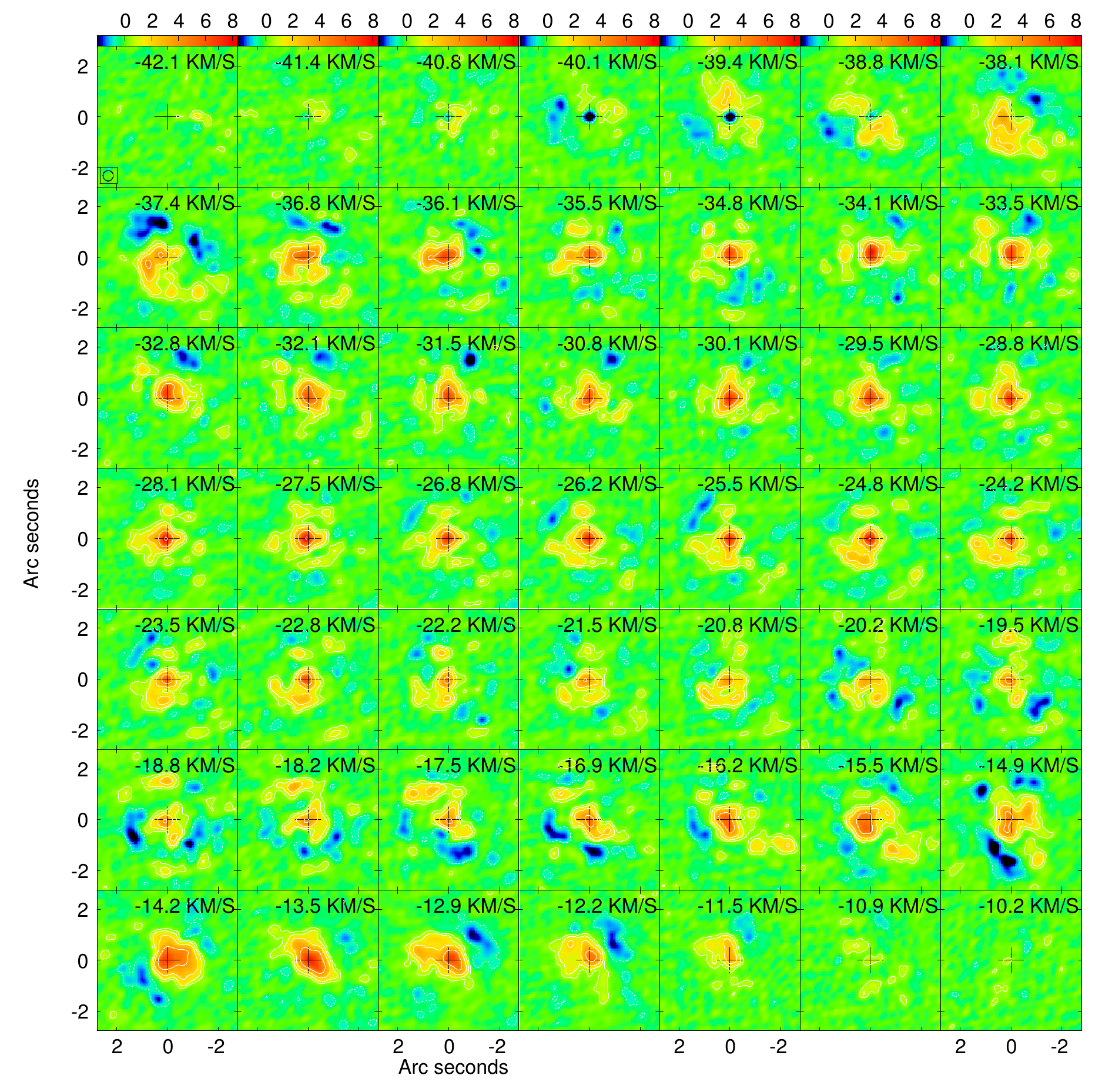}
 \caption{$^{13}$CO J=6-5 channel map, averaged over 3 channels with a natural weighting and for a circular beam of 0.4\arcsec. North is up, East is left; the black cross indicates the center of the continuum map. The flux density units are Jy/beam. The contours are at ($-$1,1,2,4,8\ldots)$\times$0.5\,Jy/beam ($\sim$3\,$\sigma_{\rm{rms}}$). Different arc-like features can be discerned. The contrast in the figure is best visible on screen.}
 \label{Fig:13CO_channel}
\end{figure*}

The $^{13}$CO J=6-5 channel map (Fig.~\ref{Fig:13CO_channel}) displays a complex structure. At different velocities, parts of a spiral-like arm (arcs) can be distinguished: around $-$17.5\,km/s one can see an arc to the North at $\sim$1.5\arcsec\ offset from the central source, around $-$24.6\,km/s at 0.7\arcsec\ offset to the south-east, around $-$27.5\,km/s at 1\arcsec\ to the south-west, and around $-$37.4\,km/s at 2\arcsec\ to the south-west. Local density enhancements, the change in (projected) velocity for different arcs, gaps in the UV-coverage (see Sect.~\ref{Sec:data_reduction}) and the fact that we resolve out flux for scales larger than $\sim$3\arcsec\ can explain why the spiral arms are not complete. At, for example, $-$37.4\, km/s, the extended emission has a distinctive curve not related to
   the beam sidelobe pattern. The negative artefact is due to the
   "bowl" effect mentioned in Sect.~\ref{Sec:data_reduction}, resulting from emission on scales
   just larger than is sampled by the interferometer.  Although the
   detailed flux distribution at radii $\ga1\farcs5$ is unreliable, it
   does represent the general shape of the brighter emission at each
   velocity.
   
\begin{figure}[htp]
 \includegraphics[width=.48\textwidth]{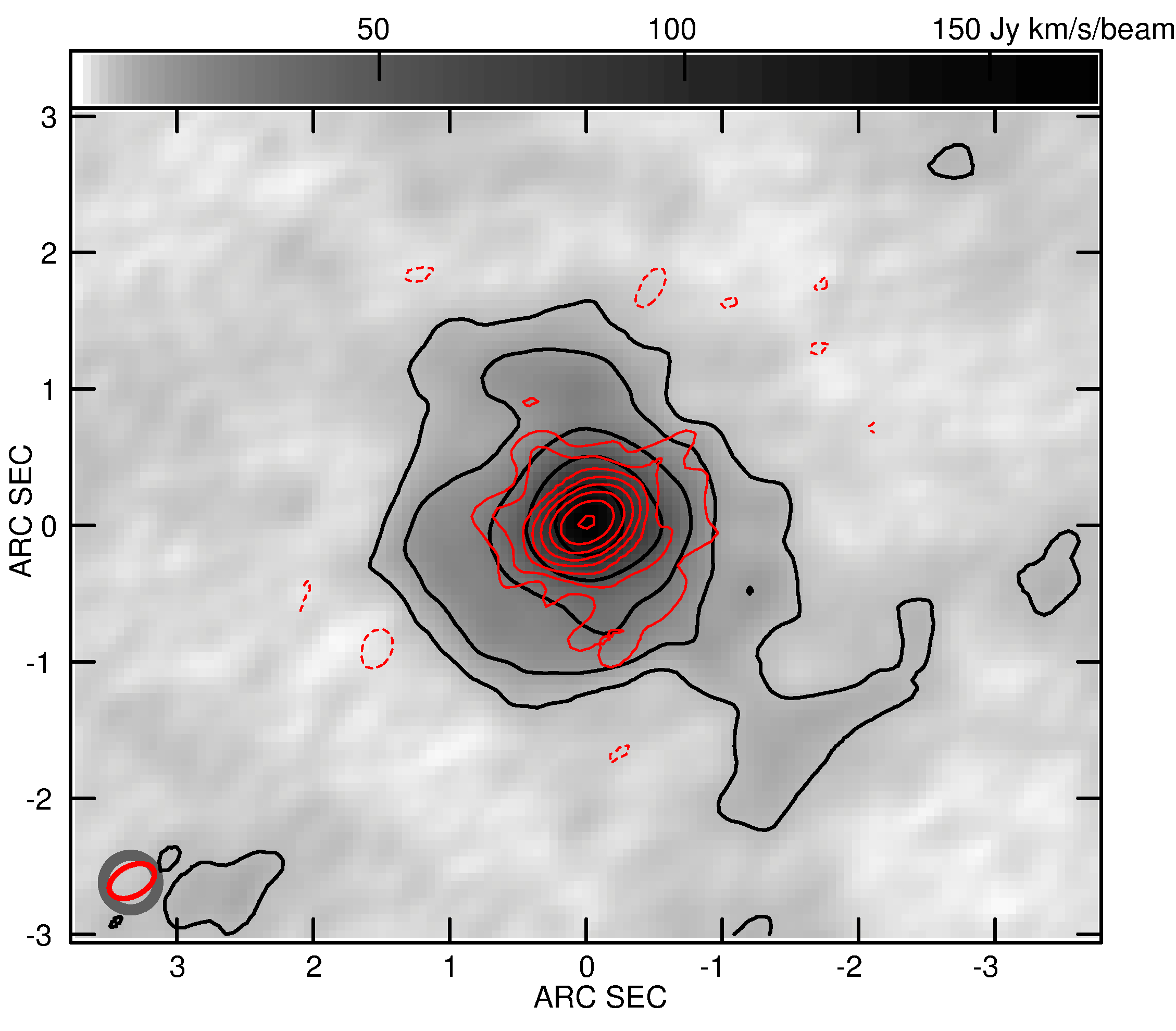}
 \caption{\textit{Gray scale:} Integrated $^{13}$CO J=6-5 emission (cut-off level of 0.02 Jy/beam), with contour levels at [$-$1, 1, 2, 4, 8, 16]$\times$6.82\,Jy$\cdot$\,km/s per beam. \textit{Red contours:} dust emission with contours levels are at [$-1$, 1, 2, 3, 4, 8, 16, 28]$\times$15\,mJy beam$^{-1}$ (cf.\ Fig.~\ref{Fig:Flux-4}). In the bottom left corner, the beam ellipse for the integrated $^{13}$CO J=6-5 emission is shown in gray and for the dust continuum in red.}
 \label{Fig:13CO_integrated_dust}
\end{figure}

The local density enhancements/spiral arcs are seen best in the maps of $^{13}$CO, as it is the lowest-excitation and most abundant transition in our ALMA observations, mitigating the over-resolution of its extended emission. Some of the less-abundant, intermediate-excitation lines have a 
similar LAS for integrated emission  in the ALMA data, but are too faint to resolve extended details, spectrally or spatially. Fig.~\ref{Fig:13CO_integrated_dust} compares the integrated $^{13}$CO J=6-5 emission (zeroth moment) and dust emission. Bright arcs at 1\arcsec\ to the North and 1\arcsec\ and 1.6\arcsec\ to the South-west can clearly be discerned in the $^{13}$CO emission, but the dust emission at 650\,GHz is below our detection threshold at radii beyond 1\arcsec. Important to note is that it is the first time that we can confirm that the extended continuum and molecular emission are both centred on the continuum peak position.

\subsection{Position-velocity diagrams and morphology} \label{Sec:PV}

\begin{figure*}[htp]
\sidecaption
     \begin{minipage}[t]{5.8cm}
        \centerline{\resizebox{\textwidth}{!}{\includegraphics{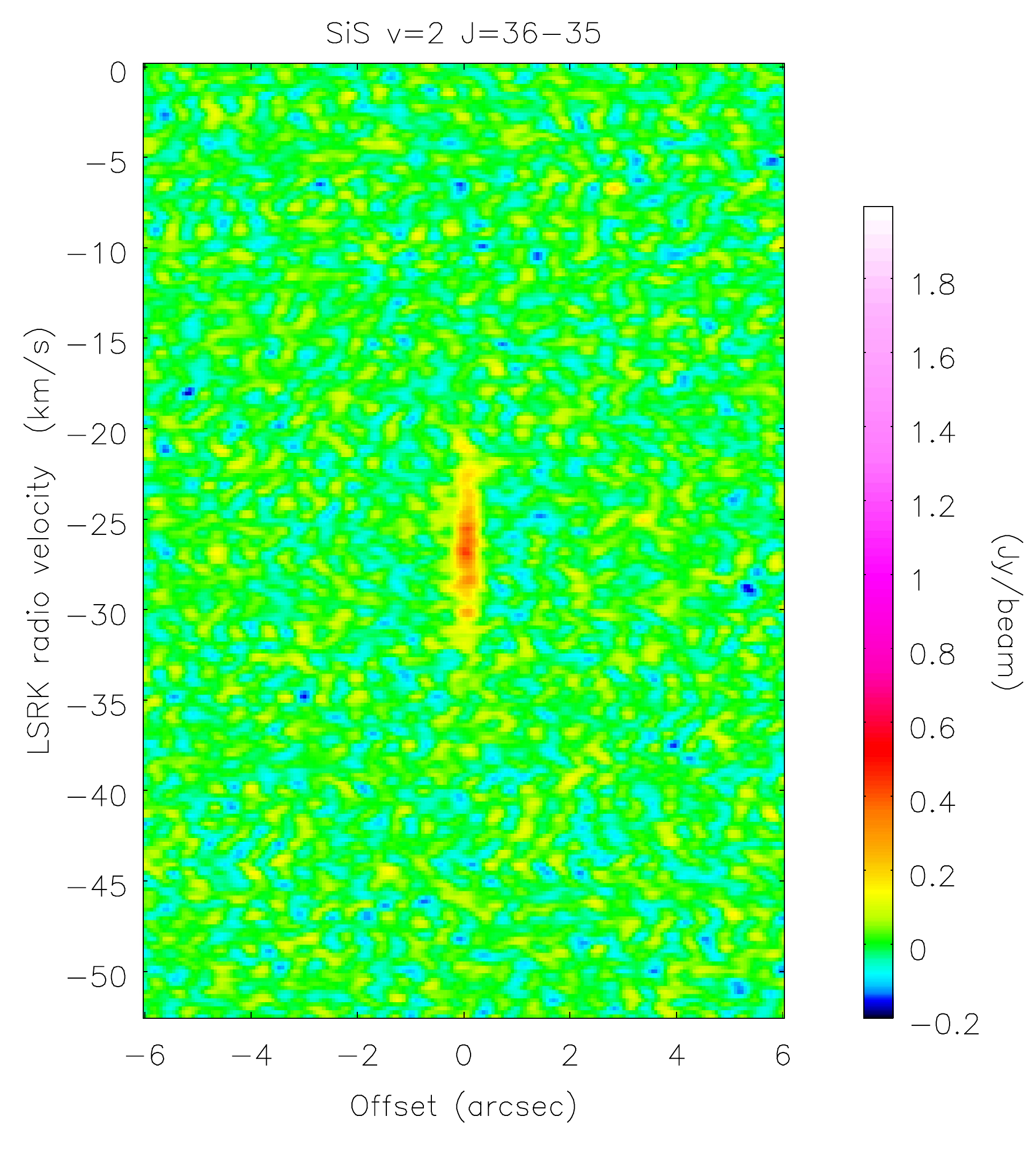}}}
    \end{minipage}
    \begin{minipage}[t]{5.8cm}
        \centerline{\resizebox{\textwidth}{!}{\includegraphics{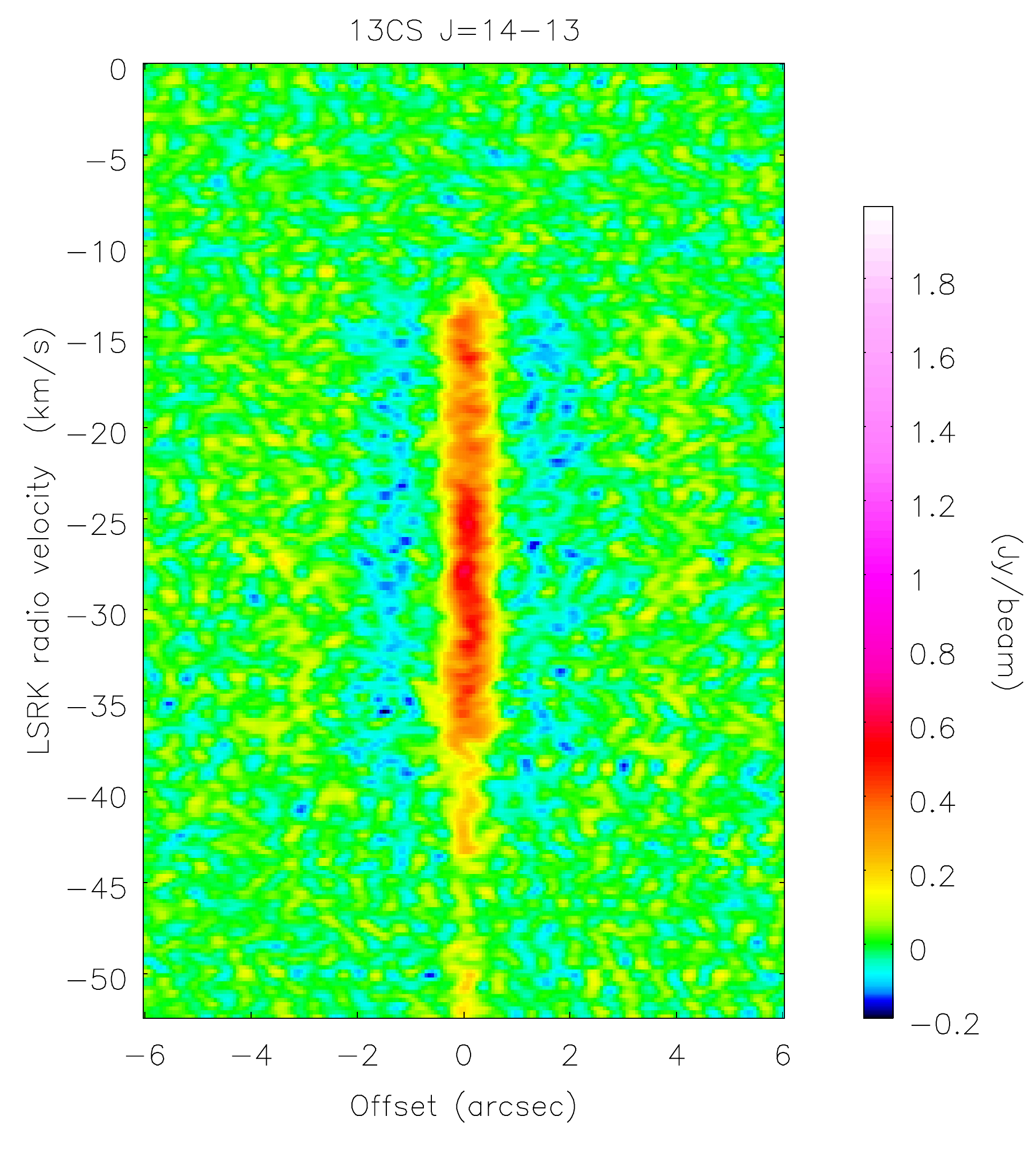}}}
    \end{minipage}
\caption{Position-velocity diagram in right ascension for SiS v=2 J=36-35 at 646.100\,GHz \textit{(left)} and $^{13}$CS J=14-13 at 647.076\,GHz \textit{(right)}. Note that an offset to the East is negative, and to the West positive. Sidelobe-effects are seen in the right image, as well as the fact that the $^{13}$CS J=14-13 line is blended in the blue wing with a 'U'-line. }
\label{Fig:PV_other_excitation}
\end{figure*}

Using the {\sc CASA} task {\sc impv} position-velocity diagrams have been calculated for all transitions. In the standard set-up, 201 slits with a width of 1 pixel (\,=\,0.060\arcsec) and a length of 101 pixels (\,=\,6.06\arcsec) were taken.  Each slit is then centered perpendicular to the direction slice (i.e.\ in this case 3.03\arcsec\ above and 3.03\arcsec\ below the central pixel) and the average for each slit is taken. A PV along the right ascension axis hence has PA=90\deg, and along the declination axis PA=0\deg. The strength of PV plots is that one can correlate structure at different spatial offsets, which is otherwise a challenging task from channel maps alone.

As discussed in Sect.~\ref{Sec:data_reduction}, the PV diagrams are not free from artifacts. (1)~For lines with an integrated line peak flux larger than 0.4\,Jy/beam, sidelobe effects are clearly visible at $-$4\arcsec\  and $+$4\arcsec\ offset (see right panel in Fig.~\ref{Fig:PV_other_excitation}). (2)~The instrument resolves out some flux for most of the lines, which is especially true for the emission of the CO isotopologues (see Sect.~\ref{Sec:Herschel_ALMA}). In some PV diagrams, one can see emission extending beyond the $-10$ to $-40$\,km/s range (covering the radially outflowing wind), which is due to 
blending with molecular line transitions at slightly different frequencies (see right panel in Fig.~\ref{Fig:PV_other_excitation}).

For all transitions, the channel maps and PV diagrams indicate that (some part of) the emission originates from the region just above the stellar photosphere, i.e. within $\sim$0.2\arcsec\ or 10\,\Rstar\ (see online Appendix). The ALMA data prove that SiO is formed in the inner wind region indicating an active photochemical region where shock-driven chemistry \citep{Cherchneff2011A&A...526L..11C} and/or photodissociation of molecules in a clumpy medium \citep{Agundez2010ApJ...724L.133A} dictate the chemical balance.

\begin{figure*}[htp]
\begin{minipage}[t]{12cm}
 \includegraphics[width=5.8cm]{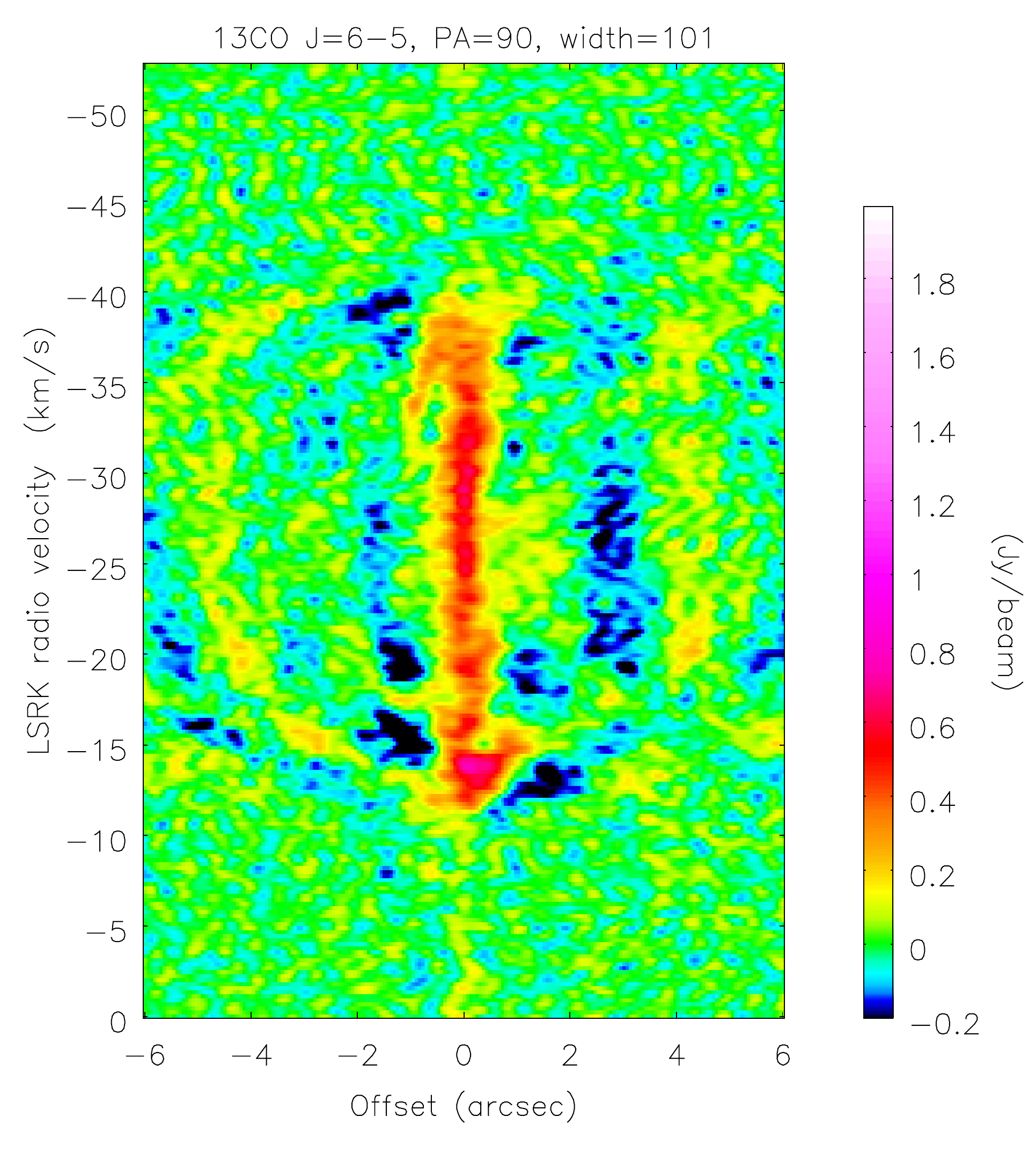}
\includegraphics[width=5.8cm]{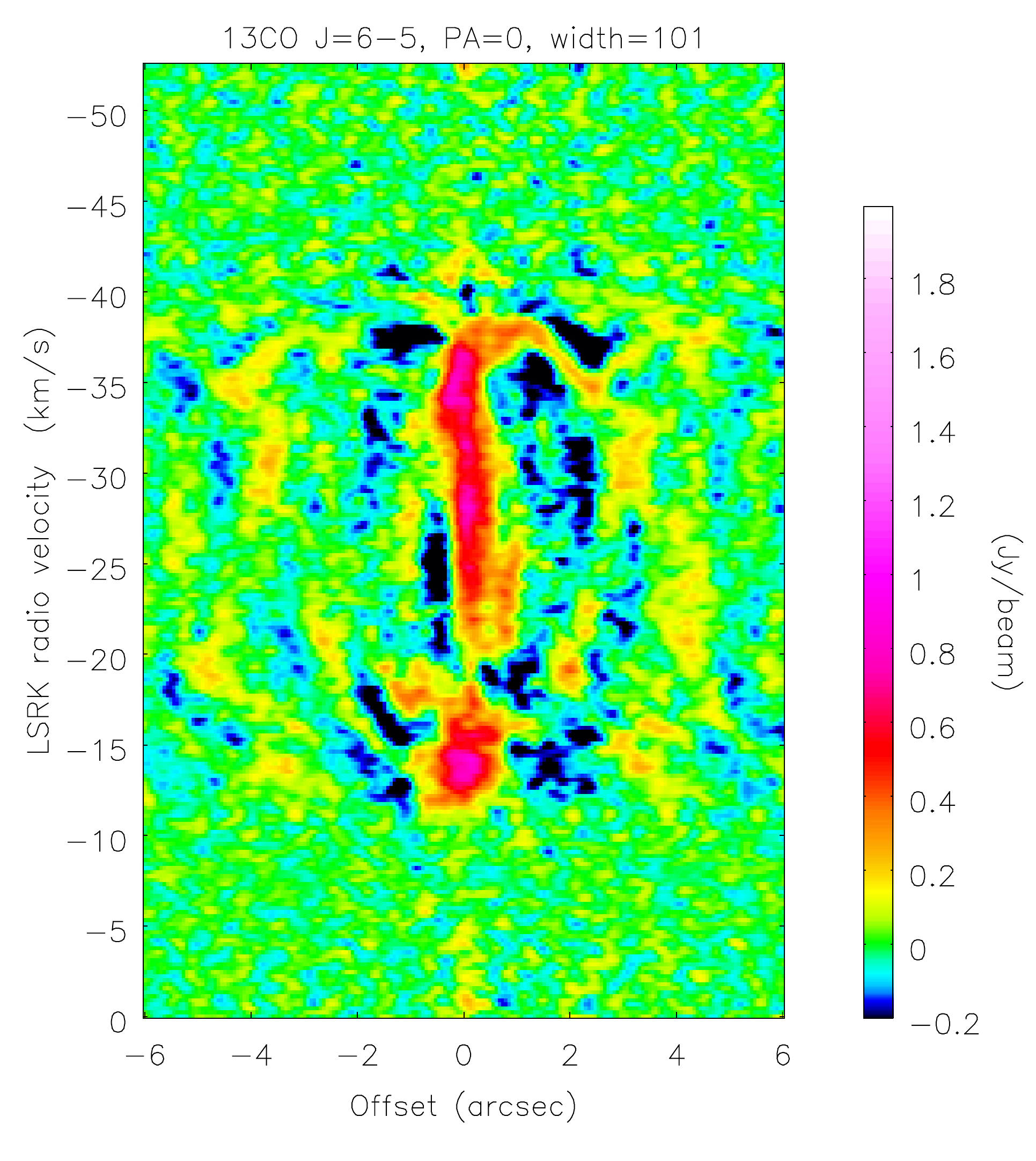}
\end{minipage}
\newline
\begin{minipage}[b]{12cm}
\includegraphics[width=5.8cm]{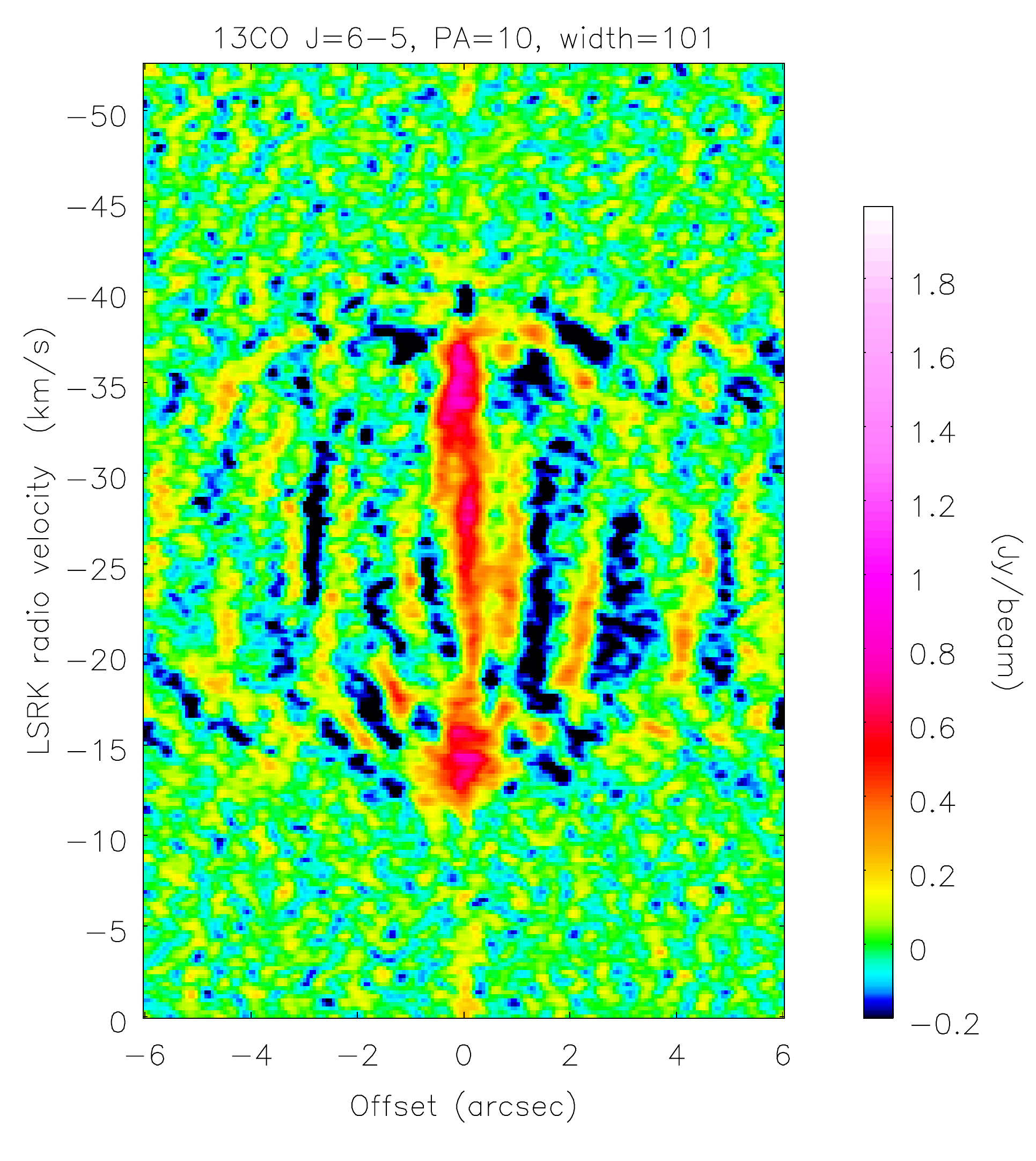}
\includegraphics[width=5.8cm]{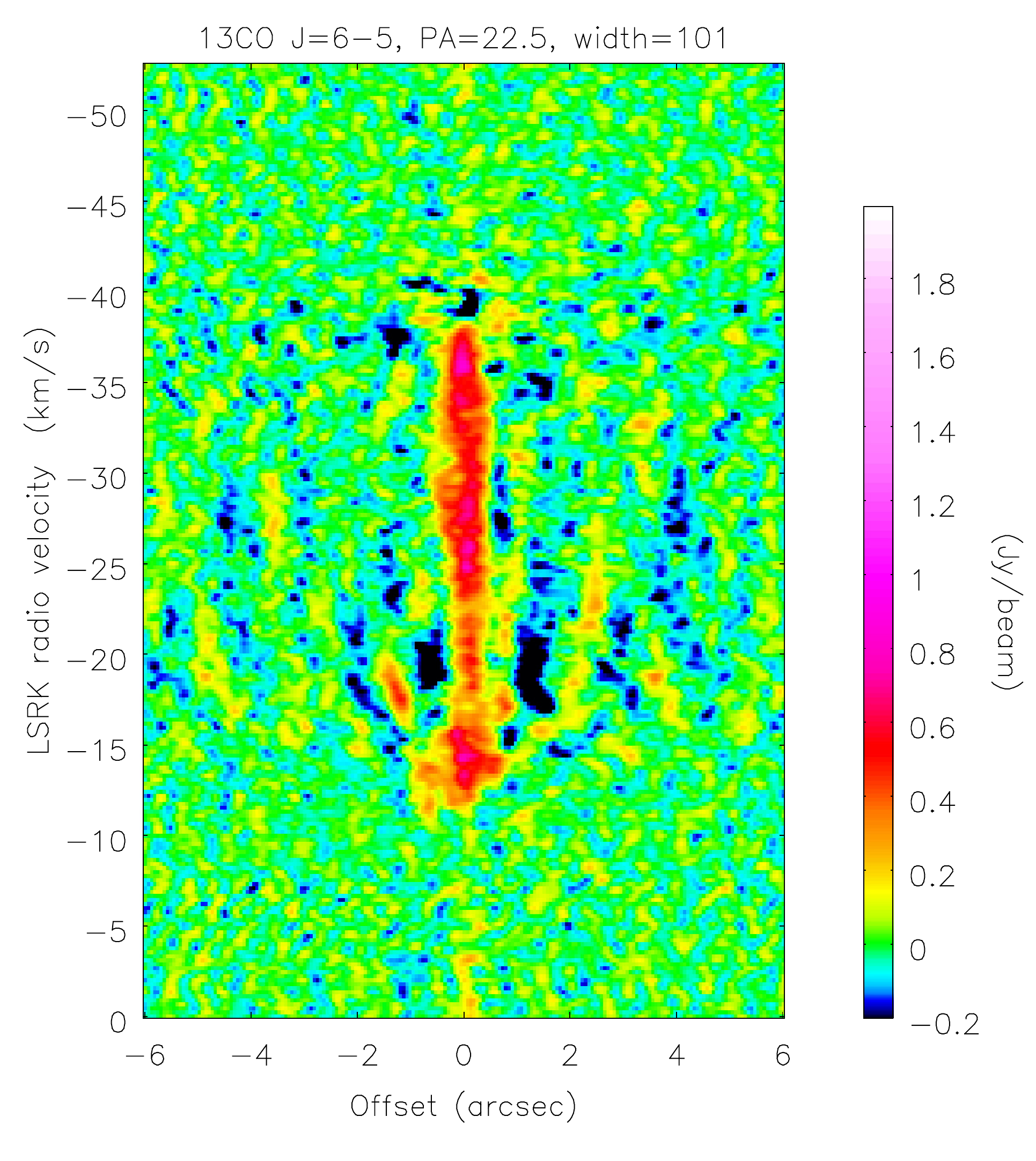}
\end{minipage}
\hfill
\begin{minipage}[b]{6cm}
\vspace*{-2ex}
 \caption{Position-velocity diagram of $^{13}$CO J=6-5 at 661.067\,GHz for a slit width of 101 pixels at a position angle (PA, measured from North to East) of (\textit{top left:}) 90\deg\ (i.e., along the right ascension; offset to the East is negative, and to the West positive), (\textit{top right:}) 0\deg\ (i.e., along the declination; offset to the North is negative and to the South positive), (\textit{bottom left:}) 10\deg, and (\textit{bottom right:}) 22.5\deg.}
\label{Fig:PV_13COJ6}
\end{minipage}
\end{figure*}

\begin{figure*}[htp]
\sidecaption
\includegraphics[width=12cm]{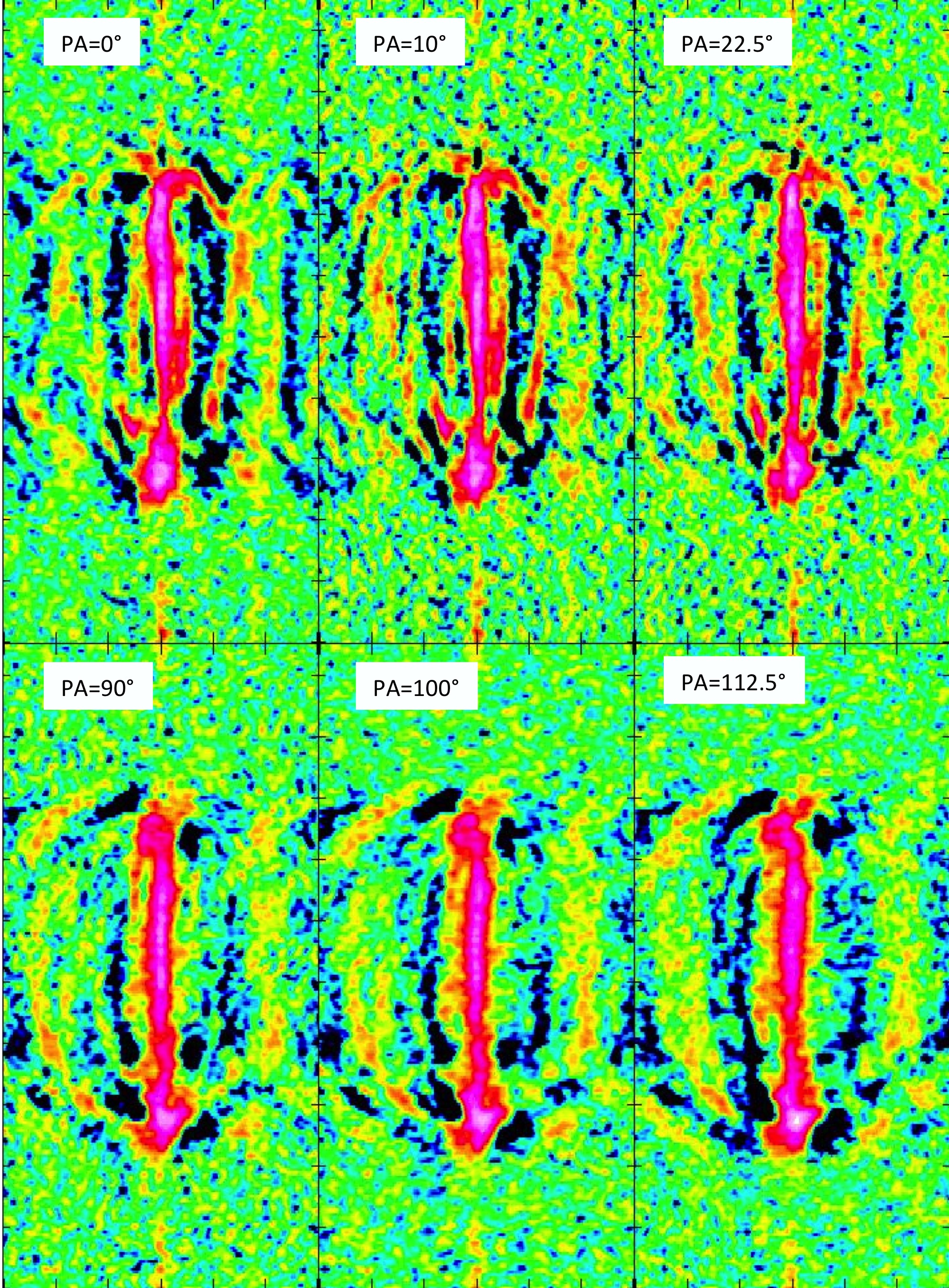} 
\caption{Position-velocity diagram of $^{13}$CO J=6-5 for a slit width of 51 pixels at different position angles, indicating in the top left corner of each panel. The same color wedge has been used as in Fig.~\ref{Fig:PV_13COJ6}. The position angles at top and bottom row differ by 90\deg. The offset (x-axis) ranges from $-$6\arcsec to 6\arcsec, the velocity (y-axis) from 0 to $-$55\,km/s.}
\label{Fig:PV_13COJ6_other_angles}
\end{figure*}

Fig.~\ref{Fig:PV_13COJ6} shows the PV diagrams of  the $^{13}$CO J=6-5 channel maps for a direction slice in right ascension or in declination. These PV diagrams display a more complex morpho-kinematical structure compared to the PV diagrams in Fig.~\ref{Fig:PV_other_excitation}, or any other line. This is not unexpected as the $^{13}$CO J=6-5 line is the strongest in the frequency range we covered, as discussed in Sect~\ref{Sec:channel_maps}. It is clear that the curved distribution of the $^{13}$CO PV maps represents the genuine morpho-kinematical shape, but that the actual distribution of the emission on scales larger than $\sim$3\arcsec\ is artificially fragmented due to the gaps in UV spacing (see Sec.~\ref{Sec:data_reduction}). By changing the position angle of the PV diagram, one can diagnose changes in the morpho-kinematical structure. The most revealing $^{13}$CO J=6-5 PV-diagram is obtained for position angles around 
10-25\degree\ to the North-East (see Fig.~\ref{Fig:PV_13COJ6}). To show and enhance the effect of the small-scale structure, we have also calculated these PV diagrams for a slit of 51 pixels (instead of 101 pixels in the standard setup); see bottom panels in Fig.~\ref{Fig:PV_13COJ6_other_angles}. In all panels of Fig.~\ref{Fig:PV_13COJ6_other_angles}, one can clearly distinguish correlated curved structures ranging from $-40$ to $-12$\,km/s and the distribution of the brightest emission around 0\arcsec\ offset displays an `{$S$}'-shape.

\section{Qualitative interpretation of the $^{13}$CO J=6-5 ALMA data} \label{Sec:qualitative}

The $^{13}$CO J=6-5 PV diagrams in Fig.~\ref{Fig:PV_13COJ6_other_angles} show a very characteristic shape with correlated structure at an interval of $\sim$1\arcsec\ (around the systemic velocity) for position angles around 10-20\deg. These structures are almost absent in the PV diagrams for PA$\sim$112.5\deg. In almost all PV diagrams a typical `{$S$}'-shape is seen around zero offset.
Based on the simulations of \citet{Kim2013ApJ...776...86K}, we postulate that this type of morphology hints towards the presence of a spiral structure induced by a binary companion. In contrast to R~Scl for which the ALMA data have recently shown the presence of a spiral seen almost face-on \citep{Maercker2012Natur.490..232M}, the ALMA data of CW~Leo are reminiscent of a spiral seen almost edge-on, with the orbital axis at an angle of $\sim$22.5\deg\ to the North-East (see Sect.~\ref{Sec:Similarities}). 
Our ALMA images capture fragments of the spiral arms, seen almost edge-on, with a typical width of $\sim$300\,mas.  We cannot be sure whether the fragmentation is an instrumental artifact, but in the case of CIT~6 the 0\farcs7-resolution data analysed by \citet{Kim2013ApJ...776...86K}
show that the structure in the spiral arms is not smooth.  High spatial resolution observations 
\citep{Tuthill2000ApJ...543..284T, Weigelt2002A&A...392..131W, Menut2007MNRAS.376L...6M} and hydrodynamical models 
\citep{Woitke2006A&A...452..537W} have shown that gas and dust clumps are created in the inner envelope by the wind formation mechanism. These density inhomogeneities will remain in the inner spiral-arm structures, until they are eventually dispersed.

\begin{figure*}[hbtp]
\sidecaption
\includegraphics[width=12cm]{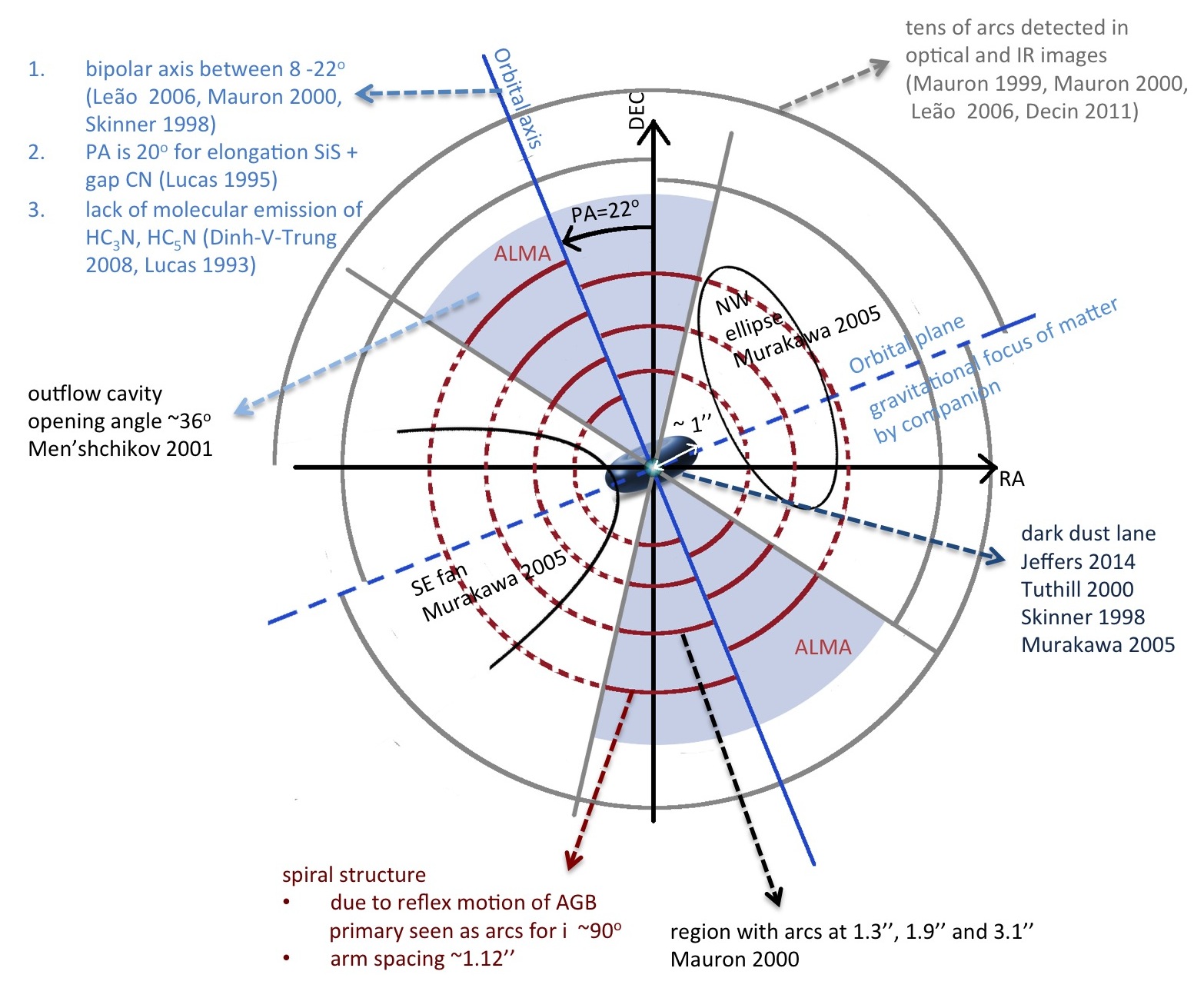}
 \caption{Sketch of the inner wind region of CW~Leo indicating the different observations/results which signal the presence of a binary companion. From the modeling of the ALMA data, we infer an orbital axis with a PA of $\sim$20\deg\ (Sect.~\ref{Sec:Similarities}). The reflex motion of the primary AGB star results in a one-armed spiral structure that is seen almost edge-on and with an extent almost reaching the orbital axis (full red arcs). The dust lane with a radius of $\sim$0.5--1\arcsec\ situated in the orbital plane partly impedes the expansion of the spiral shock in the orbital plane direction (illustrated by the dashed red arcs).}
 \label{Fig:sketch}
\end{figure*}

No direct evidence has yet been found on the presence of a binary companion. However, the current ALMA data in combination with a plethora of other observational diagnostics (schematized in Fig.~\ref{Fig:sketch}) suggest that CW Leo is part of a binary system. (1)~\citet{Guelin1993A&A...280L..19G} have suggested that CW~Leo is a binary since the clumpy shells observed in MgNC, C$_4$H, and other carbon-chain molecules and radicals appear off-center from the continuum source by 2--3\arcsec\ to the East. The cause of such a drift could be an acceleration of the star due to a companion. (2)~As mentioned in Sect.~\ref{Sec:introduction}, the central regions around CW~Leo show the presence of an axi-symmetric peanut- or bipolar-like structure with the major axis lying at a position angle of $\sim$8--22\deg\ \citep[see Fig.~\ref{Fig:sketch};][]{LeBertre1989Msngr..55...25L, Skinner1998MNRAS.300L..29S, Mauron2000A&A...359..707M, Kastner1994ApJ...434..719K}. This is an indirect evidence of the possible existence of an 
equatorial density 
enhancement of dust and gas, with the equatorial plane being located perpendicular to the major axis of the nebula, enhancing scattered star light emission through the bi-conical openings. 
A direct detection of an equatorial dust lane seen almost edge-on was recently presented by 
\citet{Jeffers2014}, corroborating the results of \cite{Murakawa2005A&A...436..601M}. The inferred radius of the dust lane is $\sim$0.5--1\arcsec. The elongated East-West structure seen in our ALMA continuum images at a PA of $\sim$ 128\deg\ with a length of 1.8\arcsec\ across (Fig.~\ref{Fig:Flux-4} and Fig.~\ref{Fig:13CO_integrated_dust}) might be the signature of this equatorial density enhancement\footnote{Although we note that this direction is close to the direction of elongation of the natural synthesised beam, and may be coincidence (see Sect~\ref{Sec:dust}).}, while the extended emission at a PA of $\sim$20\deg\ and $\sim$200\deg\ might reflect the biconical cones. This type of morphology can be explained in terms of a binary system in which angular momentum provides a natural way to define the equatorial plane and polar axis. Interaction between the secondary and expanding red giant primary star will create a bipolar nebula and the dissipative tidal interaction might eventually lead to 
a merger of the two stars in a common envelope system \citep{Morris1987PASP...99.1115M}. 
(3)~Optical and infrared data proved the presence of multiple, almost concentric, shells (or arcs) out to 320\arcsec\ from the central star \citep{Mauron1999A&A...349..203M, Leao2006A&A...455..187L, Fong2003ApJ...582L..39F, Decin2011A&A...534A...1D}. The shells are incomplete and cover $\sim$30\deg--90\deg\ in azimuth. These shells might represent the limb-brightened spiral arms. The shell-intershell density contrast deduced from the observations is about a factor 3--10, in agreement with recent hydrodynamical simulations for binary induced spiral patterns as calculated by \citet{Kim2012ApJ...759...59K}.
(4)~Previous observations by \citet{Guelin1993A&A...280L..19G}, \citet{Lucas1999IAUS..191..305L} and \citet{Dinh2008ApJ...678..303D} of molecular shells show a lack of molecular emission at position angles of $\sim$0\deg--30\deg. This position angle aligns with our inferred model rotational axis of the spiral structure (see Sec~\ref{Sec:Similarities}). As shown by \citet{Mastrodemos1999ApJ...523..357M} a binary induced spiral structure has a lower density along the orbital axis, which might result in a lack of molecular emission along the inferred rotational axis.

\citet{Mauron2000A&A...359..707M} argued against the binary scenario based on the irregular arc spacing seen in the optical and infrared images. However, one has to realize that the shells are projections in the plane of the sky of a complex 3D structure. A binary companion in a smooth envelope structure would indeed create a spiral structure with regular arm spacing. However, the wind and envelope creation does not occur in a homogeneous way, adding complexity to the morpho-kinematical structure. Magnetic cool spots  during the active phase \citep{Soker1999MNRAS.307..993S} and/or large scale photospheric convective cells \citep{Freytag2008A&A...483..571F} can locally reduce the gas temperature, which might locally enhance the dust formation and thus lead to gas and dust clumps. The magnetic cool spots cover only a fraction of the stellar surface, and this inhomogeneity might explain the clumpiness seen on smaller spatial scales and in the arcs\footnote{The ALMA instrumental effects also result in a 
superficial fragmentation of the spiral structure (see Sect.~\ref{Sec:PV_radial} and \citet{Maercker2012Natur.490..232M}).}. Star-spots have not yet been observed on AGB stars, although recent observations of 
linear polarization in CO J=3-2, SiS J=19-18 and CS J=7-6 suggest a complex magnetic field configuration in the wind of CW~Leo with a strength around 50--300\,mG at $\sim$3\arcsec\ offset of the central source \citep{Girart2012ApJ...751L..20G} and recent interferometric data show strong evidence for inhomogeneities in the molecular photospheric layers of AGB stars \citep{Wittkowski2011A&A...532L...7W}. In addition, modulations in the density structure might be a natural outcome from the dust formation process during which a complex non-linear interplay between gas-grain drift, grain nucleation, radiation pressure, and envelope hydrodynamics create density irregularities. On top of that, flow instabilities (as, e.g., Rayleigh-Taylor instabilities) have time to fragment the outward moving density structures and can produce numerous small-scale cloud-like sub-structures \citep{Woitke2006A&A...452..537W}.

 \section{Modeling a binary-induced spiral structure} \label{Sec:Shape}
 A full reconstruction of the intensities, the 3D shell pattern and its kinematics using the ALMA $^{13}$CO channel maps and PV diagrams requires detailed hydrodynamical simulations and radiative transfer calculations. This is beyond the scope of the current paper, but will be presented in a forthcoming paper (Homan et al., in prep.).
 In this section, we want to demonstrate how different kinds of density fluctuations impact a PV diagram and how the morphology in the current ALMA PV data points towards the influence of a binary companion in the shaping of the wind structure around CW~Leo. The simulations are obtained using the {\sc Shape} software developed by \citet{Steffen2006RMxAA..42...99S} and \citet{Steffen2011}.
 
 In this section, we first shortly describe the impact of a binary companion on the wind envelope structure (Sect.~\ref{Sec:binary}). Then, we present  {\sc Shape} simulations, where we gradually increase the complexity of the envelope structure (Sect.~\ref{Sec:Shape_models}). In Sect.~\ref{Sec:Similarities}, we demonstrate the similarities between the ALMA $^{13}$CO PV diagrams and PV diagrams based on spiral structures.

 \subsection{Binary-induced spiral structures}\label{Sec:binary}
 
As shown in the seminal work of \citet{Mastrodemos1999ApJ...523..357M}, a binary companion might produce a spiral pattern in the circumstellar wind material. \citet{Kim2012ApJ...744..136K} and \citet{Kim2012ApJ...759...59K} have studied the separate effects of the orbital motions of the individual stars and have shown that two types of spiral patterns are created. Firstly, a more flattened spiral pattern confined within a very limited height from the orbital plane is created due to the companion's motion \citep{Kim2012ApJ...744..136K}. Secondly, a spiral-shell-shaped pattern is created due to the orbital motion of the mass-losing star around the center of gravity \citep{Kim2012ApJ...759...59K}.  These two spiral patterns are different in several ways. (1)~The companion's wake is attached to the companion, while the pattern due to the motion of the mass-losing star has a stand-off radius defined by the orbital and wind velocity \citep[see Eq.~7 in][]{Kim2012ApJ...759...59K}. (2)~Secondly, the propagation 
speeds of the patterns determining the shape are different. However, in the case of AGB stars, this difference is very small since the propagation speed of all patterns is close to the wind speed, which dominates over the orbital and sound speed. (3)~Thirdly, while the arc pattern due to the reflex motion of the mass-losing star nearly reaches the orbital axis and introduces an oblate-shaped flattening of the circumstellar envelope density, the direct effect of the companion results in a spiral structure confined toward the orbital plane. When combining both types of spiral structures, the hydrodynamical models show the presence of clumpy structures within the vertical extension limit of the companion's wake due to shocks. 

\citet{Kim2013ApJ...776...86K} presented PV diagrams at different inclinations for simulations including both types of spirals for the carbon star CIT~6. Note that a spherical central region of 2\arcsec\ was carved out in their simulations to mimic the central hole seen in the observed HC$_3$N data. The PV diagrams in their simulations are dominated by the spiral induced by the reflex motion. In Sect.~\ref{Sec:Shape_models}, we will show that for spiral structures quite confined towards the orbital plane and for position angles significantly different than the orbital axis or plane, an $S$-type feature around offset zero appears in a PV diagram around offset zero when seen almost edge-on; a feature also seen in the ALMA data (see Fig.~\ref{Fig:PV_13COJ6}). 

 \subsection{Morpho-kinematical simulations}\label{Sec:Shape_models}
 
To understand the complex morpho-kinematical structure seen in the $^{13}$CO PV diagrams and link the ALMA data to a 3D shell pattern, we have used the {\sc Shape} modeling tool \citep{Steffen2006RMxAA..42...99S, Steffen2011}. {\sc Shape} is a flexible interactive 3D morpho-kinematical modeling application for astrophysics, that is publicly available. By interactively defining 3D structures, one can calculate intensity maps, PV diagrams, channel maps and spectra. While eulerian 3D grid-based hydrodynamic simulations are possible, in this work we use a purely mathematical description of the object structure (see below). A 3D mesh is constructed, which serves as a container of the emissivity and velocity field. If a radiation transfer computation is done, the information in the mesh is then transfered to a regular cartesian 3D grid, which is used to compute the radiation transfer. The asset of {\sc Shape} is that it is computationally very fast, which facilitates a first broad screening of the huge 3D 
kinematical and morphological parameter space. The deduced model parameters can then be used as input for a more detailed (hydrodynamical) simulation. Recently, a non-LTE (non local-thermodynamic equilibrium) radiative transfer solver ({\sc Shapemol}) has been added to compute molecular excitation levels \citep[][Santander-Garci\'ia et al.\ (2014), submitted]{Santander2012A&A...545A.114S}. This solver is based on the well-known LVG \citep[Large Velocity Gradient,][]{Castor1970MNRAS.149..111C} approximation, which significantly simplifies the radiative transfer problem. The current version of {\sc Shapemol} still has limitations when calculating accurately the molecular level populations in an AGB wind: it only includes collisional rate constants for temperatures up to 1000\,K and H$_2$ densities between 1$\times$10$^{13}$ to 1$\times$10$^{8}$\,m$^{-3}$, rotational levels in the vibrational excited states are not included,  and the tabulated densities are sometimes too low for the high densities 
encountered in the inner wind \footnote{Tabulated H$_2$ densities in {\sc Shapemol} range from  1$\times$10$^{13}$ to 1$\times$10$^{8}$\,m$^{-3}$, while the wind of CW~Leo has a H$_2$ density around 1$\times$10$^{15}$\,m$^{-3}$ at 2.5\,\Rstar\ (or 0\farcs05), reaching  6.6$\times$10$^{10}$\,m$^{-3}$ at 6\arcsec. Since the interest of this paper is not to reproduce the $^{13}$CO intensity maps but to analyse the morpho-kinematical structure in the inner envelope, we have divided the overall density by a factor 100.}. Nonetheless, it can still be used for a first morpho-kinematical interpretation of the structure seen in the ALMA PV data of $^{13}$CO. 

Based on the results of \citet{Decin2010A&A...518L.143D} and \citet{DeBeck2012A&A...539A.108D}, we chose as input for our basic models a distance of 150\,pc, a constant gas mass-loss rate of 1.5$\times$10$^{-5}$\,\Msun/yr, a fractional abundance [CO/H$_2$] of 6$\times$10$^{-4}$, an isotopologue ratio  $^{12}$CO/$^{13}$CO of 30, a stellar radius of 4.55$\times$10$^{13}$\,cm (or 0.02\arcsec\ at 150\,pc) and a wind velocity of 14.5\,km/s.  The gas kinetic temperature is assumed to follow a power law
\begin{equation}
  T(r) = \Teff \left(\frac{R_{\star}}{r}\right)^{\zeta}\,,
\end{equation}
with \Teff\ the effective temperature of 2330\,K and $\zeta \approx 0.5$. The molecular line data (energy levels, Einstein A coefficients and collisional rates) have been taken from the LAMDA database \citep{Schoier2005A&A...432..369S}.

We produced models of the inner 6\arcsec\ (i.e., total width of 12\arcsec), at $\sim$47\,mas resolution, for
spectral channels separated by 0.4\,km/s. We simulated the effect of the ALMA setup using the CASA task {\sc simobserve}, with antenna
positions and atmospheric conditions matching those used for our
observations. In order to obtain the same total duration (and range of
$uv$ spacings sampled) the simulations have slightly longer on-target duration.
The simulated  model images and PV plots therefore provide comparable sampling, including missing spacings, as the original observations. The noise in the simulated images is slightly lower, not only due to the longer time on-source, but because although atmospheric and instrumental noise affecting the target is included, the limitations due to other imperfections in phase calibration are not. The plots of simulated data do have similar dynamic range limitations and artefacts due to missing flux (where the input model has large-scale structure) as the observed images.  For example, the absolute value of the most negative emission in the simulated data in the right panels of Fig.~\ref{Fig:PV_radial_outflow} is $\sim30\%$ of the peak.  The look-up tables in some plots have been curtailed in order to highlight the most relevant features for comparison. We used the same {\sc clean} parameters
as for our observations to produce image cubes and position-velocity
plots.

\subsubsection{PV diagram for a radially outflowing wind} \label{Sec:PV_radial}

\begin{figure*}[htp]
      \begin{minipage}[t]{.3\textwidth}
\centerline{\resizebox{\textwidth}{!}{\includegraphics{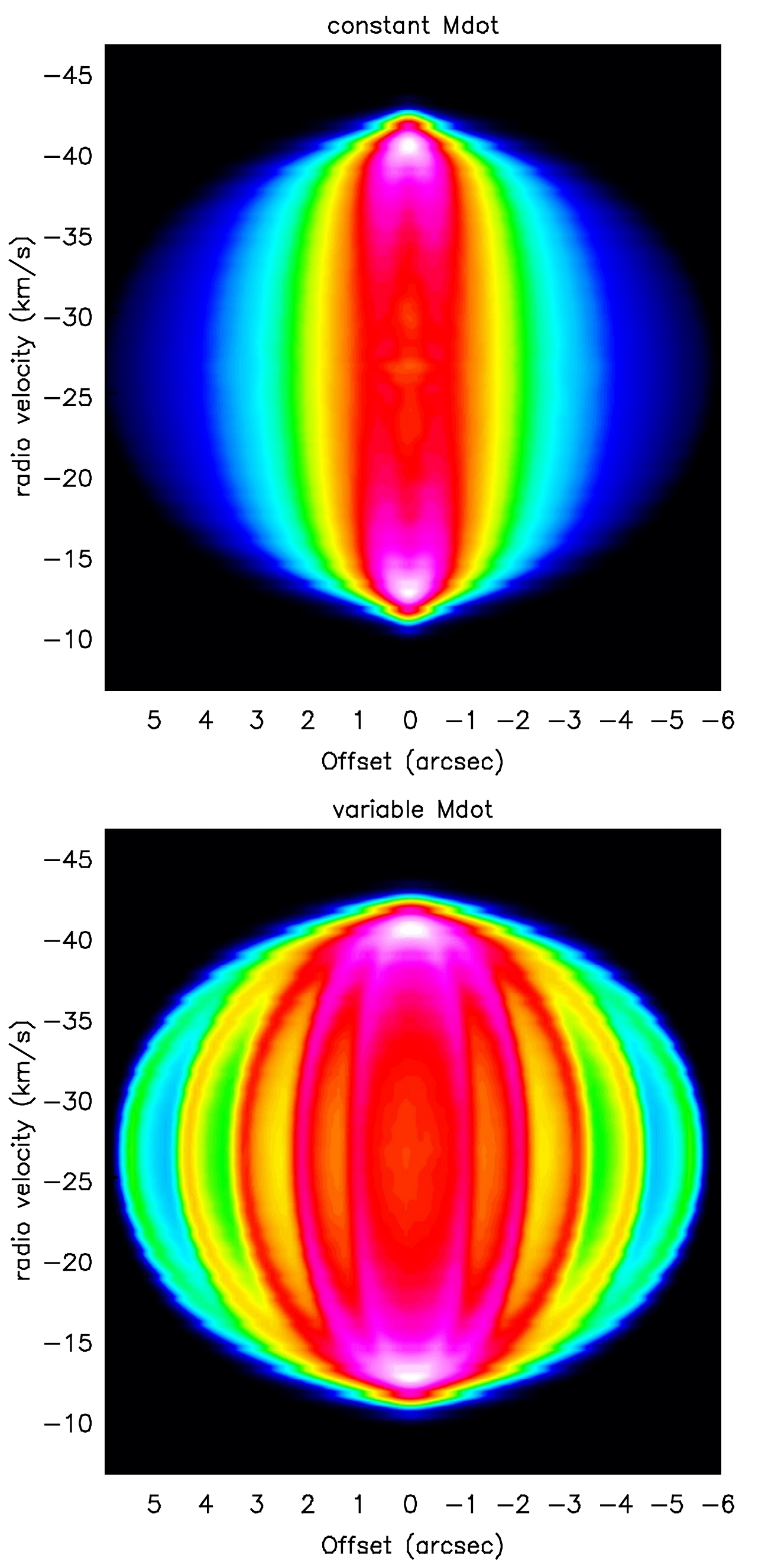}}}
    \end{minipage}
    \hfill
    \begin{minipage}[t]{.69\textwidth}
\centerline{\resizebox{\textwidth}{!}{\includegraphics{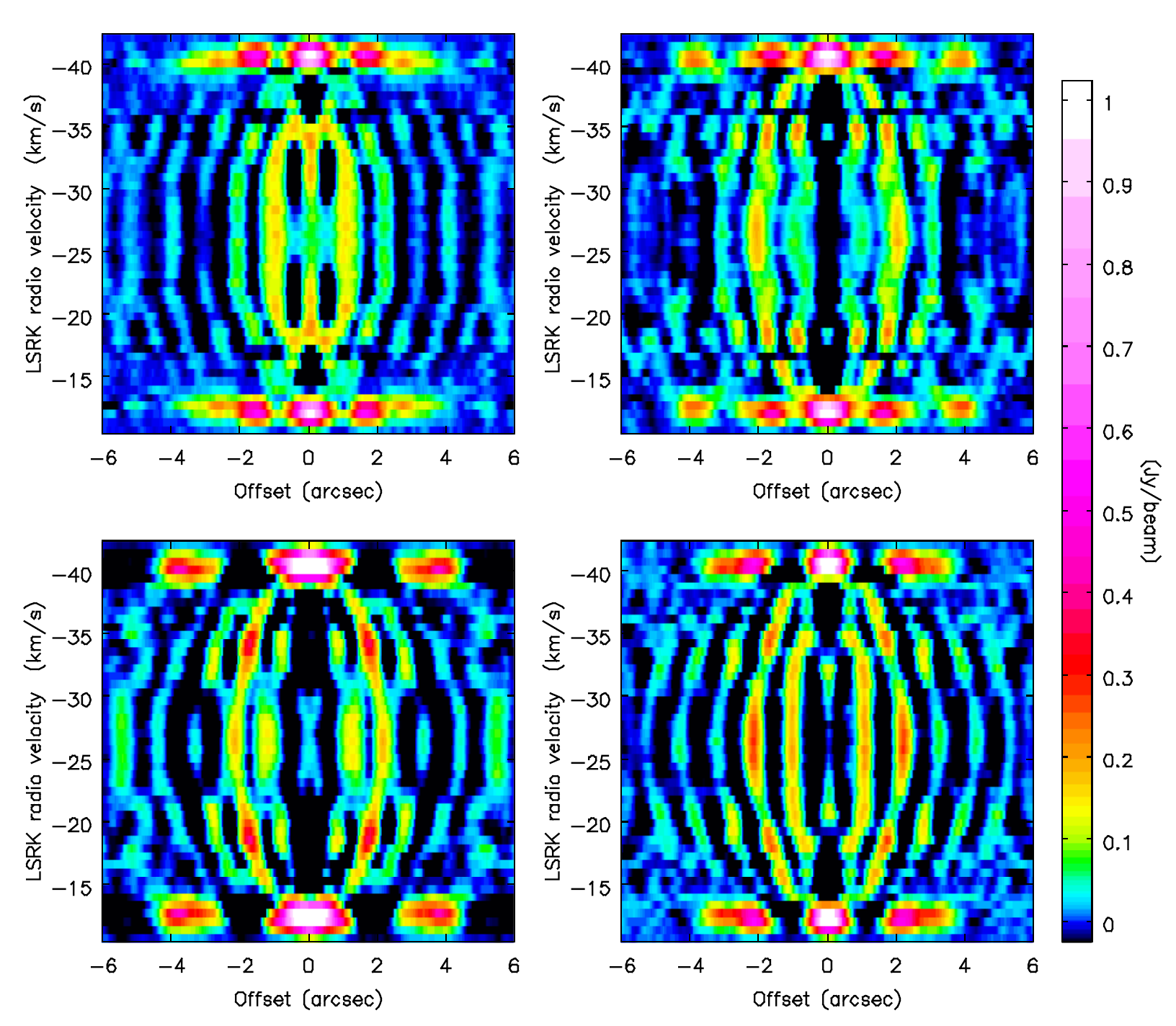}}}
    \end{minipage}
\caption{Simulated $^{13}$CO J=6-5 PV diagrams for a beam of 0\farcs43$\times$0\farcs23.
\textit{Top left:} Model for radially outflowing wind for a constant mass-loss rate ($\rho \propto r^{-2}$).  \textit{Bottom left:} Model for a radial outflowing wind with varying wind density, with the shell-intershell density contrast being 10. The density-enhanced shells are placed with an interval of 1\arcsec. No instrumental effects were taken into account. \textit{Right:} Simulations for the PV-diagram for the density-enhanced shells taking the ALMA instrumental effects into account: in the four sub-panels, the slit is varied to create a PV diagram along the right ascension coordinate, the declination coordinate, and 22.5\deg\ to the north-east (NNE) or to the north-west (NNW).}
\label{Fig:PV_radial_outflow}
\end{figure*}

We first simulate two simple examples for a radially outflowing wind at a constant expansion velocity of 14.5\,km/s. In the first case, the mass-loss rate is assumed to be constant (or density $\rho \propto r^{-2}$; see top left panel in Fig.~\ref{Fig:PV_radial_outflow}); while in the second example (bottom left and right panels in Fig.~\ref{Fig:PV_radial_outflow}) we demonstrate the effect of density-enhanced shells  \citep{Cordiner2009ApJ...697...68C,Decin2011A&A...534A...1D} with a shell-intershell density contrast of a factor 10. The wind region close to the stellar surface is responsible for the bright bar at zero offset in the PV-diagrams, while the outer density-enhanced shells create well-distinguished  correlated curved structures at larger offset in the PV-diagram. Including the ALMA instrumental effects results in a (virtual) fragmentation of the shells (see right panels in Fig.~\ref{Fig:PV_radial_outflow}). Due to the short hour-angle coverage of the simulated observations in 
Fig.~\ref{Fig:PV_radial_outflow} (right panels), the simulated PV diagrams at different position angles are not identical\footnote{Note that for the real ALMA observations this effect is not so strong since they were spread out more in hour-angle due to the longer gaps between scans spent on the calibration sources.}.

The simulated PV diagrams  for the density enhanced shells taking the ALMA instrumental effects into account show some resemblance with the ALMA $^{13}$CO J=6-5 PV diagram for a position angle around 22\degree\ to the North-East (Fig.~\ref{Fig:PV_13COJ6_other_angles}). However, a spherical symmetric geometry and isotropic density wind structure as applied in the current setup can not explain the different morphologies seen in the PV diagram with position angles 90\deg\ apart from each other  and the asymmetry when reflecting around zero-offset (Fig.~\ref{Fig:PV_13COJ6_other_angles}).

\subsubsection{PV-diagram for binary-induced spiral wind structure} \label{Sec:Shape_PV_binaries}

In this section, we simulate the effect on the PV diagram of a spiral structure. Both the spiral structure due to the companion's wake as due to the reflex motion of the mass-losing AGB star are simulated. To do so, we follow the results described by \citet{Kim2012ApJ...744..136K} and \citet{Kim2012ApJ...759...59K} by defining an Archimedean spiral $r = a_p \theta$, with $r$ the radial coordinate, $\theta$ the longitudinal coordinate, and $a_p$ a constant controlling the distance between the successive turnings.

The spiral structure is generated by a thin shell mesh with its position given as the position vector $\vec{r}$ in spherical coordinates $(r,\theta,\phi)$ as a mapping of the unit sphere (1,$\theta$,$\phi$), with $\phi$ the latitude ($\phi=90$\deg\ being the equator). The coordinates of the mesh vertices are given by
\begin{equation}
\vec{r}(r,\theta,\phi) = \left(a_p \theta, \theta, \phi \right)\,,
\label{Eq:spiral}
\end{equation}
with $r$ in units of meter, $\theta$ between 0 and 360\deg, and $\phi$ between 0 and 180\deg. The spacing between two successive turnings is hence $a_p \times 360 \equiv a$, with $a$ in units of arcsec for a distance of 150\,pc.

The overall radial density is taken $\propto r^{-p}$ \citep[with $p\sim2$,][]{Kim2012ApJ...759...59K}, which is then modified by the longitudinal and latitudinal dependencies. To simulate both types of spiral arms (confined toward the orbital plane and extending almost completely toward the orbital axis), we define the latitudinal dependence of the density as
\begin{equation}
 \frac{w^2}{(\phi - 90\deg)^2 + w^2}\,,
 \label{Eq:w}
\end{equation}
in which case small values of $w$ (in units of degrees) generate a spiral arm confined toward the orbital plane. The thickness $\beta$ of the spiral arms is given as a function of distance $r$ from the center as $\beta(r) = b r + c$, where $b$ and $c$ are parameters that determine the growth rate and the initial thickness of the spiral shell.

The different free parameters ($a, p, b, c$, inclination $i$, and $w$) have been changed to study their effect with the aim to understand the morpho-kinematical structure visible in the $^{13}$CO J=6-5 PV diagrams. In Homan et al. (in prep.) a much more detailed study will be presented. Although limited by the short duration of the observations, we estimate a value for the spiral arm spacing $a$ in the range of 1--2\arcsec\ using Fig.~\ref{Fig:13CO_channel}.

\begin{figure}
 \includegraphics[width=.48\textwidth]{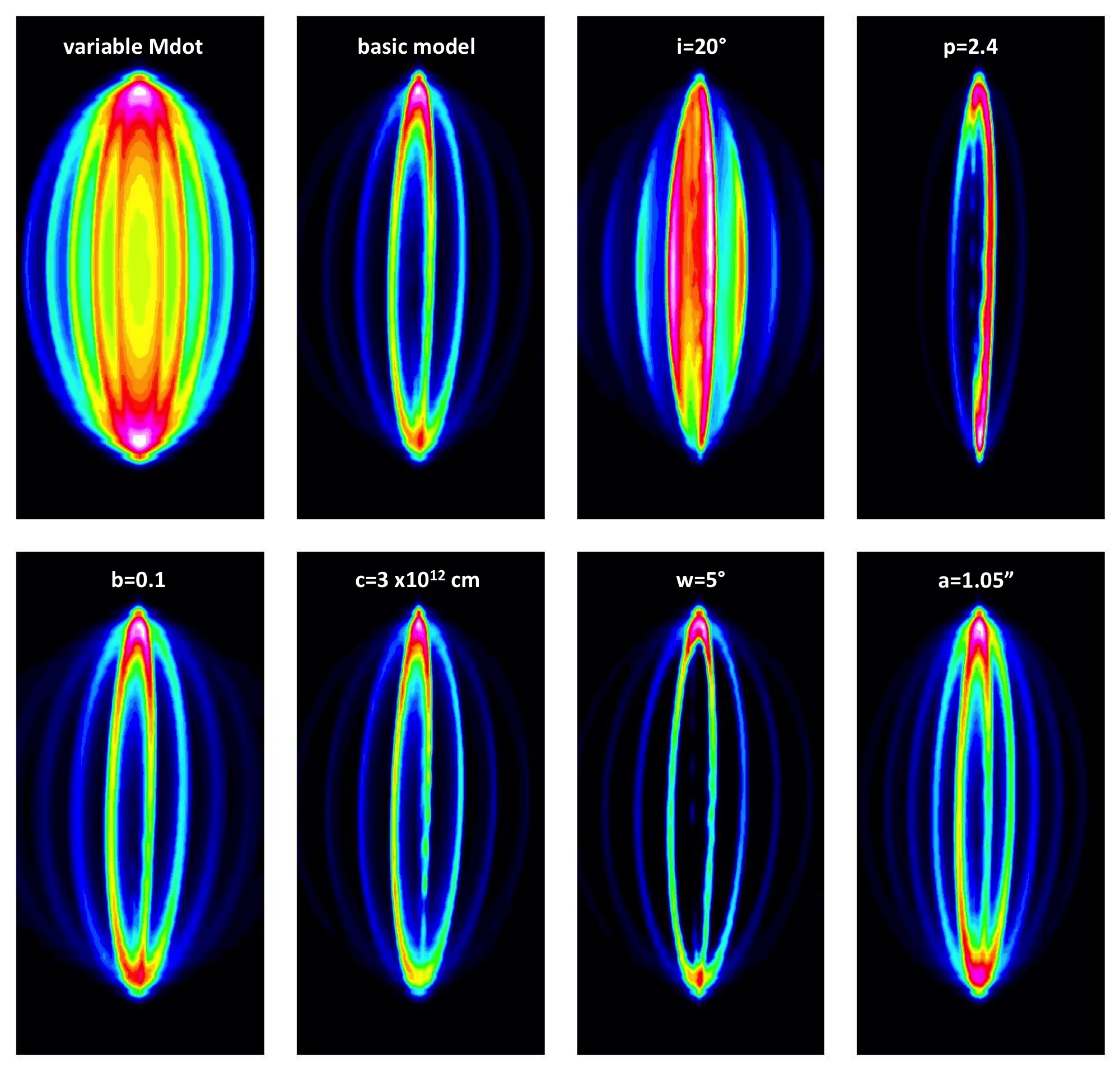}
 \caption{$^{13}$CO J=6-5 PV diagram for different model setups for a spiral structure, with the x-axis going from $-$6 to 6\arcsec and the y-axis from $-20$ to $20$\,km/s: \textit{panel~1:}~radial outflowing wind with density enhanced shells, for a shell-intershell contrast of a factor 10, \textit{panel~2:}~basic model for the spiral wind structure, with parameters $a=1.45$, $p=2$, $b=0$, $c=8\times 10^{12}$\,cm (or 16\,\Rstar), and $w=50$ at a PA along the orbital plane, \textit{panels~3--8:} same as basic model but with one parameter changed, as indicated on the top of each panel. For each panel, the color wedge goes from zero to maximum intensity (in Jy/beam), with redder colors indicating a higher intensity.}
 \label{Fig:PV_overview}
\end{figure}

\begin{figure}
 \includegraphics[width=0.48\textwidth]{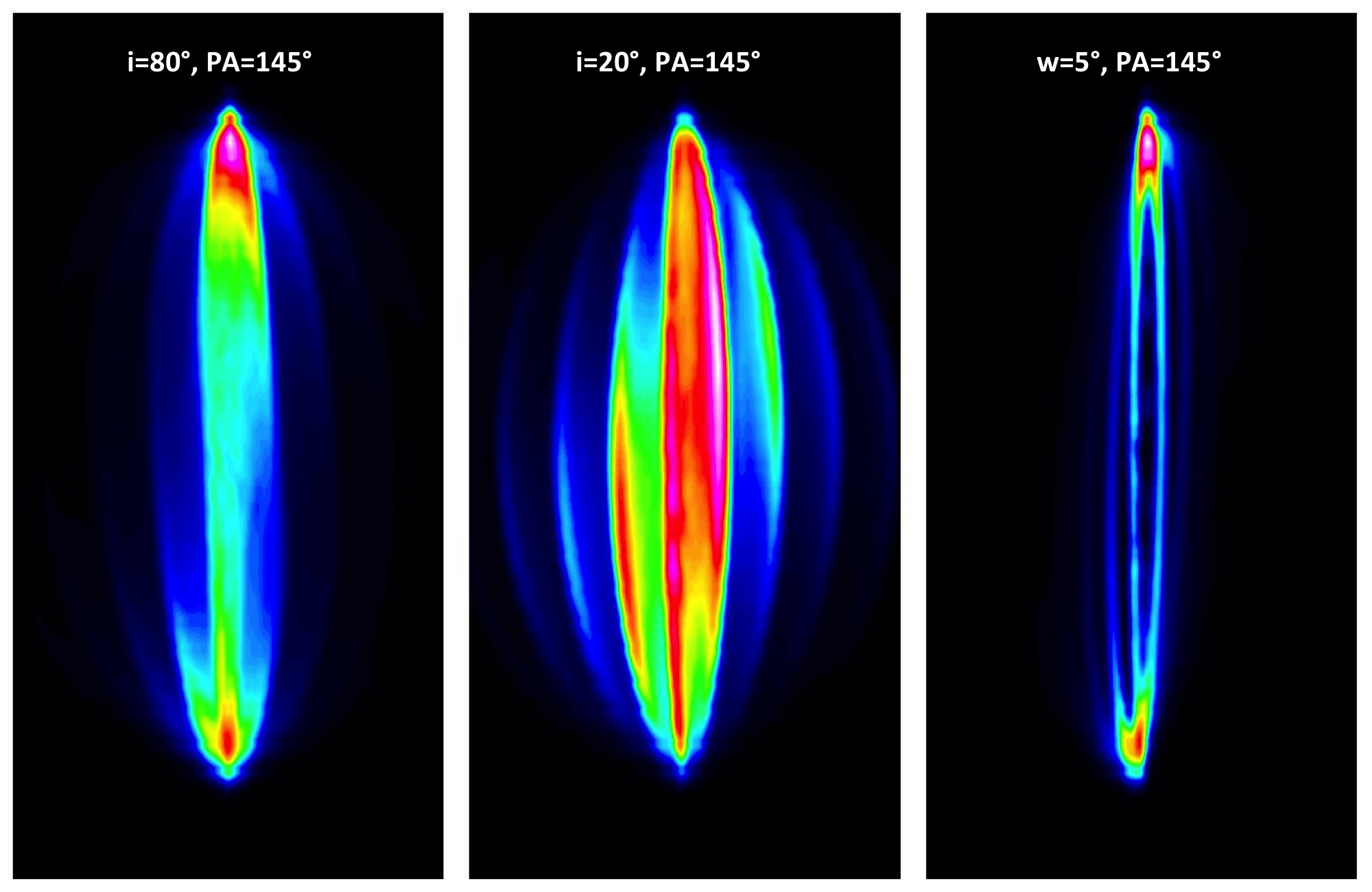}
 \caption{$^{13}$CO J=6-5 PV diagram for a PA of 145\deg\ for the basic model at $i=80$\deg\ (\textit{left}), at $i=20$\deg\ (\textit{middle}), and for $w=5$\deg (\textit{right}). }
 \label{Fig:diff_inc}
\end{figure}

As setup for our basic spiral model, we chose as free parameters $i=80$\deg\ \citep{Jeffers2014}, $a=1.45$\arcsec, $p=2$, $b=0$, $c=8\times 10^{12}$\,cm (or 16\,\Rstar), and $w=50$. Each of the parameters is then changed one after the other to demonstrate its effect (see Fig.~\ref{Fig:PV_overview}). The PV diagram for the basic model of a one-armed spiral at a position angle along the orbital plane (second panel in Fig.~\ref{Fig:PV_overview}) shows very regular structures due to the formation of higher density arcs in a plane perpendicular to the orbital plane \citep[see also Fig.~11 in][]{Mastrodemos1999ApJ...523..357M}, although its signature is slightly more asymmetric as compared to the PV diagram for the radial outflowing wind with density-enhanced shells (first panel in Fig.~\ref{Fig:PV_overview}). At position angles not aligning with the orbital axis or orbital plane, the difference with the radial outflowing wind is much more pronounced, as can be seen in the first panel of Fig.~\ref{Fig:diff_inc}.
At a PA along the orbital plane, the morphological difference between the models for an inclination $i=80$\deg\ and $i=20$\deg\ is mainly seen around zero offset for which a higher intensity contrast is 
reached for the case of $i=20$\deg. For position angles not aligning with the orbital axis or orbital plane, a more distinct morphological difference is seen for different inclination angles (see Fig.~\ref{Fig:diff_inc}): at inclination angle $\ga$60\deg\ an '$S$-type' structure around 
zero-offset is seen in the PV-diagram. This $S$-type signature is also seen in the  $^{13}$CO J=6-5 ALMA images of CW Leo (see Fig.~\ref{Fig:PV_13COJ6}); a feature that can not be reproduced by a model assuming a radial outflowing wind with density-enhanced shells (see Fig.~\ref{Fig:PV_radial_outflow}). The difference in PV diagrams for different position angles at different inclinations will be used to constrain the inclination angle of the spiral arms in CW~Leo (Sect.~\ref{Sec:Similarities}).

As demonstrated by \citet{Kim2012ApJ...759...59K}, the mean density in the binary-induced stellar wind is proportional to $r^{-p}$, with $p$$\sim$2. Fig.~\ref{Fig:PV_overview} shows the effect for increasing the value of $p$ from 2 to 2.4 (although one has to realize that the mass is not conserved for this latter simulation). In the case of $p$$\sim$2.4, the density is more concentrated in the inner arm spirals, clearly visible in the PV diagram.

When the parameter $b$ is larger than zero, the thickness of the spiral arms increases with radial distance from the star, simulating the effect of dispersion. This leads to a lower mean density in the outer spiral arms. In contrast, lowering the initial thickness of the spiral shells from $8\times10^{12}$\,cm to $3\times10^{12}$\,cm results in a higher mean density in the complete spiral structure due to the conservation of mass. Decreasing the parameter $a$ leads to a decrease in the distance between the successive arm turnings of the spiral.

A value for the parameter $w$ around 50 results in the mean density being rather uniformly distributed in latitude, while the much smaller value of $w=5$ results in the latitudinal dependence of the density being more concentrated toward the orbital plane, hence mimicking the effect of gravitational focusing. While at a PA along the orbital axis, the difference is not so large, the latitudinal dependence is more pronounced at PA=145\deg\ (see Fig.~\ref{Fig:diff_inc}).

\begin{figure}[htp]
 \begin{minipage}[b]{.28\textwidth}
 \vspace*{-.3cm}
  \centerline{ \includegraphics[width=\textwidth]{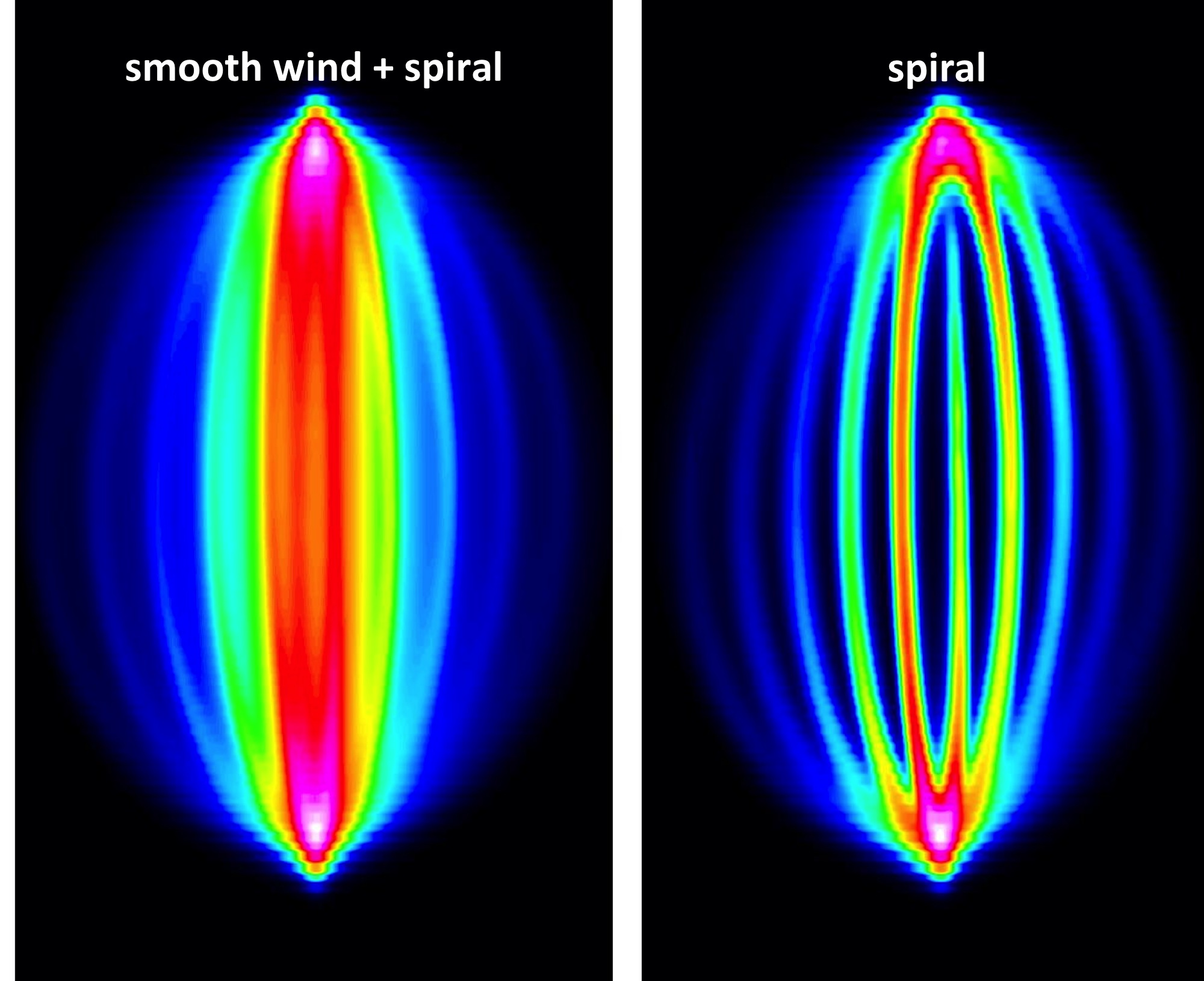}}
  \end{minipage}
  \hfill
 \begin{minipage}[b]{.2\textwidth}
 \centerline{\includegraphics[height=5ex]{blanco.pdf}}
 \caption{PV diagram for a spiral structure with parameters $a=1.12$, $p=2$, $b=0$, $c=3\times 10^{12}$\,cm, and $w=10\deg$, and $i=80$\deg. In the right panel only the spiral structure is modeled, while in the left panel a smooth wind is included with a density contrast of a factor 4.}
 \label{Fig:wind}
 \end{minipage}
\end{figure}

As discussed by \citet{Mastrodemos1999ApJ...523..357M}, the effect of a binary companion is to focus some fraction of the material into a spiral structure, the remainder still forming a quasi-spherical envelope. This smooth wind structure is well-visible in the ALMA PV diagrams as the dominant maximum around zero spatial offset covering the full velocity range (see Fig.~\ref{Fig:wind}). As will be shown by Homan et al.\ (in prep), the flux contrast between the correlated arcs and bright inner bar can be used to deduce the density contrast between the smooth wind and spiral structure.

\section{Discussion}\label{Sec:Discussion}

\subsection{Morphological comparison between ALMA $^{13}$CO PV diagrams and the theoretical simulations} \label{Sec:Similarities}

\begin{figure*}
 \includegraphics[width=\textwidth]{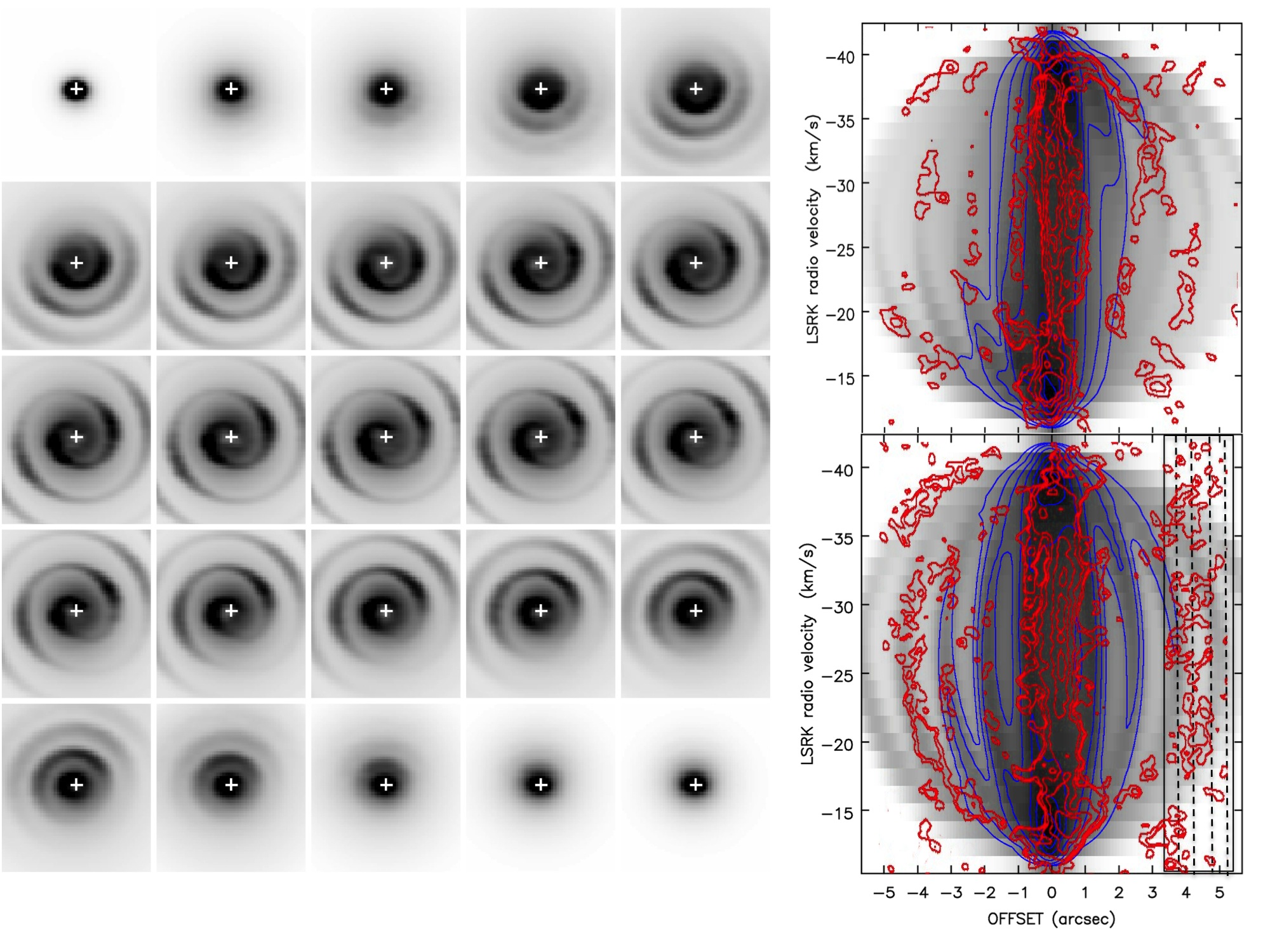}
 \caption{\textit{Left:} Simulated channel maps for a spiral wind structure with parameters $a=1.12$, $p=2$, $b=0$, $c=3\times 10^{12}$\,cm, $w=30\deg$, and $i=60$\deg, including a smooth wind structure with a spiral/wind density contrast of a factor 4. The central velocity of the panel at the top left is $-$16\,km/s (w.r.t. v$_{\rm{LSR}}$), at the bottom right corner +16\,km/s, hence $\Delta v$ is 1.33\,km/s for adjacent panels. The field-of-view is 5\arcsec$\times$5\arcsec. The ALMA instrumental effects are not taken into account in this simulation. \textit{Right:} Corresponding simulated PV diagrams (in grey-scale) for a slit along the declination axis (top) and along the right ascension axis (bottom). The contours of the simulated PV diagrams are shown in blue and for the ALMA data in red, each time at (0.05,0.1,0.2,0.4,0.6,0.8) times the maximum intensity. The shaded region in the bottom panel indicates the place where sidelobe effects deteriorate the quality of the ALMA PV data.}
 \label{Fig:comparison_Shape_ALMA}
\end{figure*}

With an on-source integration time of only 17\,minutes, an in-depth understanding of the $^{13}$CO J=6-5 channel maps (Fig.~\ref{Fig:13CO_channel}) is challenging. However, it is clear that the different PV diagrams (Fig.~\ref{Fig:PV_13COJ6} and Fig.~\ref{Fig:PV_13COJ6_other_angles}) display correlated structures which can not be interpreted with a simple spherical symmetric wind model. While the wind model with the density enhanced shell structures show some resemblance to the ALMA PV data, the asymmetry when reflecting around zero-offset  seen in the  ALMA PV diagrams  (Fig.~\ref{Fig:PV_13COJ6_other_angles}) can not be explained by this simple model, neither can it explain the difference in the PV diagrams with 90\deg\ difference in PA. A possible scenario to explain the structures seen in the ALMA PV diagrams is a binary-induced spiral structure. This interpretation is corroborated by other data and analyses already presented in the literature (see Sect.~\ref{Sec:qualitative}).  Using the simple 
mathematical 
prescription for a binary-induced spiral structure described in Sect.~\ref{Sec:Shape_models}, we 
derive from the ALMA PV diagrams at PA$\sim$10-20\deg\ that the spiral parameter $a$ is $\sim$1.1\arcsec. Different combinations for $p$, $b$, and $c$ give a good fit to the data, but the current limitations in the ALMA sensitivity of the current data and the limited UV-coverage inhibit a strict constraint. We therefore opt to use $p$=2 (mass conservation) and $b$=0 (the thickness of the spiral arms stays constant). A good fit to correlated structures in the ALMA PV data is obtained for the spiral parameters $a=1.12\arcsec$, $p=2$, $b=0$, $c=3\times 10^{12}$\,cm, and $w=30\deg$, and $i=60-80$\deg\ (see Fig.~\ref{Fig:comparison_Shape_ALMA}). The best constrained parameter is the spiral arm distance $a$, with an uncertainty of only $\sim$10\%. The general morphology is reproduced quite nicely, but the fit to the relative intensities can be improved. The latter is not unexpected taking into account the simple approach for modeling the spiral structure. An in-depth study describing the results of a 
large parameter grid, including a variation in radial temperature structure of the spiral, density contrast, etc.\ will be presented in Homan et al.\ (in prep).

It is clear that the {\sc Shape} simulations shown in Fig.~\ref{Fig:comparison_Shape_ALMA} do not capture the clumpiness seen in the ALMA data. Part of the fragmentation is artificial and comes from the poor UV coverage of the data (see Fig.~\ref{Fig:PV_radial_outflow}). However, part of the clumpiness seems also inherent to the wind structure of CW~Leo and indicates a non-smooth gas (and dust) density distribution within the spiral arms. Magnetic cool spots, photospheric convective motions, flow instabilities, etc.\ may result in localized gas and dust clumps (see the discussion in Sect.~\ref{Sec:qualitative}). The fact that we see the arc structure extending toward the orbital axis (see illustration in Fig.~\ref{Fig:sketch}) implies that  the reflex motion of the primary AGB star (CW~Leo itself) acts as a cause for the arc-like signature detected in the ALMA PV, Herschel and 
Hubble Space Telescope images.

\subsection{Orbital period and binary separation} \label{Sec:period}

The model fitting of the ALMA $^{13}$CO J=6-5 PV data indicates a spiral arm spacing of $a\sim$1.12\arcsec, which is in agreement with the arc spacing seen in the HST wide-V band image of the inner 10\arcsec\ region of CW~Leo \citep[see Fig.~3 in][]{Mauron2000A&A...359..707M}. 
The arm spacing is determined by the product of the binary orbital period $T_p$ and the pattern propagation speed in the orbital plane, which following \citet{Kim2012ApJ...759L..22K} is given by 
\begin{equation}
 \Delta r_{\rm{arm}} = \Big(\langle V_w \rangle + \frac{2}{3} V_p \Big) \times \frac{2 \pi r_p}{V_p}\,,
 \label{Eq:Tp}
\end{equation}
with $\langle V_w \rangle$ the wind velocity, $V_p$ the orbital velocity, and $r_p$ the orbital radius of the primary, and the orbital period $T_p$ given by the last term ($ 2 \pi r_p/V_p$). The first term in the right-hand part of Eq.~\ref{Eq:Tp}, $\langle V_w \rangle + \frac{2}{3} V_p$, is the pattern propagation speed throughout the orbital plane, which is close to the wind speed if $V_w$ dominates over the orbital and sound speeds \citep[which is the case for almost all AGB binary simulations, see][and \textit{Kim priv.\ comm.}]{Kim2012ApJ...759...59K}. At a distance of 1\arcsec, the wind has already reached its terminal velocity of 14.5\,km/s (see Sect.~\ref{Sec:vgas}). The derived arm spacing hence results in an orbital period of $\sim$55\,yr.

The derived orbital period is lower than the orbital periods derived for some other carbon-rich AGB stars, of which a binary companion is thought to be the cause of the detected spiral arm structure. \citet{Maercker2012Natur.490..232M} derived for R~Scl a period of 350\,yr on the basis of CO J=3-2 ALMA data, for CIT~6 \citet{Dinh2009ApJ...701..292D}  found a period of $\sim$600\,yr and for AFGL~3068 an orbital period of 830\,yr was obtained by \citet{Mauron2006A&A...452..257M}. The longer orbital periods for these AGB stars indicates that the companion must be on a wide orbit around the AGB star; typically around 50--70\,au. Using Kepler's third law, and assuming a primary initial mass of 4\,\Msun\ for CW~Leo \citep{Guelin1995A&A...297..183G} and the mass of the secondary being lower than that of the primary, the mean binary separation, $d$, for CW~Leo is $\sim$25$\pm$2\,au (or $\sim$8.2\,\Rstar=0.17\arcsec), i.e.\ a  binary system that can reside within the inferred dust lane of $\sim$0.5--1\arcsec\ radius (
see 
Fig.~\ref{Fig:sketch}). In the case that the primary has already lost some 2\,\Msun\ through its stellar wind \citep{Decin2011A&A...534A...1D}, the binary separation would reduce to $\sim$19\,au.

 \subsection{Companion mass} \label{Sec:companion_mass}

According to the method outlined by \citet{Kim2012ApJ...759L..22K}, one needs the (projected) separation between the primary star and its companion to derive the binary mass ratio. However, based on an extensive search throughout the literature, no direct identification of the potential binary companion was possible. We are hence lacking diagnostics to constrain the mass of the companion. However, the fact that the ALMA PV data around PA$\sim$10-20\deg\ might be explained by a spiral shock caused by the reflex motion of the mass-losing primary star implies that the mass of the secondary can not be very low, i.e. the secondary can not be a Jupiter-like planet or a brown dwarf. 
Although the simulations of \citet{Mastrodemos1999ApJ...523..357M} are not fine-tuned toward the specific situation of CW~Leo (i.e., the radius of CW~Leo is larger than the input radius for the model simulations, and the estimated mass of CW~Leo is around 4\,\Msun, while the input model mass for the primary, $M_p$, is 1.5\,\Msun), we can use their results as a guideline to estimate the mass-ratio,  $M_s/M_p$, with $M_s$ the mass of the secondary. Their model sequence M10$\rightarrow$M17$\rightarrow$M18 shows that for a decrease of the mass of the secondary (1\,\Msun$\rightarrow$0.5\,\Msun$\rightarrow$0.25\Msun) the morphology changes from bipolar$\rightarrow$elliptical$\rightarrow$quasi-spherical, i.e.\ the latter case resembling the signature in our ALMA data and multiple shells seen in the optical and infrared images \citep{Mauron2000A&A...359..707M, Decin2011A&A...534A...1D}. This would imply a mass ratio around $1/6$ or a mass for the secondary around 0.6\,\Msun. Using Eq.~(2) in \citet{
Eggleton1983ApJ...268..368E}, this implies that the effective radius of the Roche Lobe, $r_L$, is 0.54 times the orbital separation, or $r_L$ is $\sim$13.5\,au (4.5\,\Rstar).

An alternative method is offered by \citet{Huggins2009MNRAS.396.1805H}, who used the models of \citet{Mastrodemos1999ApJ...523..357M} to develop a prescription for the observed envelope shapes in terms of the binary parameters. The companion modifies the mass loss by gravitationally focusing the wind towards the orbital plane, and thereby determines the shape of the envelope at large distances from the star. They define the envelope shape parameter, $K_n$, as being the density contrast between the pole and the equator
\begin{equation}
 K_n - 1 = (n_{\rm{eq}} - n_{\rm{po}}) / n_{\rm{po}}\,
\end{equation}
with $n_{\rm{eq}}$ and $n_{\rm{po}}$ being the density at the equator and pole, respectively. Using the results of \citet{Mastrodemos1999ApJ...523..357M}, \citet{Huggins2009MNRAS.396.1805H} developed an empirical relation between the envelope shape parameter $K_n$ and the binary parameters, being the masses of the primary and secondary ($M_p$ and $M_s$), the binary separation ($d$) and the velocity of the wind at the orbit of the secondary ($V_s$). From the numerical simulations, they derive following equation (Eq.~(4) in their paper)
\begin{equation}
 \log{(K_n - 1 )} = 5.19(\pm0.66)+1.40(\pm0.21) \log{(M_s/V_s^2 d)}\,,
 \label{Eq:Kn_binary}
\end{equation}
independent of the mass of the primary $M_p$, and with $M_s$ in units of solar mass, $V_s$ in km/s and $d$ in astronomical units (au). Although the models are not finetuned toward the specific situation of CW~Leo (i.e., the radius of CW~Leo is larger than the input radius for the model simulations), we can use this relation and the value of $w$, characterizing the latitudinal dependence (see Sect.~\ref{Sec:Shape_PV_binaries}), to get a first crude estimate of the companion mass.

In Sect.~\ref{Sec:Similarities} we derive a best-fit value for $w$ of $\sim$30\deg, resulting in a value of $K_n$ around 10 (using Eq.~\ref{Eq:w}). Using Eq.~\ref{Eq:Kn_binary}, we derive that 
\begin{equation}
\frac{M_s}{V_s^2 d} = 0.8\times10^{-3}\,{\rm{\frac{\Msun}{(km/s)^{2}\,au}}}\,. 
\label{Eq_Kn_applied}
\end{equation}
For a binary separation $d$ of  20\,au, and the velocity of the wind at the orbit of the secondary, $V_s$, being $\sim$8\,km/s (see Fig.~\ref{Fig:vexp_LAS_sFWHM}) the derived mass for the secondary, $M_s$, is $\sim$1.1\,\Msun. However, the uncertainty on the estimated value for $M_s$ is significant due to the square dependence on the highly uncertain value of $V_s$ and the uncertainty of $\sim$15\deg\ in the value of $w$. 

The derived rough estimate for the companion mass puts the companion in the category of a white dwarf or an unevolved low-mass main sequence star. However, a binary with a white dwarf companion could result in a so-called 'dusty symbiotic system'. This kind of system harbours a very hot and active accretion disk, with clear signatures in the UV and the presence of many forbidden atomic lines of, e.g., [O\,III], [Ne\,III], and [Fe\,IV] in the optical spectrum. Since no signs of activity are seen in the optical/UV spectra of CW~Leo, we tentatively postulate that the companion of CW~Leo is an unevolved M-type dwarf.

\subsection{Other scenarios}
\begin{figure}[htp]
 \begin{minipage}[b]{.28\textwidth}
 \vspace*{-.3cm}
  \centerline{ \includegraphics[width=\textwidth]{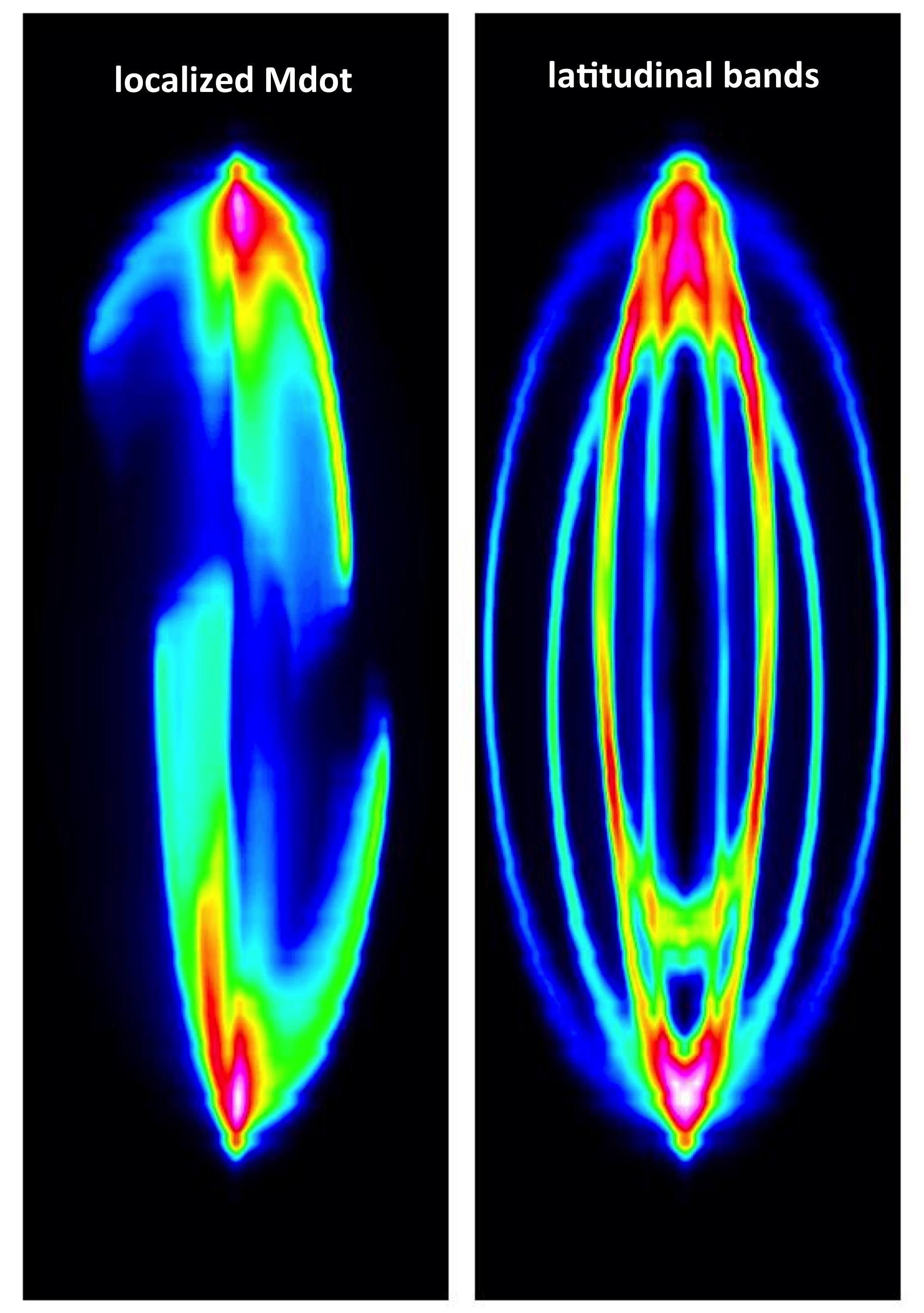}}
  \end{minipage}
  \hfill
 \begin{minipage}[b]{.2\textwidth}
 \centerline{\includegraphics[height=5ex]{blanco.pdf}}
 \caption{PV diagrams for enhanced mass-loss ejections at PA=90\deg. \textit{Left:} Simulations for localized mass-loss ejections covering $\sim$10\% of the surface at an interval of 1\arcsec \citep{Dinh2008ApJ...678..303D}.  \textit{Right:} Simulations for latitudinal bands according to the model of \citet{LeBertre1989Msngr..55...25L}.}
 \label{Fig:PV_extra}
 \end{minipage}
\end{figure}

One might wonder if localized mass-loss ejections radially driven outward, without the action of a binary companion, might result in the type of PV diagrams seen in the ALMA data. \citet{Dinh2008ApJ...678..303D} suggested that the shells seen in the VLA data of HC$_3$N and HC$_5$N at a distance of $\sim$15\arcsec\ from the central star cover some 10\% of the stellar surface at the time of ejection. One can simulate this scenario using the simple parametric description for density enhanced shells (Sect.~\ref{Sec:PV_radial}), but constraining randomly the longitudinal and latitudinal coordinate for each of the five shells so that each segment covers only $\sim$10\% of the surface. This kind of suggestion gives one a lot of (parametric) freedom to model the wind of CW~Leo. An example of such a PV diagram is shown in the left panel of Fig.~\ref{Fig:PV_extra}. While this example shows some resemblance with the ALMA PV diagrams, this scenario would need a lot of finetuning to get a model showing almost symmetrical 
signatures as seen in the ALMA PV diagram for PA$\sim$10--20\deg\ (Fig.~\ref{Fig:PV_13COJ6_other_angles}), but for which almost no structure is left in a PV diagram taken at a PA 90\deg\ apart.

A way to constrain the huge parameter space entailed by the model for the localized mass-loss ejections is offered by the model presented by \citet{LeBertre1989Msngr..55...25L}, who suggested that ejected material also moves due to the stellar rotation, creating latitudinal bands. This kind of model was recently used by \citet{Jeffers2014} to explain the ExPo images of CW~Leo. The {\sc Shape} simulations for this scenario are also based on the model for the density enhanced shells (Sect.~\ref{Sec:PV_radial}), but the latitudinal coordinates of each shell only cover $\sim$5\deg.
These simulations do not rule out the ejection of clumps with initial extent $\sim$10\% of the
   stellar surface, nor the possible effects of stellar rotation, but
   the addition of the effects of the companion in creating a spiral
   provides a more robust explanation for the asymmetry in the PV diagrams when reflecting around zero-offset, for the differences seen in the PV diagrams 90\deg\ apart and for the other diagnostics discussed in Sect.~\ref{Sec:qualitative} and shown in Fig.~\ref{Fig:sketch}.

\subsection{Further evolution}

There are numerous similarities between CW~Leo and the young post-AGB star CRL\,2688 \citep[also called the Egg Nebula;][]{Sahai1998ApJ...493..301S}. Both objects are carbon-rich, suggesting an initial mass larger than 3\,\Msun, Although CRL\,2688 is further in its evolution, both objects have an optically thick circumstellar core with biconical cavities and shell-like density enhancements are observed in scattered light as incomplete arcs. In CRL\,2688 high-speed collimated jets are present with velocities around 300\,km/s,  which seem to be only powered during the last 200\,years. According to \citet{Soker1994ApJ...421..219S}, different mechanisms can explain the presence of the collimated outflows in CRL\,2688, each of them requiring the central star to be a binary, with the collimated outflow resulting from an accretion disk. The discovery of a spiral arm structure in the ALMA data of CW~Leo, which might be induced by the presence of a binary companion, supports the suggestion that CW~Leo will evolve 
into 
an 
object similar to CRL\,2688.

 \section{Conclusions} \label{Sec:conclusions}
 
 We have presented the first ALMA band~9 data of the inner wind region of CW~Leo at a spatial resolution of 0\farcs42$\times$0\farcs24 and with a sensitivity is 0.2--0.3\,Jy beam$^{-1}$ per 488\,kHz channel. We have detected 25 emission lines, all of them centered on the dust continuum peak which has a total (star+dust) flux density within the 3$\sigma$ contour of 5.66\,Jy. The images prove that the vibrational lines are excited just above the stellar photosphere and that SiO is a parent molecule, formed close to the stellar surface, probably due to shock-induced non-equilibrium chemistry.
 The velocity traced by the line width of the emission lines suggests a steep increase of the wind velocity starting around 5\,\Rstar\ reaching almost the terminal velocity at $\sim$11\,\Rstar.
 
 Both the dust emission and the emission of the brightest lines show a clear asymmetric distribution. The position-velocity (PV) maps at different position angles of the $^{13}$CO J=6-5 line display correlated arc-like structures, which can be explained by a spiral arm. This spiral arm can be caused by the presence of a binary companion. Using the {\sc Shape} modeling tool we have modeled the $^{13}$CO J=6-5 PV diagrams. We deduce that the orbital axis lies at a position angle of $\sim$10--20\deg\ to the North-East  and that the spiral arm spacing is $\sim$1.12\arcsec. At a distance of 150\,pc, this leads to an orbital period of 55\,yr and a binary separation of $\sim$20-25\,au (or $\sim$6--8\,\Rstar). We tentatively suggest that the companion is an unevolved low mass main sequence star.
 
 The scenario of a binary system can explain (1)\,the spiral arm structure seen in the ALMA PV data, (2)\,the signature of a bipolar structure seen at arcsecond scales, and (3)\,the presence of multiple non-concentric shells detected in the outer wind, which might represent the limb-brightened edges of the spiral arms seen almost edge-on. Granted ALMA cycle~1 time will increase the signal-to-noise and UV-coverage, hence facilitating the in-depth interpretation of the ALMA data.

-------------------------------------------------------------------
\begin{acknowledgements}
 This paper makes use of the following ALMA data: 
ADS/JAO.ALMA\#2011.0.00277.S. ALMA is a partnership of ESO (representing 
its member states), NSF (USA) and NINS (Japan), together with NRC 
(Canada) and NSC and ASIAA (Taiwan), in cooperation with the Republic of 
Chile. The Joint ALMA Observatory is operated by ESO, AUI/NRAO and NAOJ.
W.S.\ acknowledges support by grant UNAM-PAPIIT 101014. The authors thank Nicholas Koning and Miguel Santander-Garc\'{\i}a for technical support with {\sc Shape}.
\end{acknowledgements}

\bibliographystyle{aa}
\bibliography{article_ALMA_CWLeo_printer}

 \Online
 \begin{appendix}

  \section{Position-velocity diagrams} \label{Sec:PV_appendix}
  Note: Appendix to be found in main paper.

\end{appendix}

\end{document}